\documentclass[12pt,a4paper]{article}
\pdfoutput=1
\usepackage[affil-sl,auth-sc]{authblk}
\usepackage{amsmath}
\usepackage{amssymb}
\usepackage{slashed}
\usepackage{cite}
\usepackage{graphicx}
\usepackage[hyperfootnotes=false,linktocpage=true]{hyperref}
\usepackage{color}
\usepackage{url}
\usepackage[utf8]{inputenc}
\usepackage[nottoc]{tocbibind} 

\evensidemargin=\oddsidemargin

\newcommand{\bea}{\begin{eqnarray}}
\newcommand{\eea}{\end{eqnarray}}
\def\beq{\begin{equation}}
\def\eeq{\end{equation}}
\newcommand{\no}{\nonumber}
\newcommand{\nn}{\nonumber\\}

\newcommand{\eq}[1]{Eq.~\eqref{#1}}

\newcommand{\Br}{\text{Br}}

\newcommand{\re}{\text{\,Re\,}}
\newcommand{\im}{\text{\,Im\,}}
 
\newcommand{\ov}{\overline}
\newcommand{\gev}{\ensuremath{\,\mbox{GeV}}}

\newcommand{\gsim}{\lower.7ex\hbox{$\;\stackrel{\textstyle>}{\sim}\;$}}
\newcommand{\lsim}{\lower.7ex\hbox{$\;\stackrel{\textstyle<}{\sim}\;$}}
\newcommand{\bra}[1]{\ensuremath{\langle #1 |}}
\newcommand{\ket}[1]{\ensuremath{| #1 \rangle }}

\numberwithin{equation}{section}

\begin{document}

\title{Lepton flavour violation in the MSSM: exact diagonalization vs
  mass expansion}

\author[1]{Andreas Crivellin}
%

\author[2]{Zofia Fabisiewicz}
\author[2]{Weronika Materkowska}

\author[3]{Ulrich Nierste}

\author[2]{Stefan Pokorski}

\author[2]{Janusz Rosiek}

\affil[1]{Paul Scherrer Institut, CH--5232 Villigen PSI, Switzerland}
\affil[2]{Faculty of Physics, University of Warsaw, Pasteura
  5,\protect\linebreak 02-093 Warsaw, Poland}
\affil[3]{Institut f\"ur Theoretische Teilchenphysik, Karlsruhe
  Institute of Technology, 76128 Karlsruhe, Germany} 

\date{February 20, 2018}
  
\maketitle


\begin{abstract}
The forthcoming precision data on lepton flavour violating (LFV)
decays require precise and efficient calculations in New Physics
models. In this article lepton flavour violating processes within the
Minimal Supersymmetric Standard Model (MSSM) are calculated using the
method based on the Flavour Expansion Theorem, a recently developed
technique performing a purely algebraic mass-insertion expansion of
the amplitudes. The expansion in both flavour-violating and
flavour-conserving off-diagonal terms of sfermion and supersymmetric
fermion mass matrices is considered. In this way the relevant
processes are expressed directly in terms of the parameters of the
MSSM Lagrangian.  We also study the decoupling properties of the
amplitudes.  The results are compared to the corresponding
calculations in the mass eigenbasis (i.e. using the exact
diagonalization of the mass matrices).  Using these methods, we
consider the following processes: $\ell \to \ell' \gamma$, $\ell \to 3
\ell'$, $\ell \to 2\ell'\ell''$, $h \to \ell\ell'$ as well as $\mu \to
e$ conversion in nuclei.  In the numerical analysis we update the
bounds on the flavour changing parameters of the MSSM and examine the
sensitivity to the forthcoming experimental results. We find that
flavour violating muon decays provide the most stringent bounds on
supersymmetric effects and will continue to do so in the future.
Radiative $\ell\to\ell^\prime\gamma$ decays and leptonic three-body
decays $\ell\to3\ell^\prime$ show an interesting complementarity in
eliminating "blind spots" in the parameter space. In our analysis we
also include the effects of non-holomorphic $A$-terms which are
important for the study of LFV Higgs decays.
\vskip -22.5cm \hfill {\normalsize TTP18-011, PSI-PR-18-04 }
\end{abstract}

\newpage

\tableofcontents

\newpage


\section{Introduction}
\label{sec:intro}

So far, the LHC did not observe any particles beyond those of the
Standard Model (SM). Complementary to direct high energy searches at
the LHC, there is a continuous effort in indirect searches for new
physics (NP). In this respect, a promising approach is the search for
processes which are absent -- or extremely suppressed -- in the SM
such as lepton flavour violation (LFV) which is forbidden in the SM in
the limit of vanishing neutrino masses. The experimental sensitivity
for rare LFV processes such as $\ell\rightarrow \ell^{\prime}\gamma$,
$\mu \to e$ conversion in nuclei and $\ell\rightarrow \ell^{\prime}
\mu^+\mu^-$ or $\ell\rightarrow \ell^{\prime} e^+e^-$ will improve
significantly in the near future, probing scales well beyond those
accessible at foreseeable colliders.  Furthermore, the discovery of
the 125 \gev{} Higgs boson $h$ \cite{Aad:2012tfa,Chatrchyan:2012ufa}
has triggered an enormous experimental effort in measuring its
properties, including studies of its LFV decays. The most recent
experimental limits {on the LFV processes} are given in
Table~\ref{tab:leplim} {in Sec.~\ref{sec:pheno}}.

Many studies of LFV processes within the MSSM (and possible extensions
of it) exist (see e.g. Refs.\cite{Borzumati:1986qx, Casas:2001sr,
  Masina:2002mv, Brignole:2004ah, Paradisi:2005fk, Fukuyama:2005bh,
  Paradisi:2005tk, Dedes:2006ni, Dedes:2007ef, Antusch:2007dj,
  Arganda:2008jj, Ilakovac:2009jf, Calibbi:2009ja,
  Altmannshofer:2009ne, Calibbi:2006nq, Calibbi:2009wk, Hisano:2009ae,
  Girrbach:2009uy, Biggio:2010me, Esteves:2010ff, Ilakovac:2012sh,
  Arana-Catania:2013ggc, Goto:2014vga, Abada:2014kba, Vicente:2015cka,
  Bonilla:2016fqd, Lindner:2016bgg} and Ref.~\cite{Calibbi:2017uvl}
for a recent review).  In this article we revisit this subject in the
light of the new calculational methods which have been recently
developed~\cite{Dedes:2015twa, Rosiek:2015jua}. These methods allow
for a systematic expansion of the amplitudes of the LFV processes in
terms of mass insertions (MI), i.e. in terms of off-diagonal elements
of the mass matrices.  We show that a transparent qualitative
behaviour of the amplitudes of the LFV processes is obtained by
expanding them not only in the flavour-violating off-diagonal terms in
the sfermion mass matrices but also in the flavour conserving but
chirality violating entries related to the tri-linear $A$-terms as
well as in the off-diagonal terms of the gaugino and higgsino mass
matrices. This procedure is useful because in the MI approximation we
work directly with the parameters of the Lagrangian and can therefore
easily put experimental bounds on them.  We compare the results of the
calculations performed in the mass eigenbasis (i.e. using a numerical
diagonalization of the slepton mass matrices) with those obtained at
leading non-vanishing order of the MI approximation, in different
regions of the supersymmetric parameter space and considering various
decoupling limits.  Of course, the MI
approximation~\cite{Gabbiani:1996hi, Misiak:1997ei} has already been
explored for many years as a very useful tool in flavour physics.
However, a detailed comparison between the full calculation and the MI
approximation is still lacking, partly because a fully systematic
discussion of the MI approximation~\cite{Dedes:2015twa} to any order
and the technical tools facilitating it~\cite{Rosiek:2015jua} have not
been available until recently.

Concerning the phenomenology, we summarise and update the bounds on
the flavour violating SUSY parameters, show their complementarity and
examine the impact of the anticipated increase in the experimental
sensitivity.  We investigate in detail the decay $h \to \mu \tau$
showing the results in various decoupling limits and analyse the role
of the so-called non-holomorphic $A$-terms~\cite{deWit:1996kc,
  Matone:1996bj, Bellisai:1997ck, Dine:1997nq, ArkaniHamed:1998wc,
  GonzalezRey:1998gz, Buchbinder:1998qd, Martin:1999hc} , which are
usually neglected in literature.
We also avoid simplifying assumptions on the sparticle spectrum and
assume neither degeneracies nor hierarchies among the supersymmetric
particles.

This article is structured as follows: in Sec.~\ref{sec:efflag} we
establish our conventions and present the results for the 2-point,
3-point, and 4-point functions related to flavour violating charged
lepton interactions in the mass eigenbasis, i.e. expressed in terms of
rotation matrices and physical masses.  Sec.~\ref{sec:lfv} contains
the formula for the decay rates of the processes under investigation.
In Sec.~\ref{sec:memi} we discuss the MI expansion and summarise
important properties of the decoupling limits $M_{\rm SUSY}\to\infty$
and $M_A\to \infty$.  In Sec.~\ref{sec:pheno} we present the numerical
bounds on LFV parameters obtained from current experimental
measurements and discuss the dependence of the results on the SUSY
spectrum.  We also discuss the correlations between the radiative
decays and the 3-body decays of charged lepton as well as the
non-decoupling effects in LFV neutral Higgs decays.  Finally we
conclude in Sec.~\ref{sec:conclusion}.  All required Feynman rules
used in our calculations are collected in appendix~\ref{app:lagr}. The
definitions of loop integrals can be found in appendix~\ref{app:loop}.
In appendix~\ref{app:ddiff} we explain the notation for the ``divided
differences'' of the loop functions used in the expanded form of the
amplitudes. The expression for the 4-lepton box diagrams and for the
MI-expanded expression of the amplitudes are given in the
appendices~\ref{app:fullbox} and ~\ref{app:miexp}, respectively.

\section{Effective LFV interactions}
\label{sec:efflag}

In this Section we collect the analytical formula in the mass
eigenbasis for flavour violating interactions generated at the
one-loop level\footnote{Note that these expressions are not valid in
  the flavour conserving case where additional terms should be
  included and renormalization is required.}.  We use the notation and
conventions for the MSSM as given in Ref.~\cite{Rosiek:1989rs,
  Rosiek:1995kg}\footnote{The conventions of~\cite{Rosiek:1989rs,
    Rosiek:1995kg} are very similar to the later introduced and now
  widely accepted SLHA2~\cite{Allanach:2008qq} notation, up to the
  minor differences summarised in the Appendix~\ref{app:lagr}.}.

In our analysis, we include the so-called non-holomorphic trilinear
soft SUSY breaking terms:
\bea
L_{nh} = \sum_{I,J=1}^3\sum_{i=1}^2 \left(A_l^{'IJ} H_i^{2\star} L_i^I
R^J+ A_d^{'IJ} H_i^{2\star} Q_i^I D^J + A_u^{'IJ} H_i^{1\star} Q_i^I
U^J + \mathrm{H.c.} \right)\,,
\eea
which couple up(down)-sfermions to the down(up)-type Higgs doublets.
Here, {as throughout the rest of the paper}, capital letters
$I,J=1,2,3$ denote flavour indices and the small letters $i=1,2$ are
$SU(2)_L$ indices.

\subsection{$\gamma-\ell-\ell^\prime$ interactions}

We define the effective Lagrangian for flavour violating couplings of
leptons to on-shell photons as
\bea
L_{\ell \gamma }^{} = - e\sum\limits_{I,J} {\left(
  {F_{\gamma}^{JI}{{\bar \ell }^J}{\sigma _{\mu \nu }}{P_L}{\ell ^I} +
    F_{\gamma}^{IJ*}{{\bar \ell }^J}{\sigma _{\mu \nu }}{P_R}{\ell
      ^I}} \right)} {F^{\mu \nu }}\,,
\label{eq:lgdef}
\eea

\begin{figure}[tbp]
\begin{center}
\includegraphics[width=0.8\textwidth]{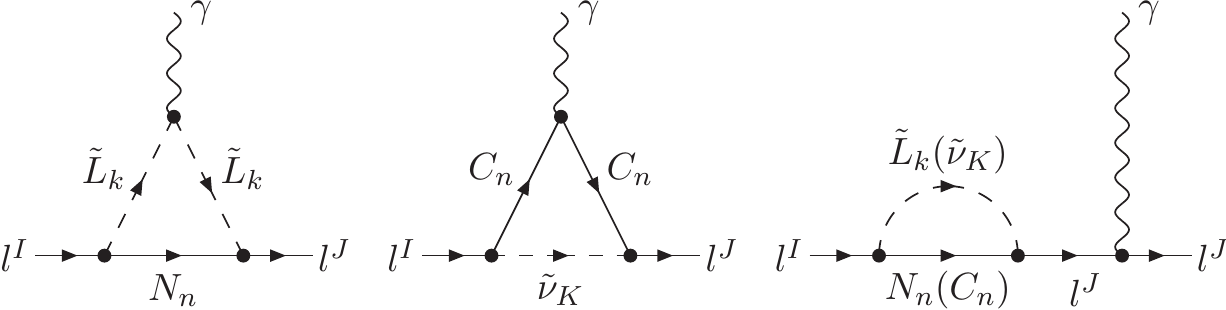}
\caption{\small One-loop supersymmetric contributions to the LF
  violating effective lepton-photon interaction (mirror-reflected
  self-energy diagram not shown).\label{fig:llg}}
\end{center}
\end{figure}

The SM contribution to $F_{\gamma}^{JI}$ is suppressed by powers of
$m_{\nu}^2/M_W^2$ and thus completely negligible. In the mass
eigenbasis the supersymmetric contributions to $F_{\gamma}^{JI}$ come
from the diagrams displayed in Fig.~\ref{fig:llg}. Let us decompose
$F_{\gamma}$ in the following way
\bea
F_{\gamma}^{JI} = F_{\gamma A}^{JI} - m_J F_{\gamma L B}^{JI} - m_I
F_{\gamma R B}^{JI}\,,
\eea
with
\bea
(4\pi)^2 F_{\gamma A}^{JI} &=& \sum_{K=1}^3 \sum_{n=1}^2
V_{\ell\tilde\nu C, R}^{JKn*} V_{\ell\tilde\nu C, L}^{IKn} \; m_{C_n}
C_{11}(m_{C_n},m_{\tilde\nu_K}) \nn
&-&\frac{1}{2} \sum_{k=1}^6\sum_{n=1}^4 V_{\ell \tilde L N,R}^{Jkn*}
V_{\ell \tilde L N,L}^{Ikn} \; m_{N_n} C_{12}(m_{\tilde
  L_k},m_{N_n})\,,\nonumber\\[3mm]
(4\pi)^2 F_{\gamma L B}^{JI} &=& - \sum_{K=1}^3 \sum_{n=1}^2
V_{\ell\tilde\nu C, L}^{JKn*} V_{\ell\tilde\nu C, L}^{IKn} \;
C_{23}(m_{C_n},m_{\tilde\nu_K}) \nn
&+& \frac{1}{2} \sum_{k=1}^6\sum_{n=1}^4 V_{\ell \tilde L N,L}^{Jkn*}
V_{\ell \tilde L N,L}^{Ikn} C_{23}(m_{\tilde L_k},m_{N_n})\,.
\label{eq:fgamab}
\eea
Here, $V$ abbreviates the tree-level lepton-slepton-neutrino and
lepton-sneutrino-chargino vertices, i.e. the subscripts of $V$ stand
for the interacting particles and the chirality of the lepton
involved. The super-scripts refer to the lepton or slepton flavour as
well as to the chargino and neutralino involved.  The specific form of
the chargino and neutralino vertices $V_{L(R)}$ is defined in
Appendix~\ref{app:lagr} and the 3-point loop functions $C_{ij}$ are
given in Appendix~\ref{app:loop}. $F_{\gamma A}\, (F_{\gamma L B}$)
denotes the parts of the amplitude which is (not) proportional to the
masses of fermions exchanged in the loop. $F_{\gamma R B}$ can be
obtained from $F_{\gamma L B}$ by exchanging $L\leftrightarrow R$ on
the RHS of~\eq{eq:fgamab}.

Gauge invariance requires that LFV (axial) vectorial photon couplings
vanish for on-shell external particles. However, off-shell photon
contributions are necessary to calculate three body decays of charged
leptons. The vectorial part of the amplitude for the $\gamma\ell\ell'$
vertex can be written as
\bea
i A_{\gamma}^{JI\;\mu} = ie q^2 \bar u_J (p_J) \left(\Gamma_{\gamma
  L}^{JI} P_L + \Gamma_{\gamma R}^{JI} P_R \right) \gamma^{\mu}
u_I(p_I)\;,
\label{eq:phvect}
\eea
where $q=p_I-p_J$ and $\Gamma_{\gamma L}^{JI}$ is at the leading order
in $p^2/M_{\rm SUSY}^2$ momentum independent and reads
\bea
\Gamma_{\gamma L}^{JI} &=& \sum_{K=1}^3 \sum_{n=1}^2 V_{\ell\tilde\nu C,
  L}^{JKn*} V_{\ell\tilde\nu C, L}^{IKn} \;
C_{01}(m_{C_n},m_{\tilde\nu_K}) \nn &-& \sum_{k=1}^6\sum_{n=1}^4
V_{\ell \tilde L N,L}^{Jkn*} V_{\ell \tilde L N,L}^{Ikn} \;
C_{02}(m_{N_n},m_{\tilde L_k})\,.
\label{eq:llgv}
\eea
$\Gamma_{\gamma R}^{JI}$ can be obtained by replacing
$L\leftrightarrow R$. Again, the loop functions $C_{01},C_{02}$ are
defined in Appendix~\ref{app:loop}.

Finally, one should note that for heavy MSSM spectrum the 2-loop
Barr-Zee diagrams~\cite{Barr:1990vd} involving the non-decoupling LFV
Higgs interactions (see Sec.~\ref{sec:hllndec}) are important and have
to be included~\cite{Chang:1993kw,Hisano:2006mj, Jung:2013hka,
  Abe:2013qla, Ilisie:2015tra,Crivellin:2015hha}.

\subsection{$Z-\ell-\ell^\prime$ interactions}

\begin{figure}[tb]
\begin{center}
\includegraphics[width=0.8\textwidth]{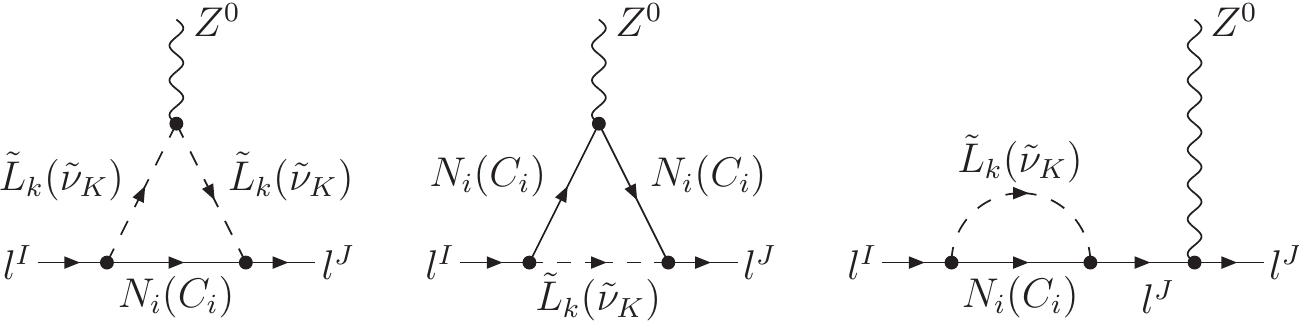}
\caption{\small One-loop supersymmetric contributions to the LFV
  effective lepton-$Z^0$ interaction (the mirror-reflected self-energy
  diagram not shown).\label{fig:zll}}
\end{center}
\end{figure}

In order to calculate the three body decays of charged leptons as to
be considered in Sec.~\ref{sec:llll} it is sufficient to calculate the
effective $Z-\ell-\ell^\prime$ interactions in the limit of vanishing
external momenta. The Wilson coefficients of the effective Lagrangian
for the $Z$ coupling to charged leptons are generated at one-loop
level by the diagrams shown in Fig.~\ref{fig:zll} and can be written
as
\bea
L_{\ell Z}^{JI} = \left(F_{ZL}^{JI}\bar\ell^J\gamma_\mu P_L \ell^I +
F_{Z R}^{JI}\bar\ell^J\gamma_\mu P_R \ell^I\right) Z^\mu\,,
\label{eq:lzdef}
\eea
with
\bea
F_{ZL}^{JI} & = & \Gamma_{ZL}^{JI} - \frac{e(1 - 2 s_W^2)}{2 s_W
  c_W}\; \Sigma_{VL}^{JI}(0)\,,\nn
F_{ZR}^{JI} & = & \Gamma_{ZR}^{JI} + \frac{e s_W}{c_W}\;
\Sigma_{VR}^{JI}(0)\,.
\eea
Here, $\Gamma_{ZL(R)}$ denote the contribution originating from the
one-particle irreducible (1PI) vertex diagram and $\Sigma_{VL(R)}$ is
the left-(right-)handed part of the lepton self-energy defined as
\bea
\Sigma^{JI}(p^2) = \Sigma_{VL}^{JI}(p^2)\, \slashed{p}\,
P_L+\Sigma_{VR}^{JI}(p^2) \,\slashed{p}\, P_R+\Sigma_{mL}^{JI}(p^2)
\,P_L+\Sigma_{mR}^{JI}(p^2) \, P_R\;.
\label{eq:sigdef}
\eea

Contrary to the left- and right-handed magnetic photon-lepton
couplings, which change chirality, the $Z\bar\ell^I\ell^J$ coupling is
chirality conserving. Therefore, the Wilson coefficients of the
left-handed and right-handed couplings are not related to each other
but rather satisfy $F_{Z L(R)}^{IJ} = F_{Z L(R)}^{JI*}$.  In the mass
eigenbasis the vectorial part of the lepton self-energy and the 1PI
triangle diagrams are given by (see Appendix~\ref{app:lagr} for
definitions of vertices $V$)
\bea
(4\pi)^2 \Sigma_{VL}^{JI}(p^2) &=& \sum_{i=1}^2\sum_{K=1}^3
V_{\ell\tilde\nu C,L}^{IKi} V_{\ell\tilde\nu C,L}^{JKi\,*}
B_1(p,m_{\tilde\nu_K},m_{C_i})\nn &+& \sum_{i=1}^4\sum_{j=1}^6 V_{\ell
  \tilde L N,L}^{Iji} V_{\ell \tilde L N,L}^{Jji\,*}
B_1(p,m_{L_j},m_{N_i})\,,
\label{eq:sigmaV}
\eea

\bea
(4\pi)^2 \Gamma_{ZL}^{JI} &=& \frac{1}{2} \sum_{i,j=1}^2\sum_{K=1}^3
V_{\ell\tilde\nu C,L}^{IKi} V_{\ell\tilde\nu C,L}^{JKj*} \, \left(
V_{CCZ,L}^{ij} C_2(m_{\tilde\nu_K},m_{C_i},m_{C_j})\right.\nn
&-&\left.  2 V_{CCZ,R}^{ij} m_{C_i} m_{C_j}
C_0(m_{\tilde\nu_K},m_{C_i},m_{C_j}) \right)\nn
&+&\frac{e}{4 s_W c_W} \sum_{i=1}^2\sum_{K=1}^3 V_{\ell\tilde\nu
  C,L}^{IKi} V_{\ell\tilde\nu C,L}^{JKi*} \,
C_2(m_{\tilde\nu_K},m_{\tilde\nu_K},m_{C_i}) \nn
&+& \frac{1}{2} \sum_{j=1}^6\sum_{i,k=1}^4 V_{\ell \tilde L N,L}^{Iji}
V_{\ell \tilde L N,L}^{Jjk\,*}\,\left(V_{NNZ,L}^{ik}
C_2(m_{L_j},m_{N_i},m_{N_k}) \right. \nn
&-&\left. 2 V_{NNZ,R}^{ik} m_{N_i} m_{N_k}
C_0(m_{L_j},m_{N_i},m_{N_k})\right)\nn
&-& \frac{1}{2} \sum_{j,k=1}^6\sum_{i=1}^4 V_{\ell \tilde L N,L}^{Iji}
V_{\ell \tilde L N,L}^{Jki\,*}\,V_{LLZ}^{jk}
C_2(m_{L_j},m_{L_k},m_{N_i})\,,
\label{eq:VZ}
\eea
at vanishing external momenta with obvious replacements
$L\leftrightarrow R$ for $\Sigma_{VR}^{JI},\, \Gamma_{ZR}^{JI}$.

\subsection{LFV Higgs interactions}

To compactify the notation, we denote the CP-even Higgs boson decays
by $H_0^K\to \bar\ell^I \ell^J$, where, following again the notation
of~\cite{Rosiek:1989rs,Rosiek:1995kg}, $H\equiv H_0^1, h\equiv
H_0^2$. As usual, we denote CP-odd neutral Higgs boson by $A_0$.

In order to study $h\to\ell\ell^\prime$ decays precisely, we keep the
terms depending on the external Higgs mass. Therefore, we assume the
following effective action governing the LFV Higgs-lepton interaction:
\bea
A_{H{\rm eff}}^{\ell} &=& \bar \ell^J(k_J) (F_{h\ell}^{JIK}(k_J,k_I)
P_L + F_{h\ell}^{IJK*}(k_J,k_I) P_R )\ell^I(k_I) H_0^K(k_I - k_J)\nn
& +& \bar \ell^J(k_J) (F_{A\ell}^{JI}(k_J,k_I) P_L +
F_{A\ell}^{IJ*}(k_J,k_I) P_R )\ell^I(k_I) A_0(k_I - k_J) \,.
\eea
In addition, to calculate the $\mu\to e$ conversion rate one needs to
include the effective Higgs-quark couplings. For this purpose, one can
set all external momenta to zero and consider the effective Lagrangian
\bea
L_{H{\rm eff}}^q &=& \bar u^J (F_{hu}^{JIK} P_L + F_{hu}^{IJK*} P_R
)u^I H_0^K + \bar d^J (F_{hd}^{JIK} P_L + F_{hd}^{IJK*} P_R ) d^I
H_0^K\,.
\label{eq:lhdef}
\eea
However, in this article we consider only the lepton sector and
therefore do not give the explicit forms of Higgs quark couplings. The
relevant 1-loop expressions in the same notation as used in the
current paper are given in Ref.~\cite{Buras:2002vd} and the formulae
that take into account also non-decoupling chirally enhanced
corrections and 2-loop QCD corrections in the general MSSM can be
found in Refs.~\cite{Crivellin:2010er, Crivellin:2011jt,
  Crivellin:2012zz}\footnote{Earlier accounts on chiral resummation
  can be found in Refs.~\cite{Hall:1993gn, Carena:1994bv,
    Carena:1999py, Bobeth:2001sq, Babu:1999hn, Isidori:2001fv,
    Dedes:2002er,Hofer:2009xb,Noth:2010jy}}.

\begin{figure}[tb]
\begin{center}
\includegraphics[width=0.8\textwidth]{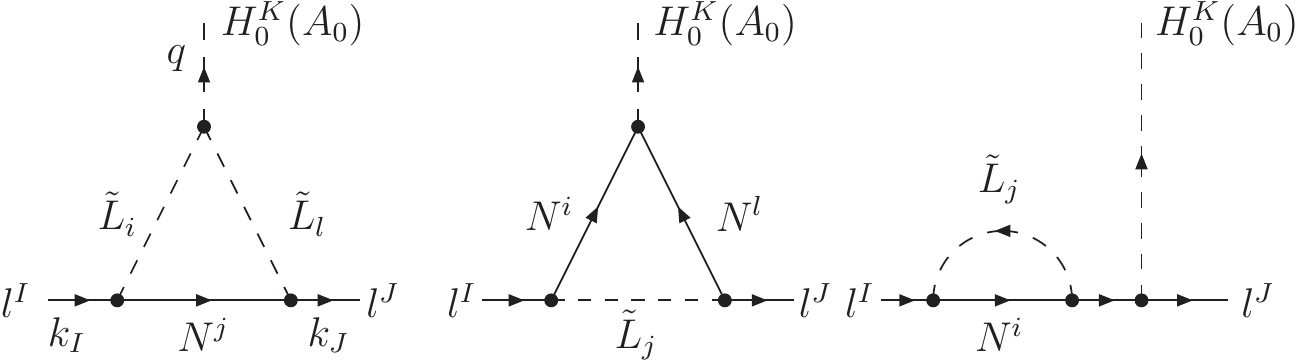}
\caption{\small Slepton-neutralino diagrams contributing to the
  $H_0^K\to \ell^I \bar \ell^J$ {and $A_0\to \ell^I \bar \ell^J$}
  decays in the MSSM (the mirror-reflected self-energy diagram is
  omitted).\label{fig:diag}}
\end{center}
\end{figure}

At the 1-loop level there are eight diagrams contributing to the
effective lepton Yukawa couplings.  The ones with slepton and
neutralino exchange are displayed in Fig.~\ref{fig:diag}, while
diagrams with the chargino exchange can be obtained by the obvious
replacements $N\to C, L\to\tilde\nu$.

The expressions for $F_h$ are obtained from 1PI triangle diagrams and
the scalar part of lepton self-energies (see \eq{eq:sigdef}) while the
chirality conserving parts of the self-energies are absorbed by a
field rotation required to go to the physical basis with a diagonal
lepton mass matrix. Therefore,
\bea
F_h^{JIK}(k_J,k_I) &=& \Gamma_{h}^{JIK}(k_J,k_I) -
\frac{Z_R^{1K}}{v_1} \,\Sigma_{mL}^{JI}(0)\;,\nn
F_A^{JI}(k_J,k_I) &=& \Gamma_{A}^{JI}(k_J,k_I) - \frac{i
  \sin\beta}{v_1} \,\Sigma_{mL}^{JI}(0)\;,
\label{eq:yukeff}
\eea
where the $Z_R$ denotes the CP-even Higgs mixing matrix (see
Appendix~\ref{app:lagr}) and the scalar self-energy contributions are
evaluated at zero momentum transfer and given by:
\bea
(4\pi)^2 \Sigma_{mL}^{JI}(0) & = & \sum_{i=1}^2\sum_{L=1}^3 m_{C_i}
V_{\ell\tilde\nu C,L}^{ILi} V_{\ell\tilde\nu C,R}^{JLi\,*}
\,\,B_0\,(0,m_{\tilde\nu_L},m_{C_i})\nn
&+& \sum_{i=1}^4\sum_{j=1}^6 m_{N_i} V_{\ell \tilde L N,L}^{Iji}
V_{\ell \tilde L N,R}^{Jji\,*}\,\,B_0\,(0,m_{L_j},m_{N_i})
\label{eq:sigmaLR}
\eea
The neutralino-slepton contributions to the 1PI vertex diagrams can be
written as (the symbols in square brackets denote common arguments of
the 3-point functions)\footnote{As we shall see later using MI
  expanded formulae (see Appendix~\ref{app:lhmi}), due to strong
  cancellations the leading order terms in Eqs.~(\ref{eq:sigmaLR},
  \ref{eq:hnn}) are suppressed by the ratios of $m_\ell/M_W$ or
  $A'_l/M_{SUSY}$. Additional terms linear in $m_\ell/M_W$, not
  included in~\eq{eq:hnn}, appear in 1PI vertex diagrams when external
  lepton masses are not neglected. We calculated such terms and proved
  explicitly that after performing the MI expansion they are
  suppressed by additional powers of $v^2/M_{SUSY}^2$ and therefore,
  {\em a posteriori}, negligible. Thus, we do not display such terms
  in~\eq{eq:hnn}.}  {
\bea
(4\pi)^2 \Gamma_h^{JIK}(k_J,k_I) &=& - \sum_{n=1}^4 \sum_{l,m=1}^6
V_{\ell\tilde L N,L}^{Jmn*} V_{\ell\tilde L N,L}^{Iln} V_{H\tilde L
  \tilde L}^{Klm} m_{N_n} C_0 [k_J,k_I-k_J,m_{N_n},m_{\tilde
    L_m},m_{\tilde L_l}]\nn
&& \hspace{-38mm} - \sum_{l,n=1}^4 \sum_{m=1}^6 V_{\ell\tilde L
  N,R}^{Jnm*} V_{\ell\tilde LN,L}^{Inl} (V_{NHN,R}^{lKm} C_2 +
V_{NHN,L}^{lKm} m_{N_l} m_{N_m} C_0 )[k_J,k_I-k_J,m_{\tilde
    L_n},m_{N_m},m_{N_l}]\;,\nn
(4\pi)^2 \Gamma_A^{JI}(k_J,k_I) &=& - \sum_{n=1}^4 \sum_{l,m=1}^6
V_{\ell\tilde L N,L}^{Jmn*} V_{\ell\tilde L N,L}^{Iln} V_{A\tilde L
  \tilde L}^{1lm} m_{N_n} C_0 [k_J,k_I-k_J,m_{N_n},m_{\tilde
    L_m},m_{\tilde L_l}]\nn
&& \hspace{-38mm} - \sum_{l,n=1}^4 \sum_{m=1}^6 V_{\ell\tilde L
  N,R}^{Jnm*} V_{\ell\tilde LN,L}^{Inl} (V_{NAN,R}^{l1m} C_2 +
V_{NAN,L}^{l1m} m_{N_l} m_{N_m} C_0 )[k_J,k_I-k_J,m_{\tilde
    L_n},m_{N_m},m_{N_l}]\;,\nn
\label{eq:hnn} 
\eea
}
while the chargino-sneutrino triangle diagram is obtained by replacing
$\tilde L\to\tilde\nu, N\to C$ and adjusting the summation limits
appropriately in vertex factors $V_{\ldots}^{\ldots}$ (see
Appendix~\ref{app:lagr}).

\subsection{Box contributions}

\begin{figure}[tb]
\begin{center}
\includegraphics[width=0.5\textwidth]{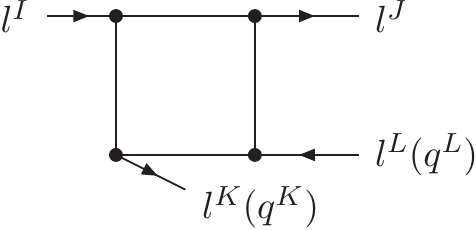}
\caption{\small Box diagrams with external charged leptons or
  quarks\label{fig:box}}
\end{center}
\end{figure}

4-fermion interactions are also generated by box diagrams. The
corresponding conventions for incoming and outgoing particles are
shown in Fig.~\ref{fig:box}. We calculate all box diagrams in the
approximation of vanishing external momenta.  The effective Lagrangian
for the 4-lepton interactions involves the quadrilinear operators
\bea
O_{VXY}^{JIKL} & = & (\bar\ell^{J}\gamma^{\mu}P_X \ell^I) \times
(\bar\ell^K\gamma_{\mu}P_Y \ell^L)\,,\nonumber\\
O_{SXY}^{JIKL} & = & (\bar\ell^{J}P_X \ell^I) \times (\bar\ell^K P_Y
\ell^L)\,,\nonumber\\
O_{TX}^{JIKL} & = & (\bar\ell^{J}\sigma^{\mu\nu}\ell^I) \times
(\bar\ell^K\sigma_{\mu\nu}P_X \ell^L)\,,
\label{eq:lboxdec}
\eea
where $X,Y$ stands for the chirality $L$ or $R$\footnote{Recall that
  $(\bar\ell^{J}\sigma^{\mu\nu}P_L\ell^I) \times
  (\bar\ell^K\sigma_{\mu\nu}P_R \ell^L)=0$.}.  The Wilson coefficients
of these operators are calculated from the box diagrams in
Fig.~\ref{fig:box} and are denoted by $B_{NXY}^{JIKL}$ with $N=V$,$S$,
or $B_{TX}^{JIKL}$.

The operator basis in \eq{eq:lboxdec} is redundant. First, we note
that
\bea
O_{NXY}^{JIKL} &=& O_{NYX}^{KLJI}\quad\mathrm{for}~N=V,S,\nn
O_{TX}^{JIKL} &=& O_{TX}^{KLJI}.
\label{eq:flip}
\eea
Second, there are Fierz relations among different operators:
\bea
O_{VXX}^{JIKL} &=& O_{VXX}^{KIJL}, \nn
O_{VXY}^{JIKL} &=& -\, 2\, O_{SXY}^{KIJL} \quad \mathrm{for}~X\neq Y,
\nn
O_{TX}^{JIKL} &=& \frac{1}{2} O_{TX}^{KIJL} -\, 6\, O_{SXX}^{KIJL},\nn
O_{SXX} ^{JIKL} &=& -\frac{1}{2} O_{SXX}^{KIJL} - \frac{1}{8}
O_{TX}^{KIJL}.
\label{eq:fierz}
\eea
Furthermore, we have 
\bea
O_{VXY}^{JIKL\,\dagger} &=& O_{VXY}^{IJLK}, \qquad\qquad
O_{SLL}^{JIKL\,\dagger} \;=\; O_{SRR}^{IJLK}, \nn
O_{SLR}^{JIKL\,\dagger} &=& O_{SRL}^{IJLK}, \qquad\qquad
O_{TL}^{JIKL\,\dagger} \;=\; O_{TR}^{JILK}.
\label{eq:herm}
\eea
Eqs.~(\ref{eq:flip}) to (\ref{eq:herm}) must be taken into account
when deriving the effective Lagrangian.

\subsubsection{Leptonic operators with $\mathbf{J\neq K}$ and
  $\mathbf{I\neq L}$\label{sec:jnotk}}

The case with both $J\neq K$ and $I\neq L$ covers the decays
$\tau^{\mp} \to \mu^{\mp} e^{\mp}\ell^{\pm} $ with $\ell=e$ or $\mu$,
but does not appear in $\mu^{\mp}$ decays.  We can therefore specify
to $I=3$ for the effective Lagrangian.  Furthermore, we can choose
either $(J,K)=(1,2)$ or $(J,K)=(2,1)$ without the need to sum over
both cases: The Fierz identities in \eq{eq:fierz} permit to bring all
operators into the form $(\ov{e}\ldots\tau) \times (\ov{\mu}\ldots
\ell)$ (corresponding to the case $(J,K)=(1,2)$) or into an
alternative form with $e$ interchanged with $\mu$. Thus we have
\bea
L_{4\ell}^{J3KL} &=&
\sum_{L=1,2} \left[ \sum_{\stackrel{N=V,S}{X,Y=L,R}} B_{NXY}^{J3KL}
  O_{NXY}^{J3KL} \, +\!  \sum_{X=L,R} B_{TX}^{J3KL} O_{TX}^{J3KL}
  \right]
\;+\; \mbox{h.c.}\nn &&\qquad\qquad\qquad \qquad\mbox{with }J\neq
K\mbox{ and }J,K,L \leq 2,
 \label{eq:lboxbasis}
\eea
as the four-lepton interaction in the Lagrangian.  Note that the
``$+$h.c.'' piece of $L_{4\ell}^{JK}$ describes $\tau^+$ decays.

The Wilson coefficients $B_{NXY}^{J3KL}$ and $ B_{TX}^{J3KL}$ in
\eq{eq:lboxbasis} are simply identical to the results of the sum of
all contributing box diagrams to the decay amplitude.  The latter is
given in \eq{eq:lllla0} with the coefficients of the spinor structure
in the right column of Tab.~\ref{tab:lbox}. The relation to the
analytic expressions in Eqs.~(\ref{eq:lboxa}) to (\ref{eq:lboxd}) is
\bea
B_{NXY}^{JIKL} &=& B_{A\, NXY}^{JIKL} + B_{B\, NXY}^{JIKL} + 
  B_{C\, NXY}^{JIKL} + B_{D\, NXY}^{JIKL},\qquad\mbox{for }N=V,S 
         \label{eq:boxsum}
\eea
and an analogous expression for $B_{TX}^{JIKL}$. 

\subsubsection{Leptonic operators with $\mathbf{J= K}$ and $\mathbf{I\neq L}$}

The case $J=K$ occurs for the decays $\mu^\pm \to e^\pm e^\pm e^\mp$
and $\tau^\pm \to \ell^\pm \ell^\pm \ell^{\mp\prime}$ with
$\ell,\ell^\prime=e,\mu$. Thanks to the Fierz identities in
\eq{eq:fierz} we may restrict the operator basis to
\bea
O_{VXX}^{JIJL},\qquad\quad 
O_{VXY}^{JIJL}\;=\; -2 O_{SXY}^{JIJL},&& \qquad 
O_{SXX}^{JIJL}\;=\; -\frac{1}{12} O_{TX}^{JIJL},\nn
&& 
\qquad \quad\mbox{with }X,Y=L,R\mbox{ and }X\neq Y.
\label{eq:jkbasis}
\eea
The four-lepton piece of the effective Lagrangian for the decay
$\ell^{I\mp} \to \ell^{J\mp} \ell^{J\mp} \ell^{L\pm} $ reads:
\bea
L_{4\ell}^{JIJL} &=& \sum_{L=1,2} \left[ \sum_{X,Y=L,R}
  \widetilde{C}_{VXY}^{JIJL} O_{VXY}^{JIJL} \, +\! \sum_{X=L,R}
  \widetilde{C}_{SXX}^{JIJL} O_{SXX}^{JIJL} \right] \;+\;
\mbox{h.c.}\nn &&\qquad\qquad\qquad \qquad\mbox{with }L,J <I.
\label{eq:lboxbasis2}
\eea
For the matching calculation it is useful to quote the tree-level
matrix elements of the operators:
\bea
\bra{l^{J-}(p_J,s_J) l^{J-}(p_J^\prime,s_J^\prime) l^{L+}(p_L,s_J) }
O_{VXX}^{JIJL} \ket{l^{I-}(p_I,s_I) } \hspace{-7cm} \nn
&=& \phantom{\; - \;} [\bar u(p_J,s_J) \gamma_{\mu} P_X u(p_I,s_I) ]
  [\bar u(p_J^\prime,s_J^\prime) \gamma^{\mu} P_X v(p_L,s_L)] \nn
&& \; - \; [\bar u(p_J^\prime,s_J^\prime) \gamma_{\mu} P_X u(p_I,s_I)
  ] [\bar u(p_J,s_J) \gamma^{\mu} P_X v(p_L,s_L)] \nn
\qquad &=& \;\; 2 \, [\bar u(p_J,s_J) \gamma_{\mu} P_X u(p_I,s_I) ]
       [\bar u(p_J^\prime,s_J^\prime) \gamma^{\mu} P_X v(p_L,s_L)]
       \no\\[2mm]
\bra{l^{J-}(p_J,s_J) l^{J-}(p_J^\prime,s_J^\prime) l^{L+}(p_L,s_J) }
O_{VXY}^{JIJL} \ket{l^{I-}(p_I,s_I) } \hspace{-7cm} \nn
&=& \phantom{\; - \;} [\bar u(p_J,s_J) \gamma_{\mu} P_X u(p_I,s_I) ]
  [\bar u(p_J^\prime,s_J^\prime) \gamma^{\mu} P_Y v(p_L,s_L)] \nn
&& \; - \; [\bar u(p_J^\prime,s_J^\prime) \gamma_{\mu} P_X u(p_I,s_I)
  ] [\bar u(p_J,s_J) \gamma^{\mu} P_Y v(p_L,s_L)] \nn
\qquad &=& \phantom{\; - \; 2\,} [\bar u(p_J,s_J) \gamma_{\mu} P_X
  u(p_I,s_I) ] [\bar u(p_J^\prime,s_J^\prime) \gamma^{\mu} P_Y
  v(p_L,s_L)] \nn
&& \; - \; 2\, [\bar u(p_J,s_J) P_X u(p_I,s_I) ] [\bar
     u(p_J^\prime,s_J^\prime) P_Y v(p_L,s_L)],\qquad\mbox{for } X\neq
   Y, \no\\[2mm]
\bra{l^{J-}(p_J,s_J) l^{J-}(p_J^\prime,s_J^\prime) l^{L+}(p_L,s_J) } 
  O_{SXX}^{JIJL} \ket{l^{I-}(p_I,s_I) } \hspace{-7cm} \nn
&=& \phantom{\; - \;} [\bar u(p_J,s_J) P_X u(p_I,s_I) ] [\bar
    u(p_J^\prime,s_J^\prime) P_X v(p_L,s_L)] \nn
&& \; - \; [\bar u(p_J^\prime,s_J^\prime) P_X u(p_I,s_I) ] [\bar
    u(p_J,s_J) P_X v(p_L,s_L)] \nn
\qquad &=& \phantom{\;-\;} \frac12 [\bar u(p_J,s_J) P_X u(p_I,s_I) ]
       [\bar u(p_J^\prime,s_J^\prime) P_X v(p_L,s_L)] \nn
&& \; - \; \frac18\, [\bar u(p_J,s_J) \sigma_{\mu\nu} P_X u(p_I,s_I) ]
   [\bar u(p_J^\prime,s_J^\prime) \sigma^{\mu\nu} P_X
     v(p_L,s_L)] \label{eq:melt}
\eea
Here we have used the Fierz transform to group the spinors into the
canonical order $[\bar u(p_J,...) ... u(p_I,...)][ \bar
  u(p_J^\prime,...) ... v(p_L,...)]$. This allows us to use the same
formula for spin-summed squared matrix elements as in the case of
$J\neq K$ of Sec.~\ref{sec:jnotk}.

To quote the Wilson coefficients $\widetilde{C}_{NXY}^{JIJL}$, $N=V,S$
in terms of the box diagrams $B_{NXY}^{JIJL}$ in \eq{eq:boxsum} we
must compare the results of the MSSM decay amplitude in
\eq{eq:lllla00} with the matrix elements in \eq{eq:melt} and read off
coefficients of the various Dirac structures. The result is
\bea
\widetilde{C}_{VXX}^{JIJL} &=& \frac{1}{2} B_{VXX}^{JIJL},\nn
\widetilde{C}_{VXY}^{JIJL} & =& B_{VXY}^{JIJL} \qquad \mbox{for }X\neq
Y, \nn
\widetilde{C}_{SXX}^{JIJL} & =& 2\, B_{SXX}^{JIJL}.
\label{eq:wc}
\eea
The Fierz identities further imply the equalities 
\bea
B_{SXY}^{JIJL} & = & -2 B_{VXY}^{JIJL} \qquad \mbox{for }X\neq Y,\nn
B_{TX}^{JIJL} & =& -\frac14 B_{SXX}^{JIJL} \; .\label{eq:feq}   
\eea

\subsubsection{Leptonic operators with $\mathbf{J= K}$ and $\mathbf{I=L}$}

These operators do not appear in lepton decays, but trigger
muonium-antimuonium transitions and describe muon or tau pair
production in $e^-$--$e^-$ collisions at energies far below
$M_{SUSY}$. Their Wilson coefficients are tiny in the MSSM.

\subsubsection{Operators with two leptons and two quarks}

The analogous Lagrangian for the 2-lepton--2-quark interactions reads
\bea
L_{2\ell2q}^{IJKL} = \sum_{N,X,Y} B^{IJKL}_{q\,NXY}O^{JIKL}_{q\,NXY}
\label{eq:qboxbasis}
\eea
where
\bea
O_{q\, VXY}^{IJKL} & = & (\bar\ell_{I}\gamma^{\mu}P_X \ell_J) \times
(\bar q_K\gamma_{\mu}P_Y q_L)\,,\nonumber\\
O_{q\,SXY}^{IJKL} & = & (\bar\ell_{I}P_X \ell_J)\times (\bar q_L P_Y
q_K)\,,\nonumber\\
O_{q\,TX}^{IJKL} & = & (\bar\ell_{I}\sigma^{\mu\nu}\ell_J) \times
(\bar q_K \sigma_{\mu\nu}P_X q_L)\,.
\label{eq:qboxdec}
\eea
Again, we consider only purely leptonic contributions here in detail
but do not give explicit expressions for the 2-lepton--2-quark box
diagrams.  The relevant expressions in the mass eigenbasis can be
found using formulae of Appendix~\ref{app:fullbox} and inserting
proper quark vertices from Refs.~\cite{Rosiek:1989rs, Rosiek:1995kg}
into these.

\section{Observables}
\label{sec:lfv}

In this Section we collect the formulae for the LFV observables in
terms of the effective interactions defined in Sec.~\ref{sec:efflag}.
All the processes listed here will be included in the future version
of the SUSY\_FLAVOR numerical library calculating an extensive set of
flavour and CP-violating observables both in the quark and leptonic
sectors~\cite{Rosiek:2010ug, Crivellin:2012jv, Rosiek:2014sia}.

\subsection{Radiative lepton decays: $\ell^I\to\ell^J\gamma$}

The branching ratios for the radiative lepton decays
$\ell^I\to\ell^J\gamma$ are given by
\bea
\Br(\ell^I\to\ell^J\gamma) &=&
\frac{48\pi^2 e^2 }{m_I^2 G_F^2} \left(|F_{\gamma}^{JI}|^2 +
|F_{\gamma}^{IJ}|^2\right) \Br(\ell^I\to e\nu\nu)\,.
\eea
Here we used $\Gamma(\ell^I\to e\nu\nu)\approx G_F^2 m_I^5/(192 \pi^3)$
for the tree-level leptonic decay width and the factors
$\Br(\mu\to e\nu\nu)\approx 1$,
$\Br(\tau\to e\nu\nu) = 0.1785 \pm 0.0005$~\cite{Agashe:2014kda} are
introduced to account for the hadronic decay modes of the $\tau$ lepton.

Even though in our numerical analyses we restrict ourselves to LFV
processes, we remind the reader that the expressions for the anomalous
magnetic moments and electric dipole moments of the charged leptons
can be also calculated in term of the quantities defined
in~\eq{eq:fgamab} and read:
\bea
\Delta a_I &=& - 4 m_I \re\left[ F_{\gamma A}^{II} - m_I
  \left(F_{\gamma L B}^{II} + F_{\gamma R B}^{II}\right)\right]\,,\\
d_l^I &=& - 2 e \im F_{\gamma A}^{II}
\eea

\subsection{$h(H)\to \bar\ell^{I}\ell^{J}$ decays}
\label{sec:hll}

The decay branching ratios for the CP-even and CP-odd Higgs bosons
read:
\bea
Br(H_0^K\to \ell^{I+} \ell^{J-}) & =&
\frac{m_{H_0^K}}{16\pi\Gamma_{H_0^K}} \left( \left|F_h^{IJK}\right|^2
+ \left|F_h^{JIK}\right|^2\right)\nn
Br(A_0\to \ell^{I+} \ell^{J-})& =& \frac{m_{A}}{16\pi\Gamma_{A}}
\left( \left|F_A^{IJ}\right|^2 + \left|F_A^{JI}\right|^2\right)
\label{eq:brhll}
\eea
with $F_h^{IJK}, F_A^{IJ}$ defined in~\eq{eq:yukeff}. Note that
summing over lepton charges in the final state, $\ell^{I+} \ell^{J-}$
and $\ell^{J+} \ell^{I-}$, would produce an additional factor of 2.

\subsection{$\ell^I\to \ell^J \ell^K\bar \ell^L$ decays}
\label{sec:llll}

The LFV decays of charged lepton into three lighter ones can be
divided into 3 classes, depending on the flavours in the final state:
\begin{itemize}
\item[(A)] $\ell\to\ell^\prime\ell^\prime\ell^\prime$: Three leptons
  of the same flavour, i.e. $\mu^\pm \to e^\pm e^+ e^-$, $\tau^\pm \to
  e^\pm e^+ e^-$ and $\tau^\pm \to \mu^\pm \mu^+ \mu^-$, with a pair
  of opposite charged leptons.
\item[(B)] $\ell^\pm \to \ell^{\prime\pm} \ell^{\prime\prime+}
  \ell^{\prime\prime-}$: Three distinguishable leptons with
  $\ell^{\prime}$ carrying the same charge as $\ell$, i.e. $\tau^\pm
  \to e^\pm \mu^+ \mu^-$ and $\tau^\pm \to \mu^\pm e^+ e^-$.
\item[(C)] $\ell^\pm \to \ell^{\prime\mp} \ell^{\prime\prime+}
  \ell^{\prime\prime-}$: Three distinguishable leptons with
  $\ell^{\prime}$ carrying the opposite charge as $\ell$,
  i.e. $\tau^\pm \to e^\mp \mu^\pm \mu^\pm$ and $\tau^\pm \to \mu^\mp
  e^\pm e^\pm$.
\end{itemize}

Class (C), representing a $\Delta L=2$ processes, is tiny within the
MSSM: it could only be generated at 1-loop level by box diagrams
suppressed by double flavour changes, or at the 2-loop level by double
penguin diagrams involving two LFV vertices. Therefore, we will not
consider these process in our numerical analysis.

\begin{figure}[tbp]
\begin{center}
\includegraphics[width=0.8\textwidth]{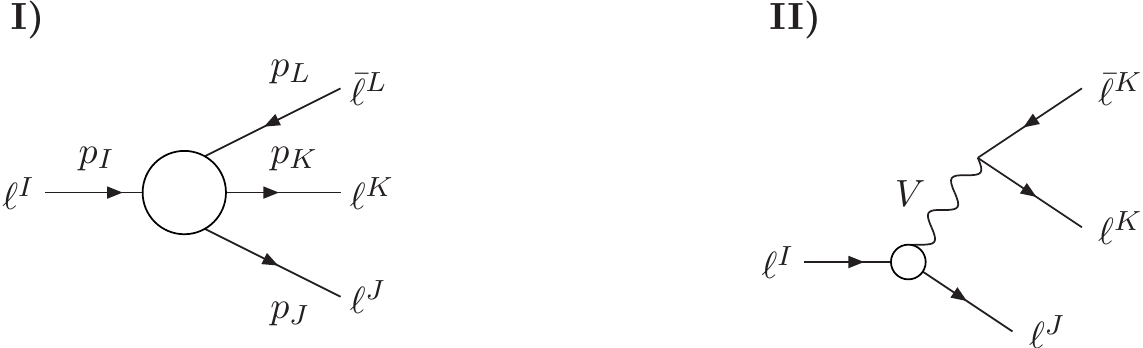}
\end{center}
\caption{\small Diagrams contributing to $\ell^I\to \ell^J\ell^K\bar
  \ell^L$ decay. I): 1PI irreducible box diagrams; II): penguin
  diagrams with $V=Z,\gamma,h,H$ or $A$. For $K=J$ crossed diagrams
  must be also included.
\label{fig:llll}}
\end{figure}

\noindent In order to calculate $\Br(\ell^I\to \ell^J\ell^K\bar
\ell^L)$ we decompose the corresponding amplitude $A$ as
\bea
A = A_0 + A_\gamma\,, \label{eq:llllampl}
\eea
The relevant diagrams are displayed in Fig.~\ref{fig:llll}.  $A_0$
contains contributions from 4-lepton box diagrams and from penguin
diagrams (including vector-like off-shell photon couplings,
see~\eq{eq:phvect}) which in the limit of vanishing external momenta
can be represented as the 4-fermion contact interactions.  $A_\gamma$
is the on-shell photon contribution originating from the magnetic
operator (see~\eq{eq:lgdef}) which has to be treated separately with
more care as the photon propagator becomes singular in the limit of
vanishing external momenta.

We further decompose $A_0$ for the two cases (A) and (B) according to
its Lorentz structure:
\bea
A_0^{(A)} &=& \sum_{Q=V,S,T} C^{(A)}_{Q,\,XY} [\bar u(p_J)
  \Gamma_Q^\prime P_X u(p_I)] [\bar u(p_J^\prime) \Gamma_Q P_Y
  v(p_L)]] \;,\label{eq:lllla00} \\
A_0^{(B)} &=& \sum_{Q=V,S,T} C^{(B)}_{Q,\,XY} [\bar u(p_J)
  \Gamma_Q^\prime P_X u(p_I)][\bar u(p_K) \Gamma_Q P_Y v(p_L)]] \;.
  \label{eq:lllla0} 
\eea
with $X,Y=L,R$. Note that the amplitude $A_0^{(A)}$ in general
contains a second term which is obtained from the one given in
\eq{eq:lllla00} by replacing $(p_J\leftrightarrow
p_J^\prime)$. However, one can use Fierz identities to reduce it to
the structure given in \eq{eq:lllla00}. The basis of Dirac
quadrilinears $\Gamma_Q$ is the same as the one used to decompose
4-lepton box diagrams in~\eq{eq:lboxdec}:
\bea
\Gamma_S=1\,, \qquad \Gamma_V = \gamma^\mu\,, \qquad \Gamma_T =
\sigma_{\mu\nu}\,,
\label{eq:qnbasis}
\eea
and $\Gamma_Q^\prime$ is obtained from $\Gamma_Q$ by lowering the
Lorentz indices.

The amplitudes originating from on-shell photon exchange are given by
\bea
A_\gamma^{(A)} & = &\! \frac{e}{(p_I-p_J)^2} [\bar u(p_J)
  i\sigma^{\mu\nu} (C_{\gamma L} P_L + C_{\gamma R} P_R) (p_I -
  p_J)_\nu u(p_I)] [\bar u(p_J^\prime) \gamma_\mu v(p_L)] \nonumber\\
&&- (p_J\leftrightarrow p_J^\prime) \nonumber\\
A_\gamma^{(B)} & = &\! \frac{e}{(p_I-p_J)^2}[ \bar u(p_J)
  i\sigma^{\mu\nu} (C_{\gamma L} P_L + C_{\gamma R} P_R) (p_I -
  p_J)_\nu u(p_I)][\bar u(p_K) \gamma_\mu v(p_L)].
\label{eq:llllag}
\eea

\begin{table}[tb]
  \begin{center}
    {\small
      \begin{tabular}{|lp{-0mm}|l|l|}
        \hline 
        \hspace{-32mm} && Decay (A) & Decay (B)  \\
        \hline
                       %
        $C_{VLL}$ && $ B_{VLL}^{JIJJ} -\frac{e(1- 2 s_W^2)}{s_W c_W M_Z^2}
                     F_{ZL}^{JI} + 2 e^2 V_{\gamma L}^{JI} $ & \scalebox{0.9}{$B_{VLL}^{JIKK} -\frac{e(1-
                                                               2 s_W^2)}{2s_W c_W M_Z^2} F_{ZL}^{JI}+ e^2 V_{\gamma L}^{JI} $}\\[2mm]
        $C_{VRR}$ && $ B_{VRR}^{JIJJ} +\frac{2 e s_W}{c_W M_Z^2} F_{ZR}^{JI}+
                     2 e^2 V_{\gamma R}^{JI} $ & $B_{VRR}^{JIKK} + \frac{e s_W}{c_W
                                                 M_Z^2} F_{ZR}^{JI}+ e^2 V_{\gamma R}^{JI} $\\[2mm]
        $C_{VLR}$ && \scalebox{0.9}{$B_{VLR}^{JIJJ} +\frac{e s_W}{c_W M_Z^2} F_{ZL}^{JI} +
                     e^2 V_{\gamma L}^{JI} + {\frac{1}{2}Y_l^J (V_H^{IJ*} - V_A^{IJ*})}
                     $} & $B_{VLR}^{JIKK} +\frac{e s_W}{c_W M_Z^2} F_{ZL}^{JI}+ e^2
                          V_{\gamma L}^{JI} $ \\[2mm]
        $C_{VRL}$ && \scalebox{0.9}{$B_{VRL}^{JIJJ} -\frac{e(1- 2 s_W^2)}{2 s_W c_W M_Z^2}
                     F_{ZR}^{JI} + e^2 V_{\gamma R}^{JI} + {\frac{1}{2}Y_l^J (V_H^{JI} -
                     V_A^{JI})} $} & \scalebox{0.9}{$B_{VRL}^{JIKK} -\frac{e(1- 2 s_W^2)}{2s_W c_W
                                     M_Z^2} F_{ZR}^{JI} + e^2 V_{\gamma R}^{JI}$} \\[2mm]
        $C_{SLL}$ && $B_{SLL}^{JIJJ} + {\frac{3}{2}Y_l^J (V_H^{JI} +
                     V_A^{JI})}$ & $B_{SLL}^{JIKK} + {Y_l^K (V_H^{JI} +
                                   V_A^{JI})}$\\[2mm]
        $C_{SRR}$ && $B_{SRR}^{JIJJ} + {\frac{3}{2} Y_l^J (V_H^{IJ*} +
                     V_A^{IJ*})}$ & $B_{VRR}^{JIKK} + { Y_l^K (V_H^{IJ*} +
                                    V_A^{IJ*})}$\\[2mm]
        $C_{SLR}$ && \scalebox{0.9}{$-2 B_{VLR}^{JIJJ} -\frac{2 e s_W}{c_W M_Z^2}
                     F_{ZL}^{JI}- 2 e^2 V_{\gamma L}^{JI} + {Y_l^J (V_H^{JI}
                     - V_A^{JI})}$} 
                                    & $B_{SLR}^{JIKK} + {Y_l^K (V_H^{JI} -
                                      V_A^{JI})}$\\[3mm]
        $C_{SRL}$ && \scalebox{0.9}{$ - 2 B_{SRL}^{JIJJ} +\frac{e(1- 2 s_W^2)}{s_W c_W M_Z^2}
                     F_{ZR}^{JI} - 2 e^2 V_{\gamma R}^{JI} + {Y_l^J (V_H^{IJ*} -
                     V_A^{IJ*})}$} 
                                    & $B_{SRL}^{JIKK} + {Y_l^K (V_H^{IJ*} -
                                      V_A^{IJ*})}$\\[2mm]
        $C_{TL}$ && $-\frac{1}{4}B_{SLL}^{JIJJ} + {\frac{1}{8}Y_l^J
                    (V_H^{JI} + V_A^{JI})}$ & $B_{TL}^{JIKK}$\\[2mm]
        $C_{TR}$ && $-\frac{1}{4}B_{SRR}^{JIJJ} + \frac{1}{8} Y_l^J
                    (V_H^{IJ*} + V_A^{IJ*}) $ & $B_{TR}^{JIKK}$\\[2mm] $C_{\gamma L}$
                       && $-2 e F_\gamma^{JI}$ & $-2 e F_\gamma^{JI}$\\[2mm] $C_{\gamma R}$
                       && $- 2 e F_\gamma^{IJ*}$ & $-2 e F_\gamma^{IJ*}$\\[2mm]
        \hline
      \end{tabular}
    }
  \end{center}
  \caption{\small Coefficients $C_N, C_\gamma$ of~\eq{eq:lllla0} and
    \eq{eq:llllag} for decay types (A) and (B). $B_{QXY}$,$B_{TX}$
    denote the irreducible box diagram contributions
    (see~\eq{eq:lboxbasis}), the terms with $F_Z$ stem from the $Z$
    penguin Lagrangian (\eq{eq:lzdef}), $V_\gamma$ is the sum of the
    vector-like photon contributions (\eq{eq:phvect}), Higgs
    contributions are defined in~\eq{eq:vhadef} and the coefficients
    $F_\gamma$ of the magnetic operator are defined
    in~\eq{eq:lgdef}. \label{tab:lbox}}
\end{table}

The full form of the coefficients $C_N^{(A,B)},\;C_\gamma$ is
displayed in Table~\ref{tab:lbox}, where we compactified the
expressions by using the following abbreviations for the Higgs penguin
contributions\footnote{Note that we define lepton Yukawa coupling
  appearing in Table~\ref{tab:lbox} to be negative, $Y_l^I =
  -\sqrt{2}m_l^I/v_1$}:
\bea
V_H^{JI} = \sum_{N=1}^2 \frac{Z_R^{1N}}{m_{H_0^N}^2} F_h^{JIN}\;,
\qquad\qquad
V_{A}^{JI} = \frac{i\sin\beta}{m_{A_0}^2} F_A^{JI}\;.
\label{eq:vhadef}
\eea
Note that in~\eq{eq:lllla0} and~\eq{eq:llllag} we do not explicitly
display flavour indices, but they are specified in
Table~\ref{tab:lbox}.

Neglecting the lighter lepton masses whenever possible, the expression
for the branching ratios can be written down as (for comparison
see~\cite{Ilakovac:2012sh}):
\bea
\mathrm{Br}(\ell^I\to \ell^J \ell^K \bar \ell^L) &=& \frac{N_c
  \Br(\ell^I\to e\nu\nu)}{32 G_F^2} \left(4 \left(|C_{VLL}|^2 +
|C_{VRR}|^2 + |C_{VLR}|^2 + |C_{VRL}|^2\right)\right.\nonumber\\
&+&|C_{SLL}|^2 + |C_{SRR}|^2+ |C_{SLR}|^2+ |C_{SRL}|^2 \nonumber\\
&+& \left.   48 \left(|C_{TL}|^2 + |C_{TR}|^2\right) + X_\gamma\right)
\label{eq:br4l}
\eea
where $N_c=1/2$ if two of the final state leptons are identical
(decays (A)), $N_c=1$ for decays (B) and $X_\gamma$ denotes the
contribution to matrix element from the photon penguin $A_\gamma$,
including also its interference with the $A_0$ part of the amplitude
($m$ denotes the mass of the heaviest final state lepton)
\bea
X_\gamma^{(A)} & =& -\frac{16e}{m_{\ell^I}}\mathrm{Re}\left[(2
  C_{VLL} + C_{VLR} - \frac{1}{2} C_{SLR} )\; C_{\gamma R}^\star +
  ( 2 C_{VRR} + C_{VRL} - \frac{1}{2} C_{SRL})\; C_{\gamma
    L}^\star \right] \nonumber\\
&& + \frac{64 e^2}{m_{\ell^I}^2} \left(\log\frac{m_{\ell^I}^2}{m^2} -
\frac{11}{4} \right)(|C_{\gamma L}|^2 + |C_{\gamma R}|^2) \nonumber\\
X_\gamma^{(B)} & =& -\frac{16e}{m_{\ell^I}}\mathrm{Re}\left[ \left(C_{VLL} +
  C_{VLR}\right) C_{\gamma R}^\star + \left(C_{VRR} + C_{VRL}\right)
  C_{\gamma L}^\star \right]\nonumber\\
&& +\frac{32 e^2}{m_{\ell^I}^2} \left(\log\frac{m_{\ell^I}^2}{m^2} -3\right)
(|C_{\gamma L}|^2 + |C_{\gamma R}|^2) \;.
\label{eq:brg}
\eea

\subsection{$\mu\to e$ conversion in Nuclei}
\label{sec:mueconv}

The full 1-loop expressions for the $\mu\to e$ conversion in Nuclei
depend on both the squark and slepton SUSY breaking terms. Thus, in
principle the resulting upper bounds on the slepton mass insertions
depend to some extent on the squark masses. Therefore, we do not
include $\mu\to e$ conversion in nuclei in our numerical
analysis\footnote{Recent discussion of interplay between the bounds on
  MI's in the slepton and squark sectors can be found in
  Ref.~\cite{Ellis:2016yje}.}.  However, for completeness we collect
here the complete set of formulae required to calculate the rate of
this process.

$\mu\to e$ conversion in nuclei is produced by the dipole, the vector,
and the scalar operators already at the tree
level~\cite{Kitano:2002mt}.  Following the discussion of
Ref.~\cite{Crivellin:2017rmk} we use the effective Lagrangian
\bea
L_{\mu\to e} &=& \sum_{N,X,Y} C_{q_Iq_I}^{N\;XY}O^{q_Iq_I}_{N\;XY} +
C^{gg}_X O^{gg}_X
\eea
where $N=V,S$ and $X,Y=L,R$ with the operators defined as
\bea
O^{q_Iq_I}_{V\;XY} &=& \left( {\bar e{\gamma ^\mu }{P_X}\mu }
\right)\left( {\bar q_I{\gamma _\mu }{P_Y}q_I} \right)\nn
O^{q_Iq_I}_{S\;XY} &=& \left( {\bar e{P_X}\mu } \right)\left( {\bar
  q_I{P_Y}q_I} \right)\nn
O^{gg}_X &=& {\alpha _s}{\mkern 1mu} {m_\mu }{G_F}\left( {\bar
  e{P_X}\mu } \right)G_{\mu \nu }^aG_a^{\mu \nu }
\eea
Using the notation introduced in previous Sections, the corresponding
Wilson coefficients can be expressed as
\bea
C_{V\;XL}^{{d_I}{d_I}} &=& C_{d\ell {\kern 1pt} VXL}^{12II} -
\frac{1}{{m_Z^2}}\frac{e}{{2{s_W}{c_W}}}\left( {1 - \frac{2}{3}s_W^2}
\right)F_{ZX}^{12} - \frac{1}{3}{e^2}V_{\gamma X}^{JI}\nn
C_{V\;XR}^{{d_I}{d_I}} &=& C_{d\ell {\kern 1pt} VXR}^{12II} +
\frac{1}{{m_Z^2}}\frac{e}{{3{s_W}{c_W}}}s_W^2F_{ZX}^{12} -
\frac{1}{3}{e^2}V_{\gamma X}^{JI}\nn
C_{V\;XL}^{{u_I}{u_I}} &=& C_{u\ell {\kern 1pt} VXL}^{12II} +
\frac{1}{{m_Z^2}}\frac{e}{{2{s_W}{c_W}}}\left( {1 - \frac{4}{3}s_W^2}
\right)F_{ZX}^{12} + \frac{2}{3}{e^2}V_{\gamma X}^{JI}\nn
C_{V\;XR}^{{u_I}{u_I}} &=& C_{u\ell {\kern 1pt} VXR}^{12II} -
\frac{1}{{m_Z^2}}\frac{e}{{{s_W}{c_W}}}\frac{2}{3}s_W^2F_{ZX}^{12} +
\frac{2}{3}{e^2}V_{\gamma X}^{JI}\nn
C_{S\;LX}^{{d_I}{d_I}} &=& C_{d\ell {\kern 1pt} SLX}^{12II} +
\frac{1}{{{{\left( {m_0^K} \right)}^2}}}F_h^{12K}F_{hd}^{IIK}\nn
C_{S\;LX}^{{u_I}{u_I}} &=& C_{u\ell {\kern 1pt} SLX}^{12II} +
\frac{1}{{{{\left( {m_0^K} \right)}^2}}}F_h^{12K}F_{hu}^{IIK}\nn
C_{S\;RX}^{{d_I}{d_I}} &=& C_{d\ell {\kern 1pt} SRX}^{12II} +
\frac{1}{{{{\left( {m_0^K} \right)}^2}}}F_h^{21K*}F_{hd}^{IIK}\nn
C_{S\;RX}^{{u_I}{u_I}} &=& C_{u\ell {\kern 1pt} SRX}^{12II} +
\frac{1}{{{{\left( {m_0^K} \right)}^2}}}F_h^{21K*}F_{hu}^{IIK}
\eea

For this process, a Lagrangian involving only quark, lepton and photon
fields is not sufficient. Instead, an effective Lagrangian at the
nucleon level containing proton and neutron fields is required. It can
be obtained in two steps. First, heavy quarks are integrated out. This
results in a redefinition of the Wilson coefficient of the gluonic
operator~\cite{Shifman:1978zn}
\bea
C^{gg}_L &\to& \tilde{C}^{gg}_L = C^{gg}_L - \frac{1}{12 \pi}
\sum_{q=c,b} \frac{C^{qq}_{S\;LL} + C^{qq}_{S\;LR}}{G_F\, m_\mu m_q}
\label{anomaly}
\eea
with an analogous equation for $C_{gg}^R$. Second, the resulting
Lagrangian is matched at the scale of $\mu_n=1$~GeV to an effective
Lagrangian at the nucleon level.  Following~\cite{Cirigliano:2009bz}
the transition rate $\Gamma_{\mu\to e}^N = \Gamma(\mu^- N \to e^- N)$
can then be written as
\bea
\Gamma_{\mu\to e}^N &=& \frac{m_{\mu}^{5}}{4} \left| -e\, C^{D}_{L} \;
F_\gamma^{12}/m_\mu + 4\left( G_F m_\mu m_p \tilde{C}^{(p)}_{SL}
S^{(p)}_N + \tilde{C}^{(p)}_{VR} \; V^{(p)}_N + (p \to n) \right)
\right|^2\nn
&+& (L\leftrightarrow R),
\label{Gconv}
\eea
where $p$ and $n$ denote the proton and the neutron, respectively. The
effective couplings in Eq.~(\ref{Gconv}) can be expressed in terms of
our Wilson coefficients as
\bea
\tilde{C}^{(p/n)}_{VR} &=& \sum_{q=u,d,s} 
\left(C^{qq}_{V\;RL} + C^{qq}_{V\;RR}\right) \; f^{(q)}_{Vp/n} \, , 
\label{tildeCVR} \\
\tilde{C}^{(p/n)}_{SL} &=& \sum_{q=u,d,s}
\frac{\left(C^{qq}_{S\;LL}+C^{qq}_{S\;LR}\right)}{m_\mu m_q G_F} \;
f^{(q)}_{Sp/n} \ + \ \tilde{C}^{gg}_L \, f_{Gp/n}
\label{tildeCSL}
\eea
with analogous relations for $L\leftrightarrow R$. The Wilson
coefficients in Eqs.~(\ref{tildeCVR}) and (\ref{tildeCSL}) are to be
evaluated at the scale $\mu_n$.

The nucleon form factors for vector operators are fixed by
vector-current conservation, i.e. $f_{Vp}^{(u)}=2$, $f_{Vn}^{(u)}=1$,
$f_{Vp}^{(d)}=1$, $f_{Vn}^{(d)}=2$, $f_{Vp}^{(s)}=0$,
$f_{Vn}^{(s)}=0$. Hence, the sum in Eq.~(\ref{tildeCVR}) is in fact
only over $q=u ,d$. The calculation of the scalar form factors is more
involved.  The values of the up- and down-quark scalar couplings
$f_{S p/n}^{(u/d)}$ (based on the two-flavour chiral perturbation
theory framework of~\cite{Crivellin:2013ipa}) can be found in
Refs.~\cite{Crivellin:2014cta, Hoferichter:2015dsa}, while the values
of the $s$-quark scalar couplings $f_{S p/n}^{(s)}$ can be borrowed
from a lattice calculation~\cite{Junnarkar:2013ac}\footnote{For
  earlier determinations of the pion-nucleon sigma terms
  see~\cite{Alarcon:2011zs,Alarcon:2012nr}}. In summary, one has
\begin{align}
f_{Sp}^{(u)}&=(20.8\pm 1.5) \times 10^{-3},\qquad
f_{Sn}^{(u)}=(18.9\pm 1.4) \times 10^{-3}, \nonumber
\\
f_{Sp}^{(d)}&=(41.1\pm 2.8) \times 10^{-3},\qquad
f_{Sn}^{(d)}=(45.1\pm 2.7) \times 10^{-3}, \nonumber
\\ f_{Sp}^{(s)}&=f_{Sn}^{(s)}=(53\pm 27)\times 10^{-3}.
\end{align}
The form factor for the gluonic operator can be obtained from a sum
rule.  In our normalisation
\begin{align}
f_{Gp/n}&=-\frac{8\pi}{9}\Big(1-\sum_{q=u,d,s} f^{(q)}_{Sp/n}\Big) \,.
\end{align}
The quantities $D_N$, $S^{(p/n)}_N$, and $V^{(p/n)}_N$ in
Eq.~(\ref{Gconv}) are related to the overlap
integrals~\cite{Czarnecki:1998iz} between the lepton wave functions
and the nucleon densities.  They depend on the nature of the target
$N$. Their numerical values can be found in Ref.~\cite{Kitano:2002mt}:
\begin{align}
D_{\rm Au} &= 0.189, &S^{(p)}_{\rm Au} &=0.0614, &V^{(p)}_{\rm
  Au}&=0.0974, &S^{(n)}_{\rm Au}&=0.0918, &V^{(n)}_{\rm Au}&=0.146;
\nn
D_{\rm Al} &= 0.0362, &S^{(p)}_{\rm Al}&=0.0155, &V^{(p)}_{\rm
  Al}&=0.0161, &S^{(n)}_{\rm Al}&=0.0167, &V^{(n)}_{\rm Al}&=0.0173;
\end{align}
for gold and aluminium, respectively.

Finally, the branching ratio is defined as the transition rate, (see
\eq{Gconv}), divided by the capture rate, the latter given in
Ref.~\cite{Suzuki:1987jf}:
\begin{align}
\Gamma^{\rm capt}_{{\rm Au}}&=8.7\times 10^{-15}~{\rm MeV},&
\Gamma^{\rm capt}_{{\rm Al}}&=4.6\times 10^{-16}~{\rm MeV}\, .
\end{align}

\section{Mass eigenstates vs. mass insertions calculations}
\label{sec:memi}

For each process, we have given the exact one-loop expressions
calculated in the mass eigenbasis (ME). These formulae are compact and
well suited for numerical computations, however, do not allow for an
easy understanding of the qualitative behaviour of the LFV amplitudes
for various choices of the MSSM parameters.  Therefore, in this
Section we expand the Wilson coefficients in terms of the ``mass
insertions'', defined as the off-diagonal elements (both flavour
violating and flavour conserving) of the mass matrices. Such an
expansion allows us to:
\begin{itemize}
\item Recover the direct analytical dependence of the results on the
  MSSM Lagrangian parameters.
\item Prove analytically the expected decoupling features of the
  amplitudes in the limit of a heavy SUSY spectrum. In the case of
  Higgs boson decays, we also identify explicitly the terms decoupling
  only with the heavy CP-odd Higgs mass $M_A$ (which also determines
  the heavy CP even and the charged Higgs masses).  The decoupling
  properties serve also as an important cross-check of the correctness
  of our calculations.
\item Test the dependence of the results on the pattern of the MSSM
  spectrum and the size of the mass splitting between SUSY particles.
\item Better understand the possible cancellations between various
  types of contributions and correlations between different LFV
  processes.
\end{itemize}

The mass insertion expansion in flavour off-diagonal terms has been
used for a long time in numerous articles on the subject. However,
often various simplifying assumptions have been made, i.e. some terms
have been neglected or a simplified pattern of the slepton spectrum
was considered.  This is understandable as a consistent MI expansion
of the amplitudes for the LFV processes in the MSSM, mediated by the
virtual chargino and neutralino exchanges, is technically challenging.
The standard approach used in literature is to calculate
diagrammatically the LFV amplitudes with the ``mass insertions''
treated as the new interaction vertices. We follow the common practice
and normalise such slepton mass insertions to dimensionless
``$\Delta$-parameters''\footnote{We assume that trilinear $A_l$,
  $A'_l$ terms scale linearly with the slepton mass scale.}:
\bea
\Delta_{LL}^{IJ}
=\frac{(M_{LL}^2)^{IJ}}{\sqrt{(M_{LL}^2)^{II}(M_{LL}^2)^{JJ}}}\,,
&\qquad&
\Delta_{RR}^{IJ}
=\frac{(M_{RR}^2)^{IJ}}{\sqrt{(M_{RR}^2)^{II}(M_{RR}^2)^{JJ}}}\,,\nn
\Delta_{LR}^{IJ} =
\frac{A_l^{IJ}}{\left((M_{LL}^2)^{II}(M_{RR}^2)^{JJ}\right)^{1/4}}\,,
&\qquad&
\Delta_{LR}^{'IJ} =
\frac{A_l^{'IJ}}{\left((M_{LL}^2)^{II}(M_{RR}^2)^{JJ}\right)^{1/4}}\,,
\label{eq:midef}
\eea
where $M_{LL}^2, M_{RR}^2, A_l, A_l'$ are the slepton soft mass
matrices and trilinear terms. 

As lepton flavour violation is already strongly constrained
experimentally, it is sufficient to expand the amplitudes up to the
first order in flavour-violating $\Delta$'s. For instance, the
{effective} vertices listed in Sec.~\ref{sec:lfv} take the
schematic form:
\bea
F^{IJ} &=& \frac{1}{(4\pi)^2}\left(F_{LL}^{IJ} \Delta_{LL}^{IJ} +
F_{RR}^{IJ} \Delta_{RR}^{JI} \right. \nn
&+& \left. F_{ALR}^{IJ} \Delta_{LR}^{JI} + F_{BLR}^{IJ}
\Delta_{LR}^{IJ*} + F_{ALR}^{'IJ} \Delta_{LR}^{'JI} + F_{BLR}^{'IJ}
\Delta_{LR}^{'IJ*}\right)\;.
\label{eq:miform}
\eea
The MSSM contributions to $F_{LL}, \ldots, F_{BLR}'$ can be classified
according to their decoupling behaviour, distinguishing the following
types ($M$ denotes the average SUSY mass scale):
\begin{enumerate}
 \item Effects related to the diagonal trilinear slepton soft terms or
   to the off-diagonal elements of supersymmetric fermion mass
   matrices, decoupling as $v^2/M^2$.
 \item Effects related to the external momenta of the (on-shell) Higgs
   or $Z^0$ bosons, decoupling as $M_h^2/M^2$ or $M_Z^2/M^2$ (we
     did not include the $M_Z$ dependence as it is not necessary for
     the considered processes).
 \item Non-decoupling effects related to the 2HDM structure of the
   MSSM. Such contributions are constant in the limit of a heavy SUSY
   scale $M$ but, in case of the SM-like Higgs boson $h$, decouple
   with the CP-odd Higgs mass like $v^2/M_{A}^2$ (the effective
   couplings of heavier $H,A$ bosons do not exhibit such a
   suppression). They are proportional either to the lepton Yukawa
   couplings or to the non-holomorphic $A'_l$ terms.
\end{enumerate}

The structure of the box diagrams is more complicated as they carry 4
flavour indices. Their MI expansion is given in
Appendix~\ref{app:lboxmi}.  All box diagram contributions decouple at
least as $v^2/M^2$.

Calculating consistently the quantities $F_{LL}, \ldots, F_{BLR}'$ to
the order $v^2/M^2$ is not trivial for chargino and neutralino
contributions.  If the MI expansion is used only for the sfermion mass
matrices but the calculations for the supersymmetric fermions are done
in the mass eigenbasis, the direct dependence on the Lagrangian
parameters is hidden and the decoupling properties of the amplitude
cannot be seen directly. However, one can also treat the off-diagonal
entries of the chargino and neutralino mass matrices as ``mass
insertions''.  With such an approach, the final result is expressed
explicitly in terms of Lagrangian parameters, but the computations can
get very complicated.  At the order $v^2/M^2$ one needs to include
diagrams with all combinations of two fermionic mass insertions (each
providing one power of $v/M_1$, $v/M_2$ or $v/\mu$) or flavour
diagonal slepton terms originating from trilinear $A$-terms (providing
powers of $v A_l/M^2$, $v A'_l/M^2$). Thus, to obtain an expansion of
the $F$'s in~\eq{eq:miform}, one needs to formally go to the 3rd order
of MI expansion, adding all diagrams with up to two flavour conserving
and one flavour violating mass insertion. Therefore, the number of
diagrams grows quickly with the order of the expansion and such a
method is tedious and prone to calculational mistakes.

In our paper, we employ a recently developed technique using a purely
algebraic MI expansion of the ME amplitudes listed in
Sec.~\ref{sec:lfv}, without the need for direct diagrammatic MI
calculations (``FET theorem'')~\cite{Dedes:2015twa}, automatised in
the specialised {\tt MassToMI} {\em Mathematica}
package~\cite{Rosiek:2015jua, Rosiek:2017cwn}.  The use of this
package and full automation of the calculations allows us to perform
the required 3rd order MI expansion for a completely general SUSY mass
spectrum, without making any simplifying assumptions. Such a result
would be very difficult to obtain diagrammatically, as in the
intermediate steps of the calculations (before accounting for the
cancellations and simplifications between various contributions) the
expressions may contain up to tens of thousand terms, even if the
final results collected in Appendix~\ref{app:miexp} are again
relatively compact. In detail:
\begin{itemize}
\item We perform the expansion always up to the lowest non-vanishing
  order in the slepton LFV terms, taking into account the possible
  cancellations.  Compared to previous analyses, we consider the
  non-holomorphic trilinear soft terms as well.
\item In the MI expanded expressions we include all terms decreasing
  with the SUSY mass scale as $v^2/M_{SUSY}^2$ (or slower), where
  $M_{SUSY}$ denotes any of the relevant mass parameters in the MSSM
  Lagrangian (apart from the soft Higgs mass terms): diagonal soft
  slepton masses, gaugino masses $M_1,M_2$ or the $\mu$ parameter.
\item
We do not assume degeneracy or any specific hierarchy for the
sleptons, sneutrinos or supersymmetric fermion masses.
\item In calculating the LFV Higgs decays we keep the leading terms in
  the external Higgs boson mass ($m_{h(H)}^2/M_{SUSY}^2$).
\end{itemize}
The full set of the expanded expressions in the MI approximation for
the photon, $Z^0$ and CP-even Higgs leptonic penguins and for the
4-lepton box diagrams is collected in Appendix~\ref{app:miexp}.

\begin{figure}[tb]
\begin{center}
  \begin{tabular}{ccc}
    \includegraphics[width=0.307\textwidth]{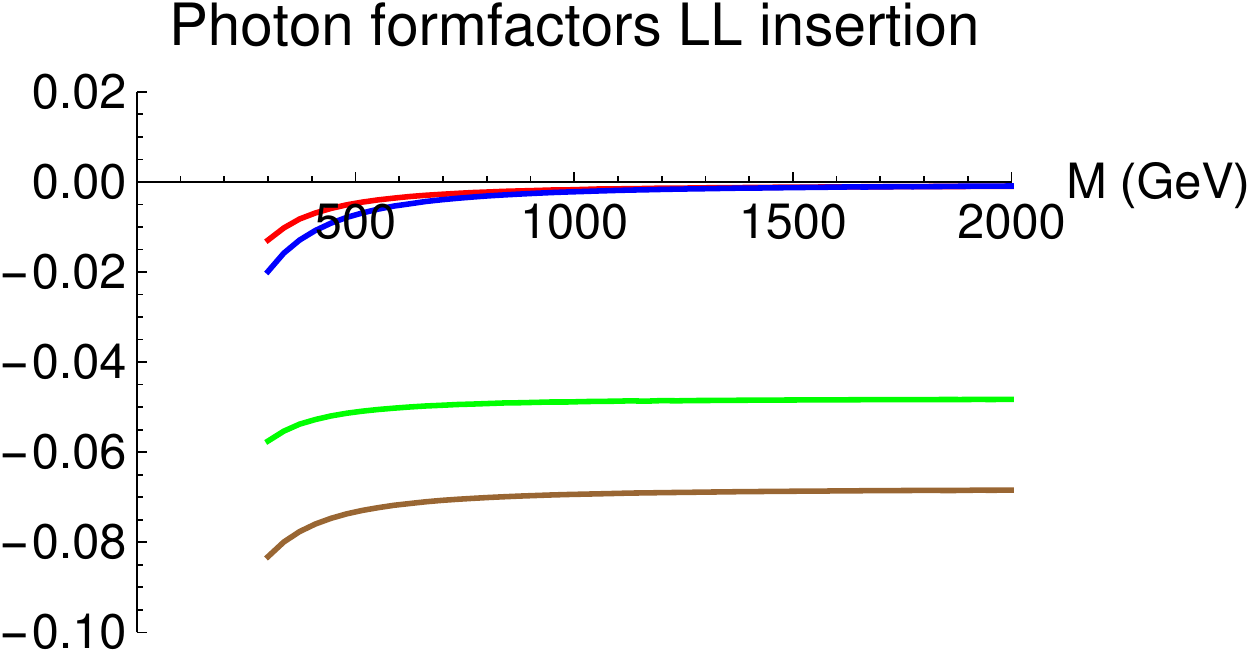}&
    \includegraphics[width=0.307\textwidth]{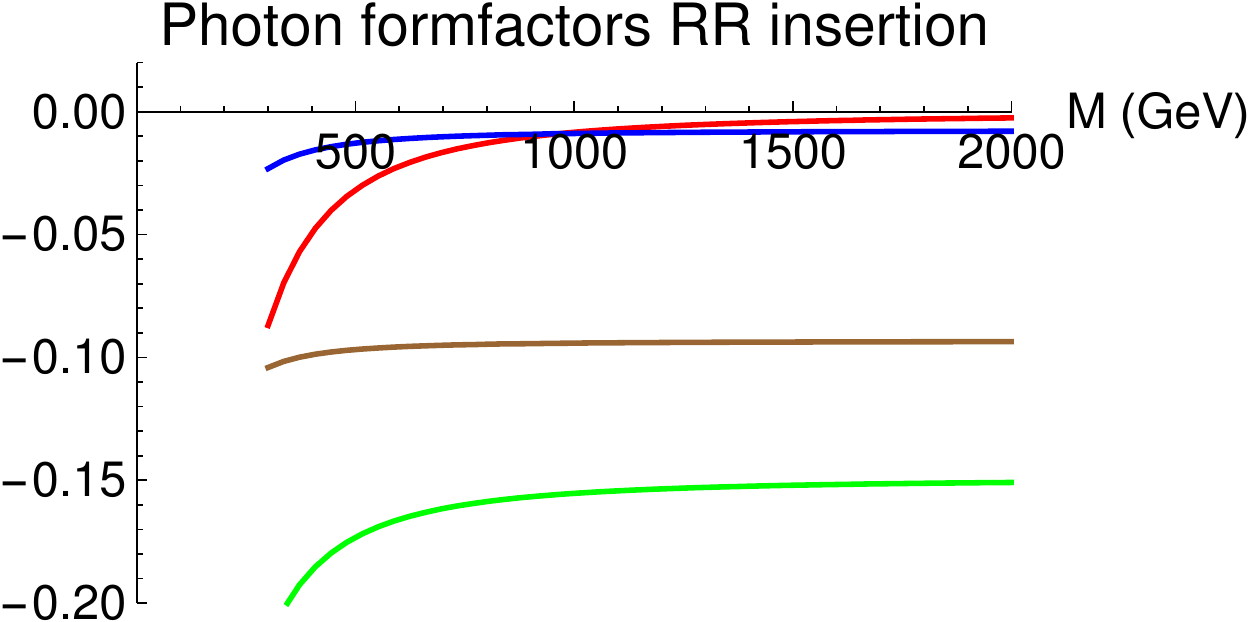}&
    \includegraphics[width=0.307\textwidth]{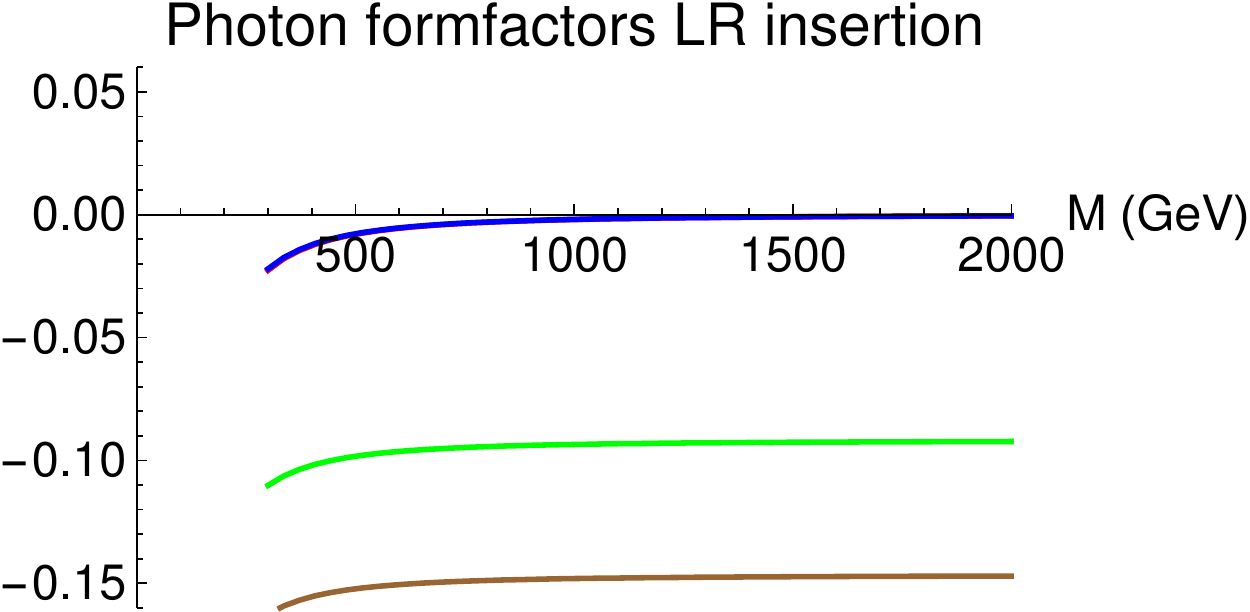}\\[3mm]
    \includegraphics[width=0.307\textwidth]{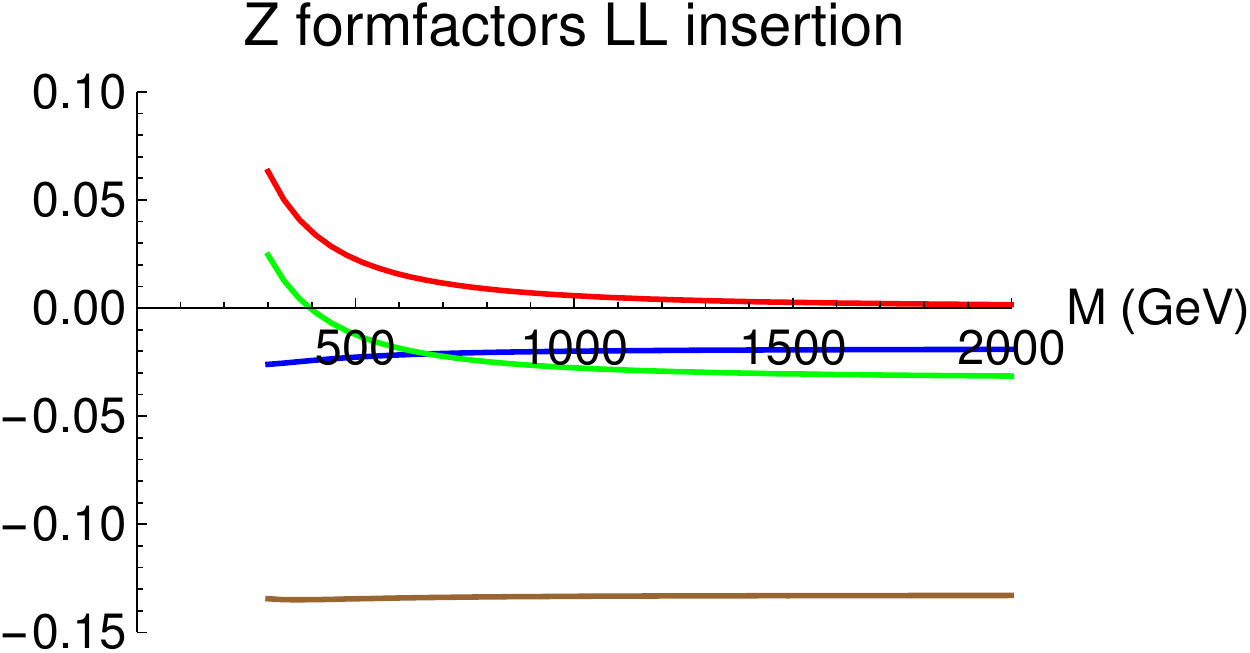}&
    \includegraphics[width=0.307\textwidth]{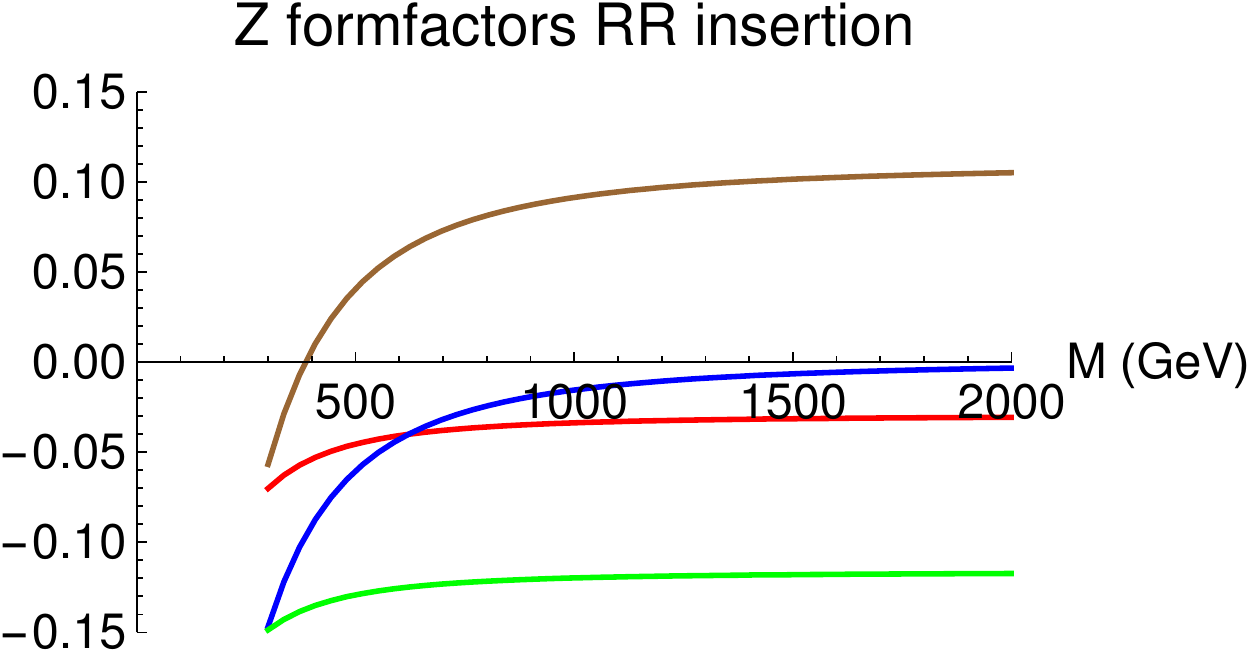}&
    \includegraphics[width=0.307\textwidth]{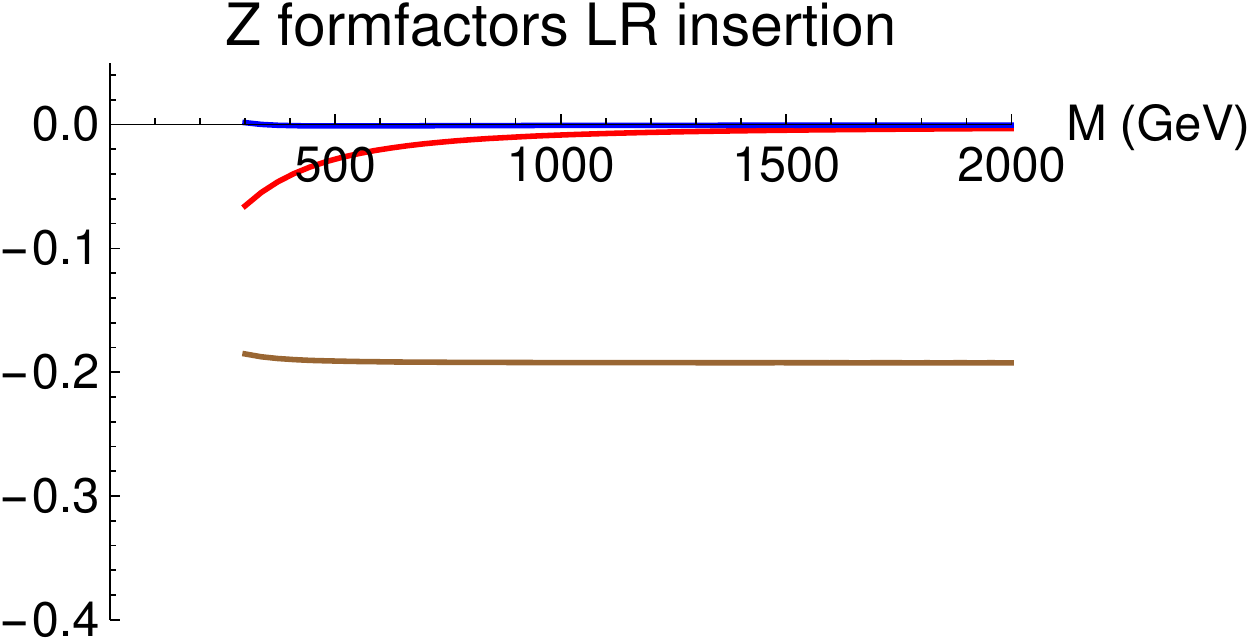}\\[3mm]
    \includegraphics[width=0.307\textwidth]{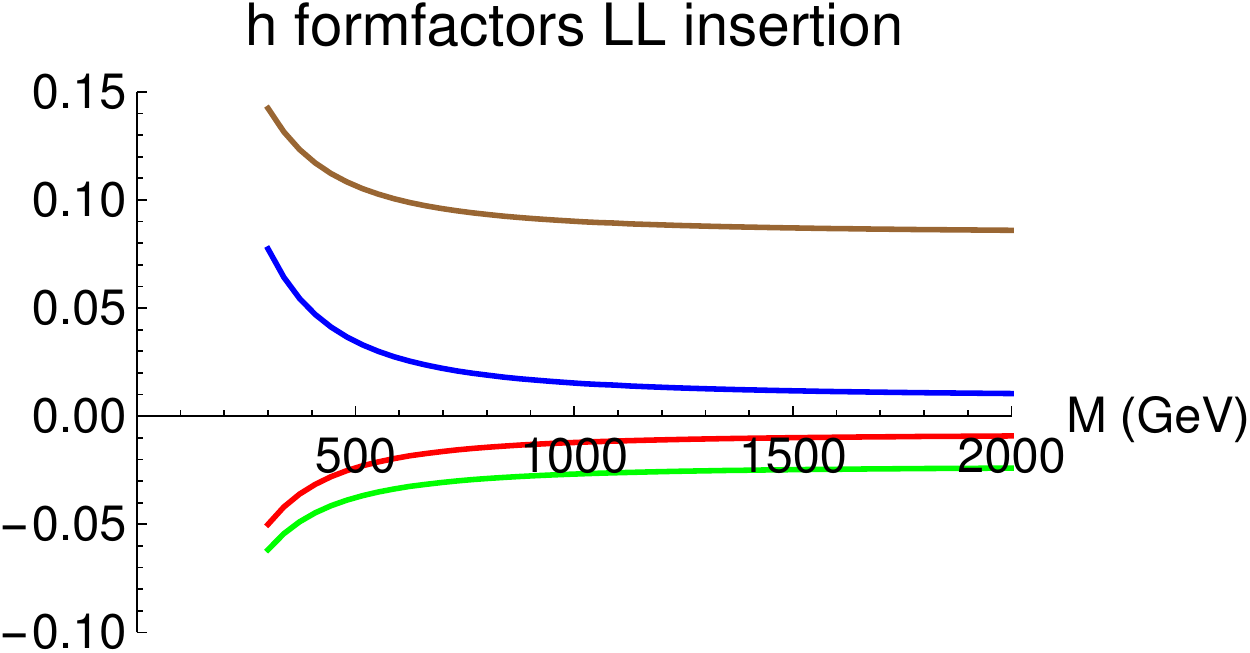}&
    \includegraphics[width=0.307\textwidth]{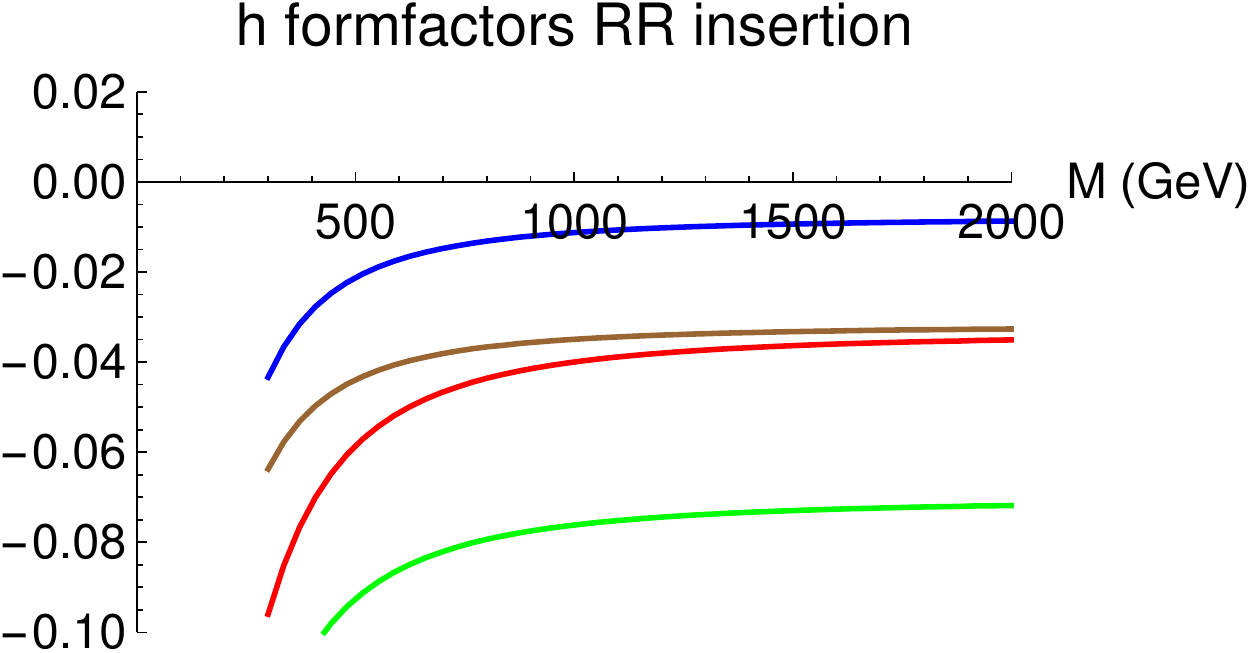}&
    \includegraphics[width=0.307\textwidth]{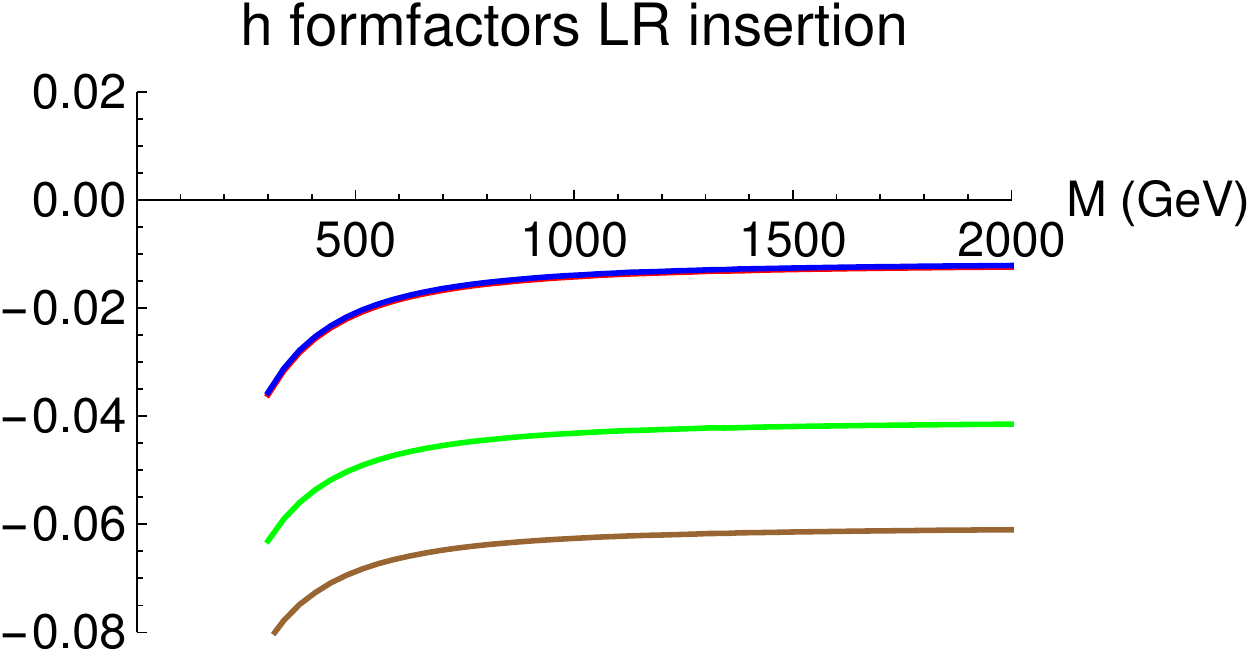}\\[3mm]
 \end{tabular}
 \caption{\small Accuracy of MI expansion for the penguin
   amplitudes. The curves show the ratio defined in~\eq{eq:miacc}.
   Red and blue lines: $\Delta F$ for left and right couplings
   assuming a spectrum of~\eq{eq:specdef} for both the MI and ME
   expressions. Brown and green lines: $\Delta F$ (again for left and
   right couplings, respectively) assuming spectrum~(\ref{eq:specdef})
   for ME expressions but an universal degenerate sfermion mass in MI
   expressions.  This assumption is inconsistent with non-zero
   off-diagonal elements of the mass matrices, which imply
   non-degenerate mass eigenstates. The plots show that the associated
   error can be numerically sizeable.  The average SUSY mass scale $M$
   (assumed to be equal to $M_2 = m_{\tilde\mu_L} = m_{\tilde\mu_R}$)
   is shown on the horizontal axis. \label{fig:acc}}
\end{center}
\end{figure}

We illustrate the accuracy of the derived MI formulae in
Fig.~\ref{fig:acc}.  The plots show the ratio of the MI expanded
couplings over the ones obtained in the mass eigenbasis with exact
diagonalization. For this purpose, we start from the following setup
where all mass parameters are given in GeV:
\bea
\begin{array}{lp{1cm}lp{1cm}l} 
\tan\beta=5 && m_{\tilde \mu_L} = 300 && A_{\mu\mu} = A'_{\mu\mu} =
  0.1 \sqrt{m_{\tilde \mu_L} m_{\tilde \mu_R}}\\
\mu = 200+100i && m_{\tilde \tau_L} = 330\\
M_1 = 150 && m_{\tilde \mu_R} = 300 && A_{\tau\tau} =
  A'_{\tau\tau} = 0.1 \sqrt{m_{\tilde \tau_L} m_{\tilde \tau_R}}\\
M_2 = 300 && m_{\tilde \tau_R} = 350 \\
\end{array}
\label{eq:specdef}
\eea
Next, to see the decoupling effects we scale this spectrum uniformly
up to slepton masses of 2 TeV.  For each of the six penguin Wilson
coefficients describing the transition between 2nd and 3rd generation,
$F_{\gamma L(R)}^{23}$ (\eq{eq:lgdef}), $F_{Z L(R)}^{23}$
(\eq{eq:lzdef}) and $F_{hL}^{23}\equiv F_h^{232}$, $F_{hR}^{23}\equiv
F_h^{322}$ (\eq{eq:lhdef}) we plot the quantity
\bea
\Delta F = \left|\frac{F_\mathrm{MI}}{F_\mathrm{ME}}\right| -1\,,
\label{eq:miacc}
\eea
as a function of the average slepton mass. The accuracy of left-handed
(right-handed) Wilson coefficients is illustrated with red(blue)
lines.  As can be seen from Fig.~\ref{fig:acc}, the accuracy of MI
expanded amplitudes is very good even for light SUSY particles and for
$M_{SUSY}>500$ GeV always better than 95\%.

Many analyses published to date for simplicity did not include the
complete set of the contributions scaling like $v/M$ order and/or
assumed a partially or fully degenerate SUSY spectrum.  This procedure
is inconsistent with non-zero off-diagonal elements of mass matrices,
because the latter enforce unequal eigenvalues of the corresponding
mass matrix.  To illustrate the numerical effects arising from the
incorrect neglect of SUSY mass splitting we plot the ratio of our
expressions in the MI approximation for penguin Wilson coefficients
calculated for degenerate slepton masses (equal to 300 GeV rescaled by
a common factor; other parameters as in~\eq{eq:specdef}) and the exact
mass eigenbasis formulae (calculated with non-degenerate sfermion
spectrum of~\eq{eq:specdef}) in Fig.~\ref{fig:acc}.  The accuracy of
left-handed (right-handed) MI expanded Wilson coefficients with
degenerate slepton spectrum is shown in green(brown). In this case
discrepancy is much larger, of the order of 10\%-40\%, and does not
disappear when increasing the total SUSY scale.

\begin{figure}[tb]
\begin{center}
  \begin{tabular}{cc}
    \includegraphics[width=0.48\textwidth]{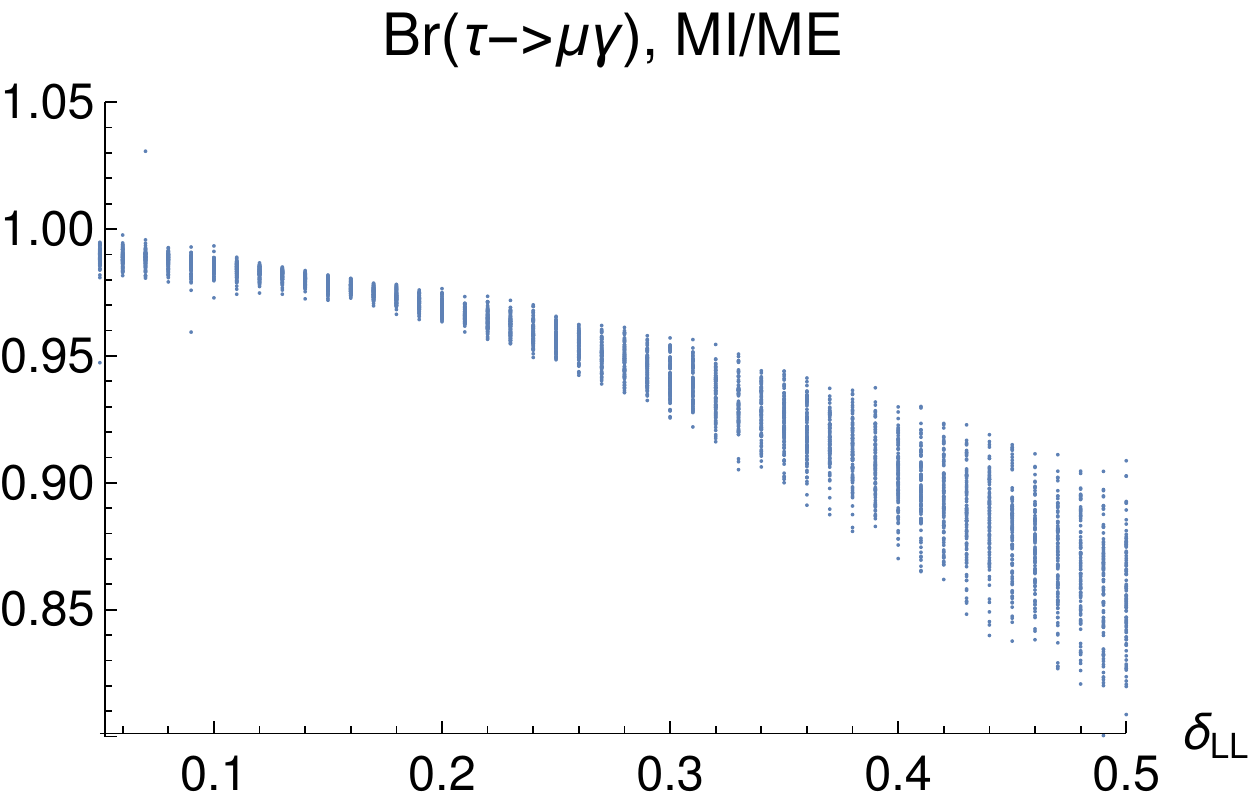}&
    \includegraphics[width=0.48\textwidth]{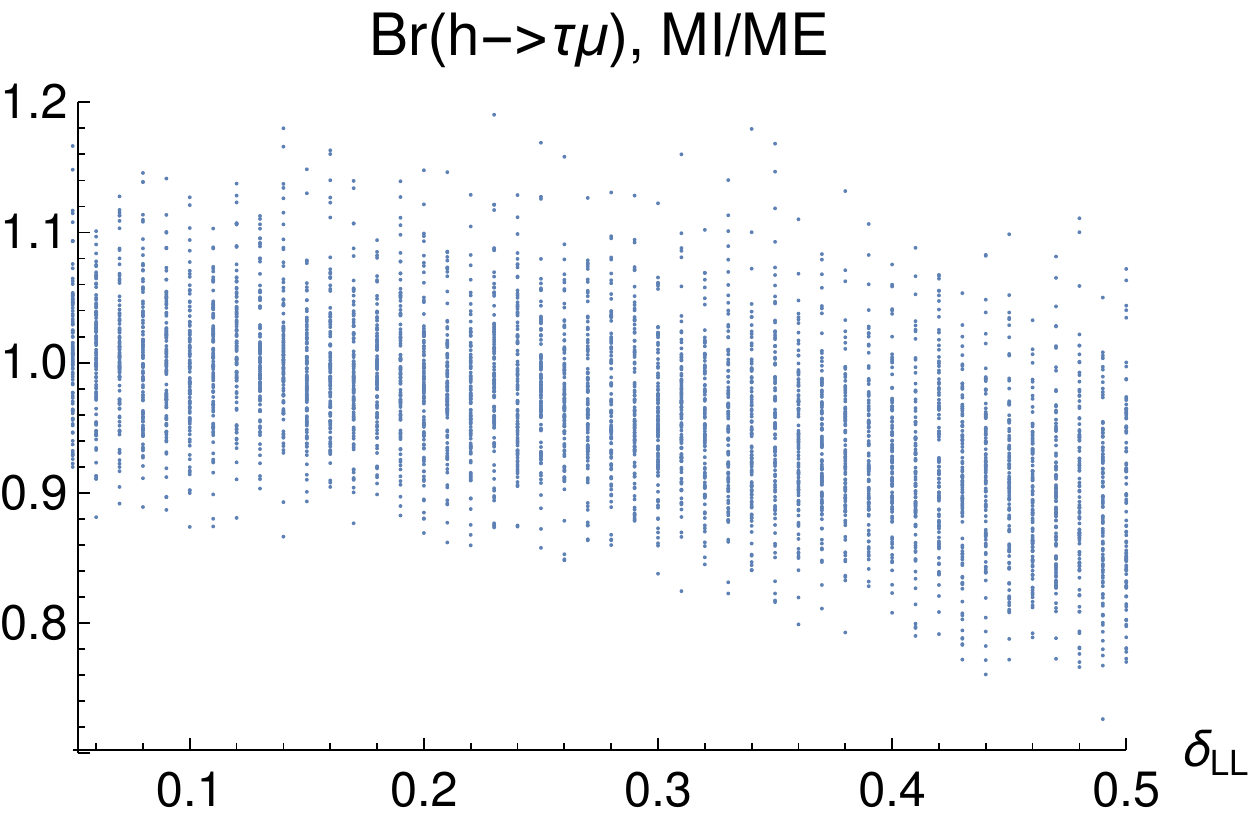}\\[3mm]
    \includegraphics[width=0.48\textwidth]{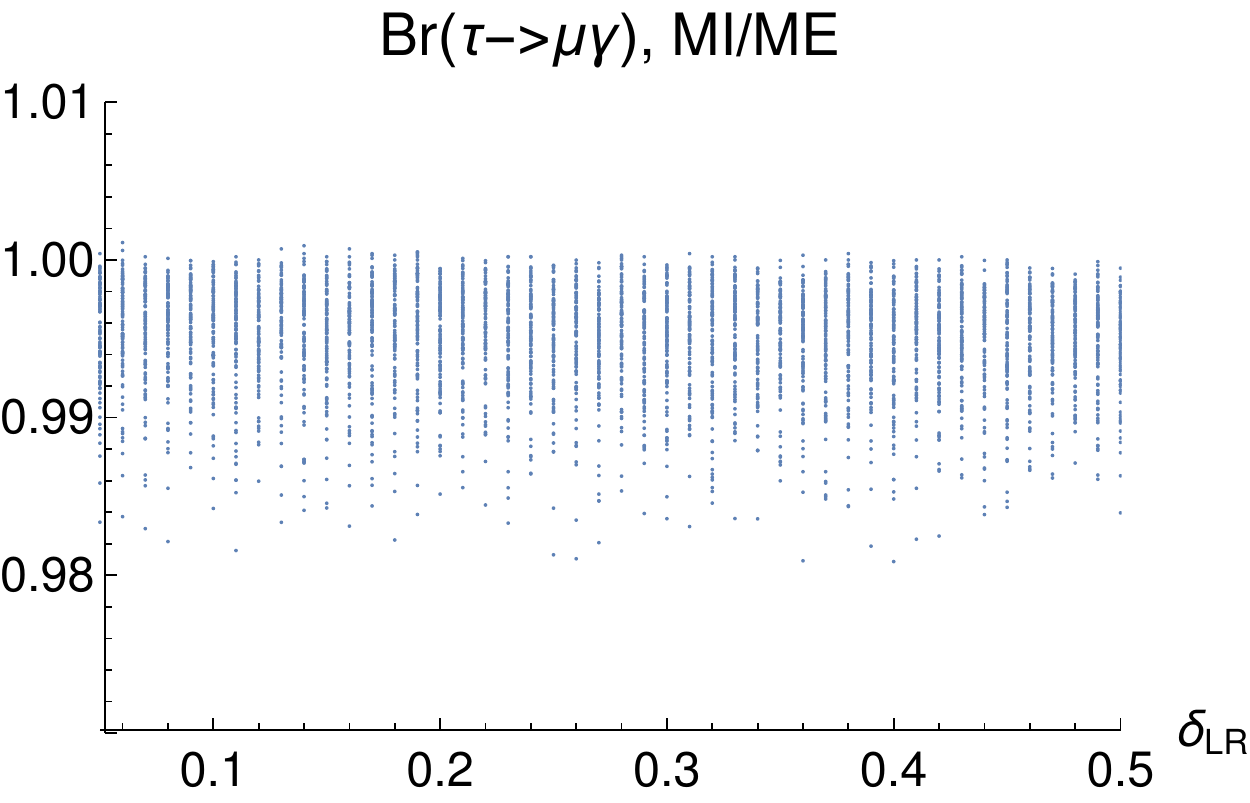}&
    \includegraphics[width=0.48\textwidth]{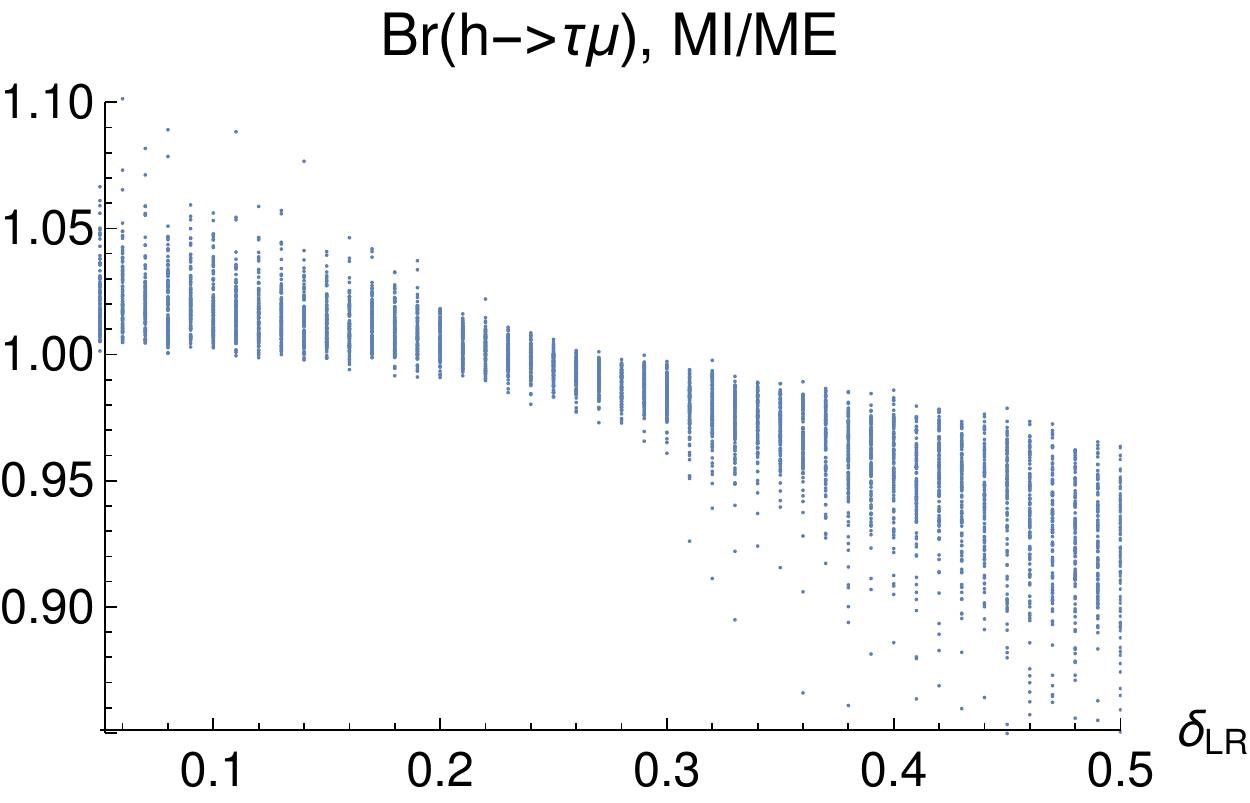}\\[3mm]
 \end{tabular}
 \caption{\small Accuracy of MI expansion for the $\tau\to\mu\gamma$
   and $h\to\tau\mu$ decay rates. Points show the ratio
   $\Br(\tau\to\mu\gamma)^{MI}/\Br(\tau\to\mu\gamma)^{ME}$ (lower and
   upper left panel) and $\Br(h\to\tau\mu)^{MI}/\Br(h\to\tau\mu)^{ME}$
   (lower and upper right panel) as a function of LL and LR mass
   insertion for $\tan\beta=5$ and random choice of other model
   parameters (see~\eq{eq:random}). \label{fig:scan}}
\end{center}
\end{figure}

Some papers on the LFV in the MSSM, like e.g.
Refs.~\cite{Paradisi:2005fk, Arganda:2015uca}, deal with general SUSY
spectra. In order to compare the accuracy of the MI approximation
derived in our analysis with previous works, we plotted in
Fig.~\ref{fig:scan} the ratios of $Br(\tau\to\mu \gamma)$ and
$Br(h\to\tau\mu)$ calculated using the exact (ME) and MI-expanded
formulae scanning over randomly chosen MSSM mass spectra.  In
particular, in Fig.~\ref{fig:scan} we assume $\tan\beta=5$,
$\alpha-\beta = -\pi/2-\pi/100$, $m_h=125$ GeV, diagonal $A$ terms
which are proportional to the lepton Yukawa couplings ($A_l^{II} =
A_l^{'II} = Y_l^I ((M_{LL}^2)_{II} (M_{RR}^2)_{JJ})^{1/4}$) and we
vary the mass parameters randomly and independently in the following
ranges (all values are given in GeV and we set $M_1= M_2/2$):
\bea
M_A &\in& (200,500) \qquad\qquad \mu, M_2,
m_{\tilde\tau_L},m_{\tilde\mu_L},m_{\tilde\tau_R},m_{\tilde\tau_R}
\in\ (500,1000),.
\label{eq:random}
\eea
As can be seen from upper left panel of Fig.~\ref{fig:scan}, even for
$\delta_{LL}^{32}=0.5$ the accuracy of our MI expansion is better than
about 15\%. This can be compared with the corresponding right panel of
Fig.~8 in ref.~\cite{Paradisi:2005fk} - there the difference between
MI and ME calculation for the same value of $\delta_{LL}^{32}=0.5$ is
20\%-70\%, also the spread of points around the parabolic shape
arising from neglected $(\delta_{LL}^{32})^2$ terms is much larger,
50\% against a maximum of 10\% in our approach. It is worth noting
that the agreement in the lower left plot of Fig.~\ref{fig:scan} is
almost perfect everywhere, which could be attributed to the fact that
terms of higher order in $\delta_{LR}$ are suppressed by additional
$v/M$ powers (with $A$ terms scaling linearly with $M$, like we
choose) and thus small even for large $\delta_{LR}$ values.

Accuracy of our MI expansion for $Br(h\to\tau\mu)$ shown in the right
upper panel of Fig.~\ref{fig:scan} is worse, up to 20\%, because we
include only non-decoupling $LL$ terms in our formulae, which is not a
fully satisfying approximation for a SUSY scale in the range of
500-1000 GeV.  Concerning the expansion in $\delta_{LR}^{32}$ (where
only decoupling terms contribute), we include them consistently and
the accuracy is much better.  This can be compared with
Ref.~\cite{Arganda:2015uca}, considering the same process. The
agreement for $\delta_{LL}^{32}=0.5$ in upper left panel of Fig.~6
(``general scenario'') in~\cite{Arganda:2015uca} is better than ours,
as they consistently include all $LL$ terms, not just non-decoupling
ones.  However, for $\delta_{LR}^{32}$ (lower left panel of Fig.~6 in
~\cite{Arganda:2015uca}) numerical accuracy of our formulae seems to
be similar or even better. In general, no significant deviations
should be expected here, as for this process our approach and the
analysis of Ref.~\cite{Arganda:2015uca} are equivalent up to the
chosen calculational technique (FET vs. diagrammatic MI calculation)
and, eventually, the selection of the included or neglected
contributions.

\section{Phenomenological analysis}
\label{sec:pheno}

\subsection{Generic bounds on LFV parameters}
\label{sec:genmi}

As outlined in the introduction, flavour violation in the charged
lepton sector is strongly constrained experimentally. In
Table~\ref{tab:leplim} we collect the current and expected future
experimental bounds on the processes discussed so far.

\begin{table}[tb]
\begin{center}  
  \begin{tabular}{|l|r|c|r|c|}
    \hline
 & Experimental upper bound & CL & Future sensitivity & CL \\
\hline
$ \tau \to e \gamma $ & $ 3.3 \times 10^{-8} $ \cite{Aubert:2009ag} &
90\% & $ 10^{-9} $ \cite{Hayasaka:2013dsa,Cavoto:2017kub} & 90\% \\
$ \tau \to \mu \gamma $ & $ 4.4 \times 10^{-8} $
\cite{Aubert:2009ag,Hayasaka:2007vc} & 90\% & $ 10^{-9} $
\cite{Hayasaka:2013dsa} & 90\% \\
$ \mu \to e \gamma $ & $ 5.7 \times 10^{-13} $ \cite{Adam:2013mnn} &
90\% & $ 6 \times 10^{-14} $ \cite{Baldini:2018nnn} & 90\% \\
\hline
$ Z \to \mu e $ & $ 7.5 \times 10^{-7} $ \cite{Aad:2014bca} & 95\% & &
\\
$ Z \to \mu \tau $ & $ 1.2 \times 10^{-5} $ \cite{Abreu:1996mj} & 95\%
& & \\
$ Z \to \tau e $ & $ 9.8 \times 10^{-6} $ \cite{Abreu:1996mj}& 95\% &
& \\
\hline
$ \mu \to e^- e^+ e^- $ & $ 1.0 \times 10^{-12} $
\cite{Bellgardt:1987du} & 90\% & $ 10^{-16} $
\cite{Blondel:2013ia,Berger:2014vba} & 90\% \\
$ \tau \to e^- e^+ e^- $ & $ 2.7 \times 10^{-8} $
\cite{Hayasaka:2010np} & 90\% & & \\
$ \tau \to \mu^- \mu^+ \mu^- $ & $ 2.1 \times 10^{-8}
$\cite{Hayasaka:2010np} & 90\% & & \\
$ \tau \to e^- \mu^+ \mu^- $ & $ 2.7 \times 10^{-8} $
\cite{Hayasaka:2010np} & 90\% & & \\
$ \tau \to e^+ \mu^- \mu^- $ & $ 1.7 \times 10^{-8} $
\cite{Hayasaka:2010np} & 90\% & &\\
$ \tau \to \mu^- e^+ e^- $ & $ 1.8 \times 10^{-8} $
\cite{Hayasaka:2010np} & 90\% & & \\
$ \tau \to \mu^+ e^- e^- $ & $ 1.5 \times 10^{-8} $
\cite{Hayasaka:2010np} & 90\% & & \\
\hline
$ h \to e \tau $ & $ 6.1 \times 10^{-3} $ \cite{CMS:2017onh} & 90\%
&&\\
$ h \to \mu \tau $ & $ 2.5 \times 10^{-3} $ \cite{CMS:2017onh} & 90\%&
& \\
$ h \to \mu e $ & $ 3.6 \times 10^{-4} $ \cite{CMS:2015udp} & 90\% &
&\\
\hline
$(\mu\to e)_{Au}$ & $7.0 \times 10^{-13}$ \cite{Bertl:2006up} & 90\%
&& \\
$(\mu\to e)_{Al}$ & & & $10^{-16}$ \cite{Abrams:2012er} & 90\% \\
\hline
\end{tabular}
\caption{\small Upper bounds on LFV decays of charged leptons. $h$
  denotes the SM-like Higgs boson\label{tab:leplim}}
\end{center}
\end{table}

Assuming the absence of fine-tuned cancellations between different
flavour violating parameters, the order of magnitude of the bounds on
a given flavour violating entry $\Delta$ can be obtained by assuming
that it is the only source of flavour violation. At the lowest order
in the MI expansion, any LFV observable $X$ scales like $\Delta^2$:
\bea
X \approx f(m_1,\ldots, m_n) |\Delta|^2\,,
\eea
where $f$ is a known (non-negative) function of diagonal mass
parameters - for any given process it can be extracted from the
expanded expressions listed in Appendix~\ref{app:miexp}.  Thus, the
experimental bound on $\Delta$ from a given measurement can be written
as:
\bea
|\Delta| \leq \sqrt{\frac{X^{\rm exp}}{f(m_1,\ldots, m_n)}}
\sqrt{\frac{X^{\rm future}}{X^{\rm exp}}} \equiv \Delta(m_1,\ldots, m_n)
\sqrt{\frac{X^{\rm future}}{X^{\rm exp}}}
\eea
where by $X^{\rm exp}$ we denote one of the current experimental
bounds listed in Sec.~\ref{sec:genmi} and $X^{\rm future}$ is the
expected future sensitivity.

To estimate the order of magnitude of the bounds on all types of mass
insertions, we assume a common mass scale $M$ for all flavour diagonal
SUSY parameters:
\bea
&& m_{\tilde e_{LI}} = m_{\tilde e_{RI}} = M_1 = M_2 = \mu = M_A = M\,,\nn
&& A_{\ell}^{II} = A_{\ell}^{'II} = Y_{\ell}^I\; M\,.
\label{eq:mpatt}
\eea
Currently, the strongest bounds on the dimensionless LFV parameters
$\Delta$ defined in~\eq{eq:miform} originate from the radiative lepton
decays $\ell\to\ell'\gamma$. We list such bounds for the parameter
setup defined in~\eq{eq:mpatt} and for the SUSY scale of $M=400$ GeV
in Table~\ref{tab:llgbounds}.

\begin{table}[tb]
\begin{center}
\begin{tabular}{|l|l|c|c|c|c|c|c|}
\hline
Process & $(I,J)$ & $\Delta_{LL}^{IJ}$ & $\Delta_{RR}^{IJ}$ &
$\Delta_{LR}^{IJ}$ & $\Delta_{RL}^{IJ}$ & $\Delta_{LR}^{'IJ}$ &
$\Delta_{RL}^{'IJ}$ \\
\hline
$\tan\beta=2$ & \multicolumn{7}{|l|}{}\\
\hline
$ \mu\to e \gamma$ & $(2,1)$ & $ 8.4 \cdot 10^{-4}$ & $ 5.0 \cdot
10^{-3}$ & $ 8.4 \cdot 10^{-6}$ & $ 8.3 \cdot 10^{-6}$ & $ 4.1 \cdot
10^{-6}$ & $ 4.1 \cdot 10^{-6}$ \\
$\tau \to \mu \gamma$ & $(3,2)$ &$ 5.3 \cdot 10^{-1}$ & 
${\cal O}(1)$ & $ 9.1 \cdot 10^{-2}$ & $ 9.1 \cdot 10^{-2}$ & $ 4.5
\cdot 10^{-2}$ & $ 4.5 \cdot 10^{-2}$ \\
$ \tau \to e \gamma$ & $(3,1)$ &$ 4.6 \cdot 10^{-1}$ &
${\cal O}(1)$ & $ 7.8 \cdot 10^{-2}$ & $ 7.8 \cdot 10^{-2}$ & $ 3.9
\cdot 10^{-2}$ & $ 3.8 \cdot 10^{-2}$ \\
\hline
$\tan\beta=20$ & \multicolumn{7}{|l|}{}\\
\hline
$ \mu\to e \gamma$ & $(2,1)$ & $ 1.0 \cdot 10^{-4}$ & $ 4.5 \cdot
10^{-4}$ & $ 7.5 \cdot 10^{-5}$ & $ 7.4 \cdot 10^{-5}$ & $ 3.7 \cdot
10^{-6}$ & $ 3.7 \cdot 10^{-6}$ \\
$ \tau \to \mu \gamma$ & $(3,2)$ & $ 6.5 \cdot 10^{-2}$ & $ 2.9 \cdot
10^{-1}$ & $ 8.2 \cdot 10^{-1}$ & $ 8.2 \cdot 10^{-1}$ & $ 4.0 \cdot
10^{-2}$ & $ 4.0 \cdot 10^{-2}$ \\
$ \tau \to e \gamma$ & $(3,1)$ & $ 5.7 \cdot 10^{-2}$ & $ 2.5 \cdot
10^{-1}$ & $ 7.0 \cdot 10^{-1}$ & $ 7.0 \cdot 10^{-1}$ & $ 3.4 \cdot
10^{-2}$ & $ 3.4 \cdot 10^{-2}$ \\
\hline
\end{tabular}
\end{center}
\caption{\small Upper bounds on the LFV parameters $\Delta$ from
  radiative charged lepton decays for the MSSM spectrum defined
  in~\eq{eq:mpatt} and a SUSY scale of $M=400$ GeV.  All bounds scale
  (i.e. weaken) like $M^2$.
\label{tab:llgbounds}}
\end{table}

The 3-body decays of charged lepton lead to bounds which are
approximately one order of magnitude weaker. In
Table~\ref{tab:l3bounds} we display the relative strength of such
bounds comparing them to the ones obtained from the radiative lepton
decays, i.e. the ratios of bounds from radiative decays over the ones
from 3-body decays. Such ratios remain constant with increasing $M$ up
to the scale where the non-decoupling Higgs penguin contributions
start to contribute. However, such effects occurs for $M\gsim 30$ TeV
for $\tau^\pm\to\mu^\pm\mu^\pm\mu^\mp$ and
$\tau^\pm\to e^\pm\mu^\pm\mu^\mp$ decays and for even higher $M$ for
the decays with electron pair in the final state. For such a large $M$
the branching ratios for all 3-body decays are, anyway, below the
current experimental sensitivities even for
${\cal O}(\Delta^{IJ})\sim 1$.

We do not display the bounds from LFV violating $Z^0$ decays as they
are much weaker (3 to 8 orders of magnitude depending on which
parameter $\Delta$ is chosen). This can be attributed to the large $Z$
boson width -- for comparable $\Gamma(Z\to \ell\ell')$ and
$\Gamma(\ell\to\ell'\gamma)$ partial decay widths the difference in
total widths leads to $\Br(\ell\to\ell'\gamma)\gg \Br(Z\to
\ell\ell')$. Thus, bounds from $\Br(Z\to \ell\ell')$ are not
competitive (nor they will be in the foreseeable future) compared to
these from other observables.

\begin{table}[tb]
\begin{center}
\begin{tabular}{|l|l|c|c|c|c|c|c|}
\hline
Process & $(I,J)$ & $\Delta_{LL}^{IJ}$ & $\Delta_{RR}^{IJ}$ &
$\Delta_{LR}^{IJ}$ & $\Delta_{RL}^{IJ}$ & $\Delta_{LR}^{'IJ}$ &
$\Delta_{RL}^{'IJ}$ \\
\hline
$\tan\beta=2$ & \multicolumn{7}{|l|}{}\\
\hline
$ \mu \to eee$ & $(2,1)$ & $ 1.7 \cdot 10^{+1}$ & $ 1.5 \cdot
10^{+1}$ & $ 1.6 \cdot 10^{+1}$ & $ 1.6 \cdot 10^{+1}$ & $ 1.6
\cdot 10^{+1}$ & $ 1.6 \cdot 10^{+1}$ \\
$ \tau \to \mu\mu\mu$ & $(3,2) $ & $ 1.5 \cdot 10^{+1}$ & $ 1.2 \cdot
10^{+1}$ & $ 1.4 \cdot 10^{+1}$ & $ 1.4 \cdot 10^{+1}$ & $ 1.4
\cdot 10^{+1}$ & $ 1.4 \cdot 10^{+1}$ \\
$ \tau \to \mu e^+ e^-$ & $(3,2) $ & $ 1.3 \cdot 10^{+1}$ & $ 1.1
\cdot 10^{+1}$ & $ 1.2 \cdot 10^{+1}$ & $ 1.2 \cdot 10^{+1}$ & $
1.2 \cdot 10^{+1}$ & $ 1.2 \cdot 10^{+1}$ \\
$ \tau \to eee$ & $(3,1) $ & $ 8.6 \cdot 10^{+0}$ & $ 8.2 \cdot
10^{+0}$ & $ 8.5 \cdot 10^{+0}$ & $ 8.5 \cdot 10^{+0}$ & $ 8.5
\cdot 10^{+0}$ & $ 8.5 \cdot 10^{+0}$ \\
$\tau \to e \mu^+ \mu^-$ & $(3,1)$ & $ 6.9 \cdot 10^{+0}$ & $ 6.7
\cdot 10^{+0}$ & $ 6.8 \cdot 10^{+0}$ & $ 6.8 \cdot 10^{+0}$ & $
6.8 \cdot 10^{+0}$ & $ 6.8 \cdot 10^{+0}$ \\
\hline
$\tan\beta=20$ & \multicolumn{7}{|l|}{}\\
\hline
$ \mu \to eee$ & $(2,1)$ & $ 1.6 \cdot 10^{+1}$ & $ 1.6 \cdot
10^{+1}$ & $ 1.6 \cdot 10^{+1}$ & $ 1.6 \cdot 10^{+1}$ & $ 1.6
\cdot 10^{+1}$ & $ 1.6 \cdot 10^{+1}$ \\
$ \tau \to \mu\mu\mu$ & $(3,2) $ & $ 1.4 \cdot 10^{+1}$ & $ 1.4 \cdot
10^{+1}$ & $ 1.4 \cdot 10^{+1}$ & $ 1.4 \cdot 10^{+1}$ & $ 1.4
\cdot 10^{+1}$ & $ 1.4 \cdot 10^{+1}$ \\
$ \tau \to \mu e^+ e^-$ & $(3,2) $ & $ 1.3 \cdot 10^{+1}$ & $ 1.2
\cdot 10^{+1}$ & $ 1.2 \cdot 10^{+1}$ & $ 1.2 \cdot 10^{+1}$ & $
1.2 \cdot 10^{+1}$ & $ 1.2 \cdot 10^{+1}$ \\
$ \tau \to eee$ & $(3,1) $ & $ 8.5 \cdot 10^{+0}$ & $ 8.5 \cdot
10^{+0}$ & $ 8.5 \cdot 10^{+0}$ & $ 8.5 \cdot 10^{+0}$ & $ 8.5
\cdot 10^{+0}$ & $ 8.5 \cdot 10^{+0}$ \\
$\tau \to e \mu^+ \mu^-$ & $(3,1)$ & $ 6.8 \cdot 10^{+0}$ & $ 6.8
\cdot 10^{+0}$ & $ 6.8 \cdot 10^{+0}$ & $ 6.8 \cdot 10^{+0}$ & $
6.8 \cdot 10^{+0}$ & $ 6.8 \cdot 10^{+0}$ \\
\hline
\end{tabular}
\end{center}
\caption{\small Ratios of upper bounds on the LFV parameters $\Delta$
  from the searches for 3-body and radiative decays of charged
  leptons. The MSSM spectrum is defined in~\eq{eq:mpatt}.
\label{tab:l3bounds}}
\end{table}

As can be seen in Table~\ref{tab:hbounds}, the bounds on $\Delta$
parameters from LFV flavour Higgs boson decay searches are much weaker
than those from the radiative charged lepton decays. However, in the
Higgs sector some effects proportional to lepton Yukawa couplings or
to the non-holomorphic terms are non-decoupling and are not weakened
by increasing $M$ like other contributions, for fixed Higgs sector
parameters.  In Table~\ref{tab:hbounds} we assume
\bea
\alpha - \beta = - \frac{\pi}{2} -\gamma \,,
\label{eq:alval}
\eea
with $\gamma = \pi/100$. Using the tree-level relations of the MSSM
Higgs sector in the limit of $\tan\beta>1$ and small values of
$\gamma$ one has
\bea
M_A = M_Z \sqrt{\frac{\sin 2(\alpha+\beta)}{\sin 2(\alpha-\beta)}}
\approx M_Z \sqrt{\frac{-\sin 4\beta}{2\gamma}}\;,
\eea
this corresponds to $M_A\sim 350$ GeV for $\tan\beta=2$ and $M_A\sim
190$ GeV for $\tan\beta=20$ (the exact value including loop
corrections may vary, depending on the squark parameters which we do
not specify here).

\begin{table}[tb]
\begin{center}
\begin{tabular}{|l|l|c|c|c|c|c|c|}
\hline
Process & $(I,J)$ & $\Delta_{LL}^{IJ}$ & $\Delta_{RR}^{IJ}$ &
$\Delta_{LR}^{IJ}$ & $\Delta_{RL}^{IJ}$ & $\Delta_{LR}^{'IJ}$ &
$\Delta_{RL}^{'IJ}$ \\
\hline
$\tan\beta=2$ & \multicolumn{7}{|l|}{}\\
\hline
$ h\to \mu e$ & $(2,1)$ & $ 1.8 \cdot 10^{+7}$ & $ 1.7 \cdot 10^{+6}$
& $ 2.6 \cdot 10^{+7}$ & $ 2.6 \cdot 10^{+7}$ & $ 1.0 \cdot 10^{+7}$ &
$ 1.0 \cdot 10^{+7}$ \\
$ h\to \tau \mu$ & $(3,2) $ & $ 4.4 \cdot 10^{+3}$ & $ 3.8 \cdot
10^{+2}$ & $ 6.3 \cdot 10^{+3}$ & $ 6.3 \cdot 10^{+3}$ & $ 2.5 \cdot
10^{+3}$ & $ 2.5 \cdot 10^{+3}$ \\
$ h\to \tau e$ & $(3,1) $ & $ 8.0 \cdot 10^{+3}$ & $ 7.5 \cdot
10^{+2}$ & $ 1.1 \cdot 10^{+4}$ & $ 1.1 \cdot 10^{+4}$ & $ 4.9 \cdot
10^{+3}$ & $ 4.9 \cdot 10^{+3}$ \\
\hline
$\tan\beta=20$ & \multicolumn{7}{|l|}{}\\
\hline
$ h\to \mu e$ & $(2,1)$ & $ 3.9 \cdot 10^{+6}$ & $ 8.3 \cdot 10^{+6}$
& $ 5.1 \cdot 10^{+7}$ & $ 5.1 \cdot 10^{+7}$ & $ 1.2 \cdot 10^{+6}$ &
$ 1.2 \cdot 10^{+6}$ \\
$ h\to \tau \mu$ & $(3,2) $ & $ 9.5 \cdot 10^{+2}$ & $ 1.9 \cdot
10^{+3}$ & $ 1.3 \cdot 10^{+4}$ & $ 1.2 \cdot 10^{+4}$ & $ 2.9 \cdot
10^{+2}$ & $ 2.9 \cdot 10^{+2}$ \\
$ h\to \tau e$ & $(3,1) $ & $ 1.7 \cdot 10^{+3}$ & $ 3.5 \cdot
10^{+3}$ & $ 2.3 \cdot 10^{+4}$ & $ 2.2 \cdot 10^{+4}$ & $ 5.3 \cdot
10^{+2}$ & $ 5.4 \cdot 10^{+2}$ \\
\hline
\end{tabular}
\end{center}
\caption{\small Ratios of upper bounds on the LFV $\Delta$ parameters
  from leptonic Higgs boson decays and from radiative decays of
  charged leptons. The MSSM spectrum is defined in~\eq{eq:mpatt} (with
  the exception of setting $A_l^{\prime II}=0$) and a SUSY scale of
  $M=400$ GeV. The ratios for $\Delta_{LL}^{IJ}$, $\Delta_{RR}^{IJ}$,
  $\Delta_{LR}^{'IJ}$, $\Delta_{RL}^{'IJ}$ decrease with $M^2$,
  assuming fixed masses and mixing angles in the Higgs sector.
\label{tab:hbounds}}
\end{table}

The bounds on $\Delta_{LL}^{IJ}, \Delta_{RR}^{IJ}, \Delta_{LR}^{'IJ},
\Delta_{RL}^{'IJ}$ from the leptonic Higgs boson decay would decouple
only if also $M_A$ is scaled up simultaneously with SUSY particle
masses (thus assuring that the Higgs decay rates do not violate the
Appelquist-Carrazone theorem~\cite{Appelquist:1974tg}).  This
interesting feature is discussed in more details in
Sec.~\ref{sec:hllndec}.

\subsection{Dependence on the mass splitting}
\label{sec:msplit}

The formulae derived in the previous Sections allow to analyse how the
bounds on LFV mass insertions depend on the splitting between
different SUSY masses. However, any process involving transition
between the generations $I$ and $J$ depends in general, even at lowest
order in the flavour violating MI's, on many mass parameters: $\mu$,
gaugino masses $M_1, M_2$, left and right diagonal slepton soft masses
$m_{\tilde e_{LI}}, m_{\tilde e_{LJ}}, m_{\tilde e_{RI}}, m_{\tilde
  e_{RJ}}$, and for the Higgs decays also on $M_A$ or on $\alpha$
angle. To simplify the discussion, we only take into account the
bounds from $\ell\to\ell^\prime \gamma$ decays, which are currently
most constraining.

\begin{figure}[tbp]
  \begin{center}
    \begin{tabular}{cc}
      $\Delta_{LL}^{12}$ & $\Delta_{RR}^{12}$ \\[3mm]
      \includegraphics[width=0.47\textwidth]{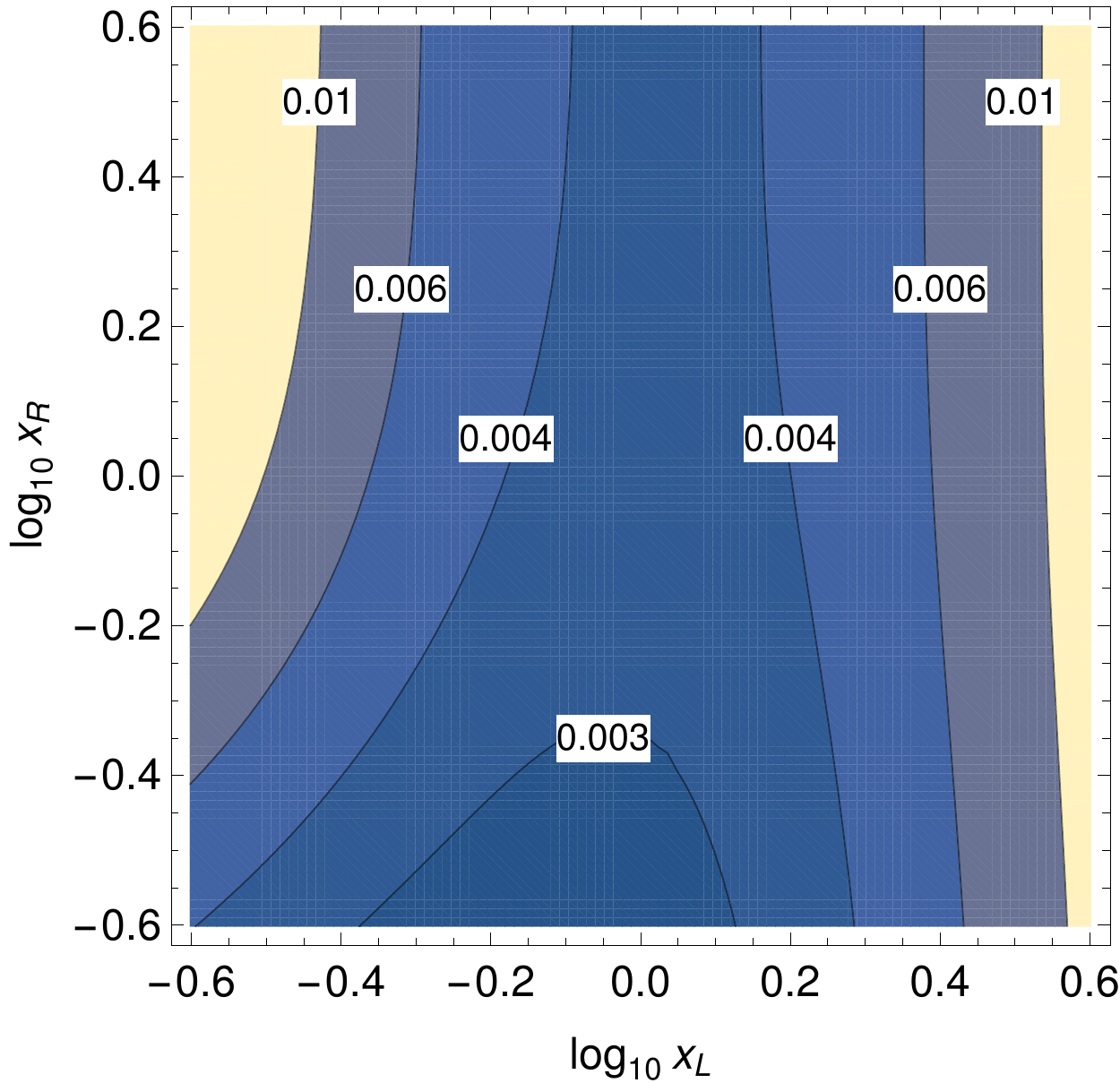}&
      \includegraphics[width=0.47\textwidth]{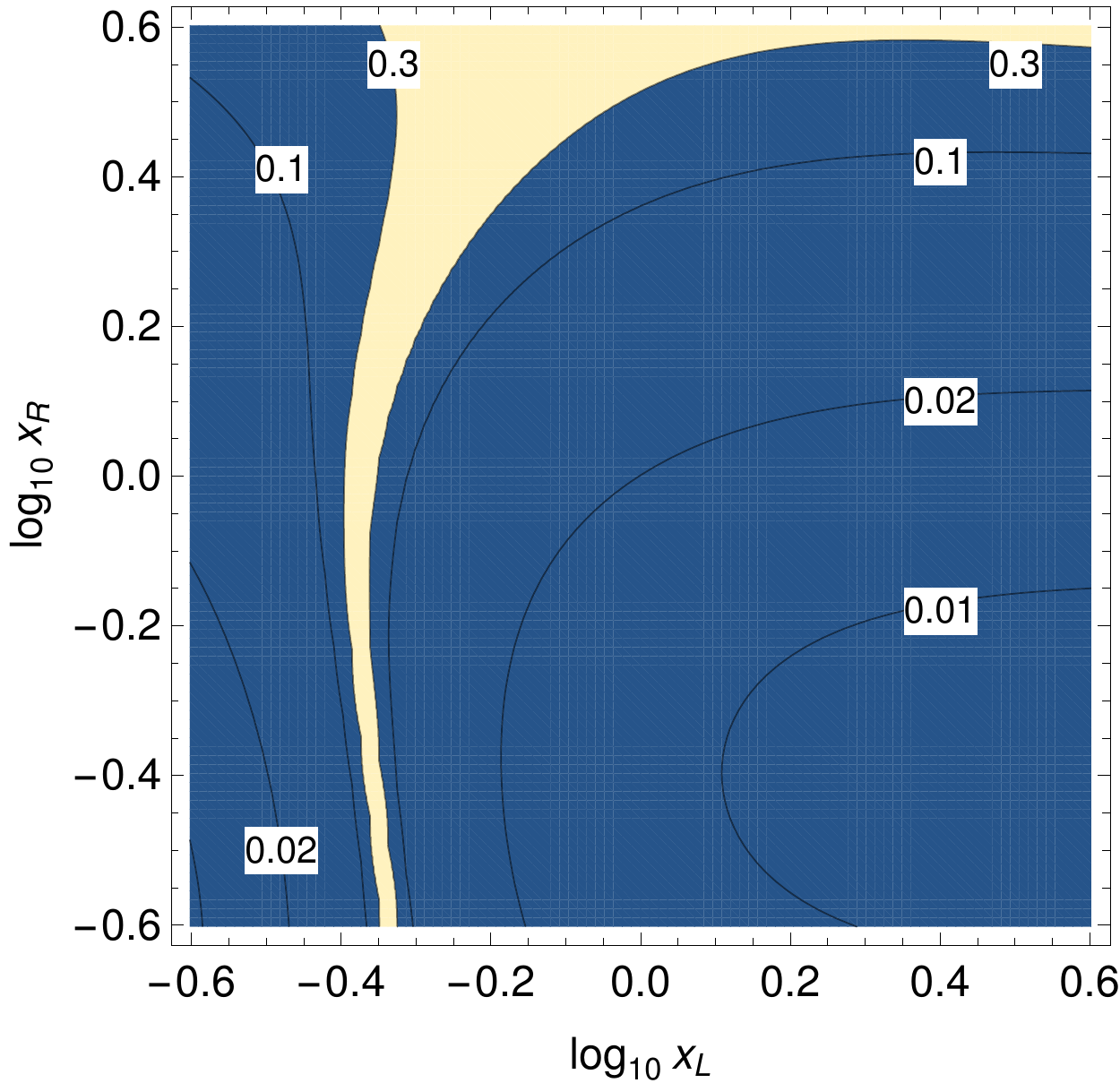}\\
      $\Delta_{LR}^{12},\Delta_{LR}^{21}$ & $\Delta_{LR}^{\prime
        12},\Delta_{LR}^{\prime 21}$ \\[3mm]
      \includegraphics[width=0.47\textwidth]{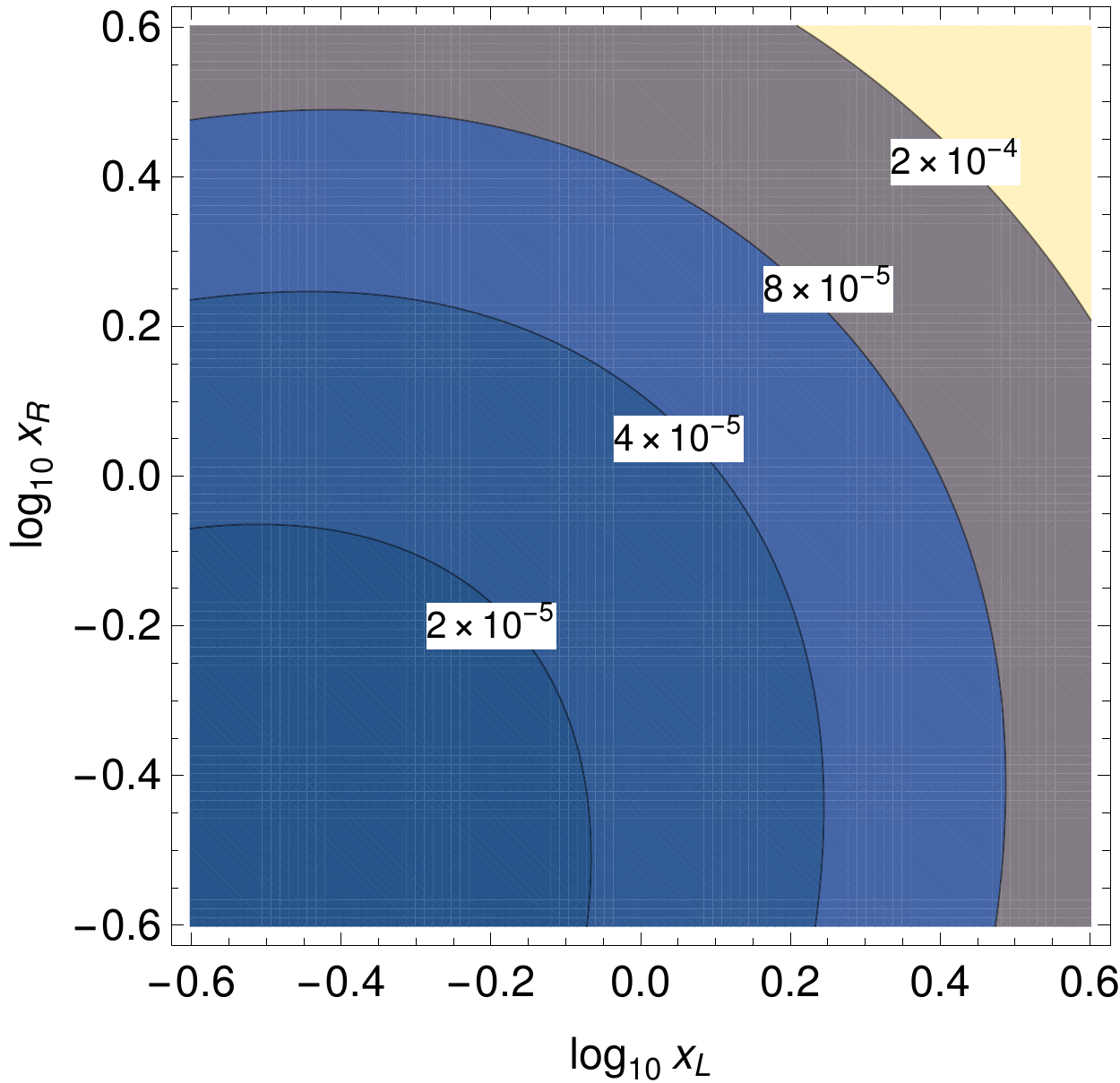}&
      \includegraphics[width=0.47\textwidth]{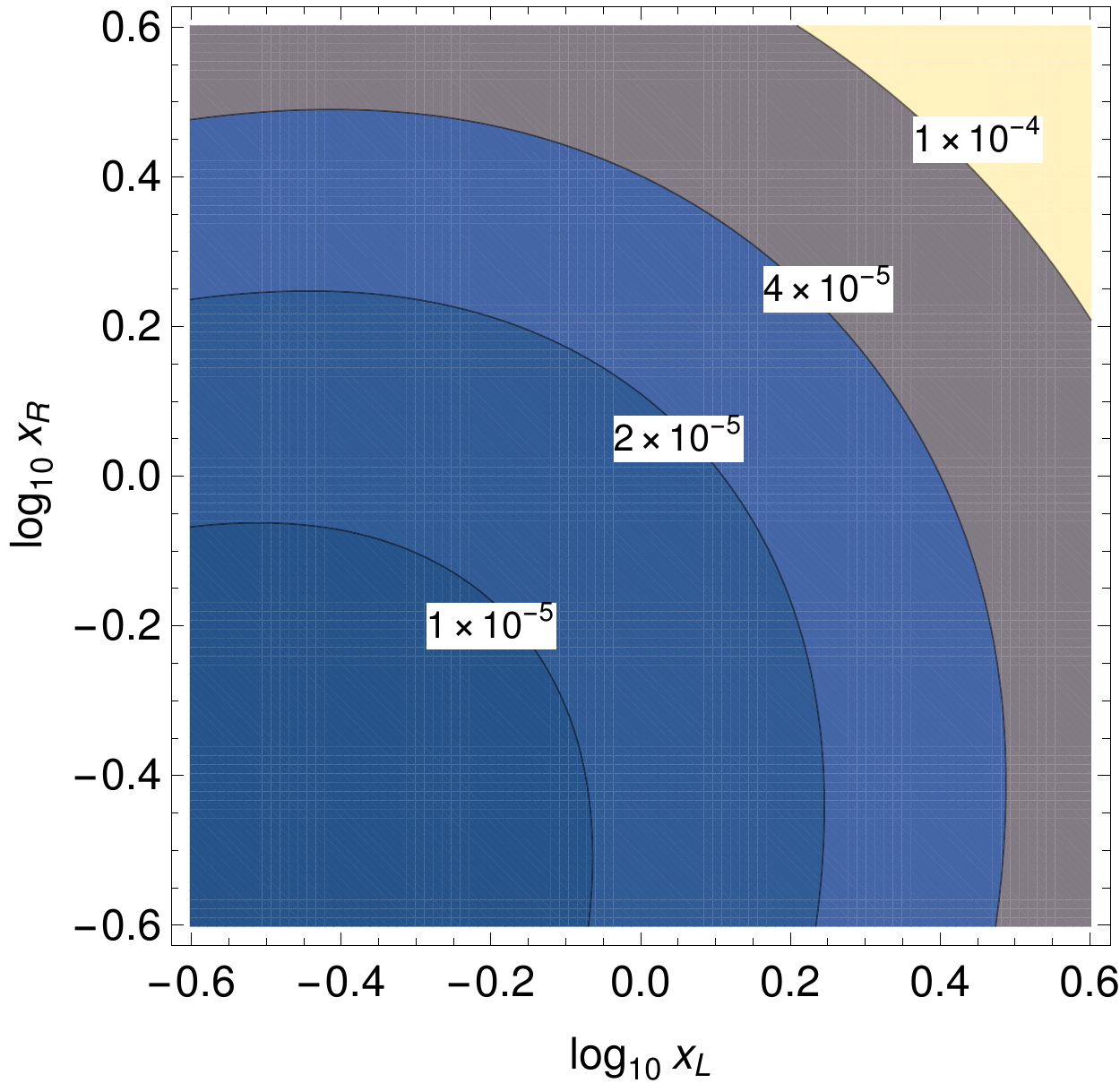}\\
\end{tabular}
\caption{\small Upper bounds on the LFV parameters from $\mu\to e
  \gamma$ for $\tan\beta=2$ and $M=800$ GeV as a function of the
  splitting between the masses of gaugino and sleptons of different
  chiralities. The normalised slepton masses $x_{L(R)}=m_{L(R)}/M$ are
  plotted on the axes.\label{fig:split_lr}}
  \end{center}
\end{figure}

In Fig.~\ref{fig:split_lr} we illustrate the dependence of the upper
bounds on the $\Delta$ parameters originating from $\mu\to e \gamma$
on the mass splitting between left and right-handed sleptons for
\begin{center}
\begin{tabular}{cc}
$\tan\beta=2$\,, & $\mu=M_1=M_2\equiv M = 800$ GeV\\[1mm] $m_{\tilde
    e_L} =m_{\tilde \mu_L} = m_L$\,, & $m_{\tilde e_R} = m_{\tilde
    \mu_R} = m_R$\,,\\[1mm]
$A_{\ell}^{ee} = A_{\ell}^{'ee} = Y_e\; \sqrt{m_L m_R}$\,,
  &$A_{\ell}^{\mu\mu} = A_{\ell}^{'\mu\mu} = Y_\mu\; \sqrt{m_L
    m_R}$\,.  \\
\end{tabular}
\end{center}
We have chosen here an average SUSY mass scale of $M=800$~GeV, higher
than $M=400$~GeV used in Tables~\ref{tab:llgbounds}-\ref{tab:hbounds},
to avoid the experimental bounds on slepton masses even in the case of
a large splitting between the left and right-handed masses.

The features of plots in Fig.~\ref{fig:split_lr} can be understood
using the expanded expressions for effective photon couplings
collected in Appendix~\ref{app:llgmi}. As an example, let us consider
the interesting cancellation between different contributions in the
case of $\Delta_{RR}^{12}$ (right upper panel of
Fig.~\ref{fig:split_lr}).  For our parameter setup, the coefficient
$X_{\gamma N2}^{e\mu}$ multiplying the RR parameter
(see~\eq{eq:llgmineut1}) can be reduced to the form
\bea
X_{\gamma N2}^{e\mu} = \frac{v_1 Y_\mu}{M^2}\; f(x_L,x_R)\,,
\eea 
where $f(x_L,x_R)$ is a known, although complicated, dimensionless,
rational and logarithmic function of mass ratios whose analytical form
can be obtained using~\eq{eq:llgmineut1}, the loop integrals collected
in Appendix~\ref{app:loop} and the definitions of divided differences
from Appendix~\ref{app:ddiff}. The properties of this function can be
examined analytically and numerically.  One finds
\begin{itemize}
\item For $x_R$ in the wide range $0.1-4$ the function $f$ vanishes
  for $x_L\sim 0.45$ (the exact value depends only weakly on
  $x_R$). As a result, the bounds on $\Delta_{RR}^{12}$ disappear
  completely for $m_L \sim 0.45 M$.
\item For large values of $x_R\gsim 5$ the position where the function
  $f$ becomes zero shifts towards bigger values of $x_L$. In addition,
  in this limit $f$ is suppressed by an overall factor $1/x_R$, thus
  the bounds on $\Delta_{RR}^{12}$ become weaker for a larger values
  of $x_R$.
\item For large values of $x_L$ the function $f$ depends on $x_R$
  only. Therefore, the contour lines become horizontal.
\item For small values of $x_L$ the function $f$ behaves like
  $1/x_L$. Thus, the bounds on $\Delta_{RR}^{12}$ become stronger.
\end{itemize}
A similar analysis can be done for the bounds on
$\Delta_{LL}^{12}$. However, the coefficient multiplying
$\Delta_{LL}^{12}$ contains contributions from both chargino and
neutralino loops and does not vanish for any mass pattern. Therefore,
there is no cancellation area in the upper left panel of
Fig.~\ref{fig:split_lr}. In this case, the bound on $\Delta_{LL}^{12}$
is strongest for $m_L\sim M$ and $m_R\lsim M$. For the case in which
the left slepton masses are much lighter or much heavier than the
masses of the SUSY fermion, the bounds become weaker.

Bounds on LR parameters, both holomorphic and non-holomorphic, are
typically 1-2 orders of magnitude stronger than for LL and RR ones.
In this case, the coefficient $X_{\gamma N1}^{e\mu}$ multiplying the
LR terms has a much simpler functional form. Therefore, it never
vanishes and in addition is explicitly symmetric (as follows from the
properties of divided differences) under the exchange of slepton mass
arguments, as visible in both lower panels of Fig.~\ref{fig:split_lr}.
Furthermore, one can see that bounds on LR parameters are strongest
for $m_L,m_R \lsim M$ and become weaker when the slepton masses are
much heavier than the chargino and neutralino masses. More
quantitatively, $X_{\gamma N1}^{e\mu}$ is proportional to the divided
difference of the function $C_{12}$, which for $x\equiv x_L = x_R$
(corresponding to the diagonal of lower plots in
Fig.~\ref{fig:split_lr}) has the simple asymptotic behaviour
\bea
C_{12}(\{m_L,m_R\},M)=\left\{
\begin{array}{lp{1cm}l}
-\frac{5}{2M^2} && x \ll 1 \\[2mm]
\phantom{-}\frac{1}{2M^2x^2} && x \gg 1 \\
\end{array}
\right.\;.
\label{eq:c12lr}
\eea
From the form of~\eq{eq:c12lr} it is immediately visible that the
bounds become constant for small $x$ and fall like $1/x^2$ for large
$x$, as illustrated in the plots.

\begin{figure}[tbp]
  \begin{center}
    \begin{tabular}{cc}
      $\Delta_{LL}^{12}$ & $\Delta_{RR}^{12}$ \\[3mm]
      \includegraphics[width=0.47\textwidth]{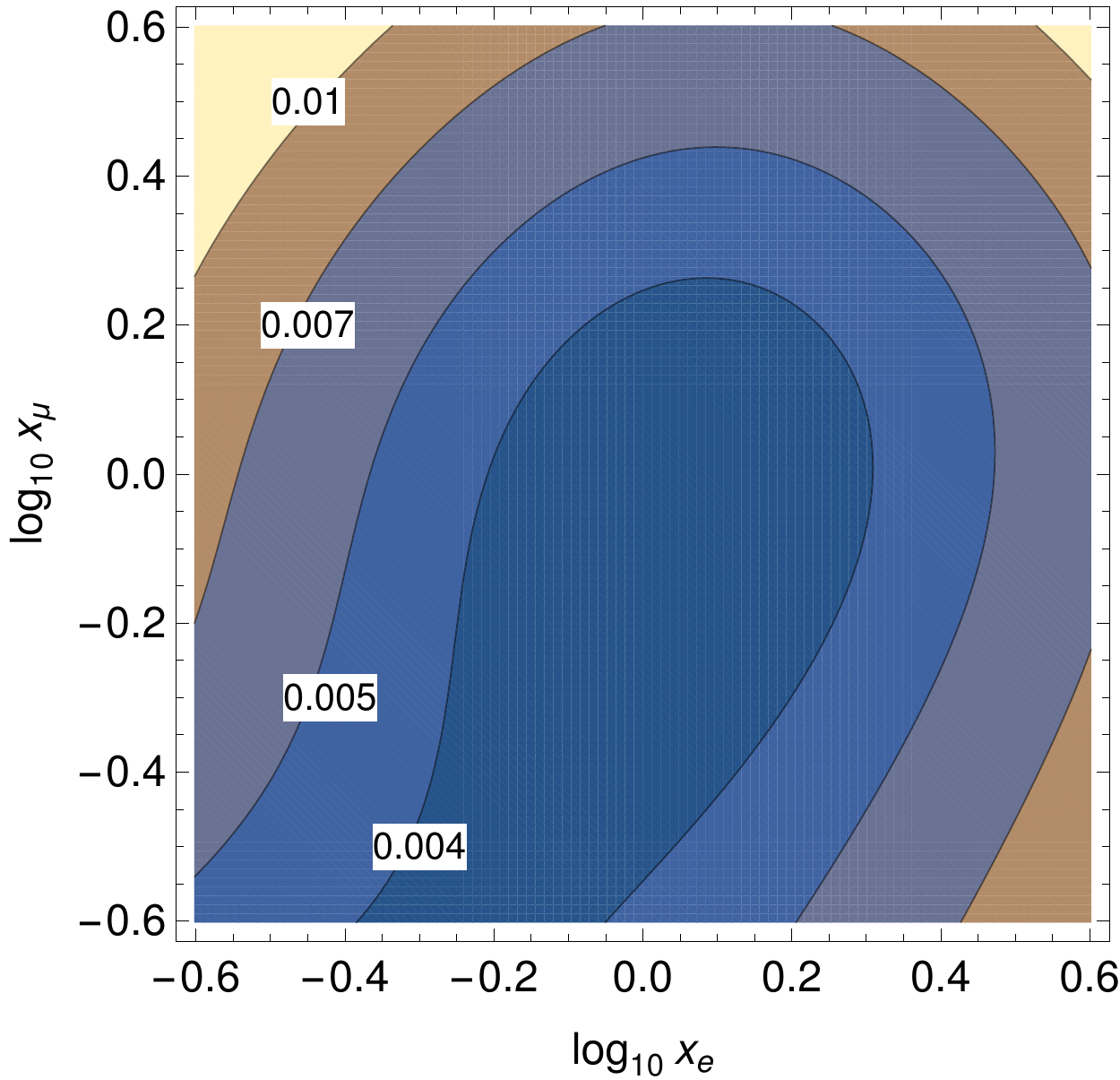}&
      \includegraphics[width=0.47\textwidth]{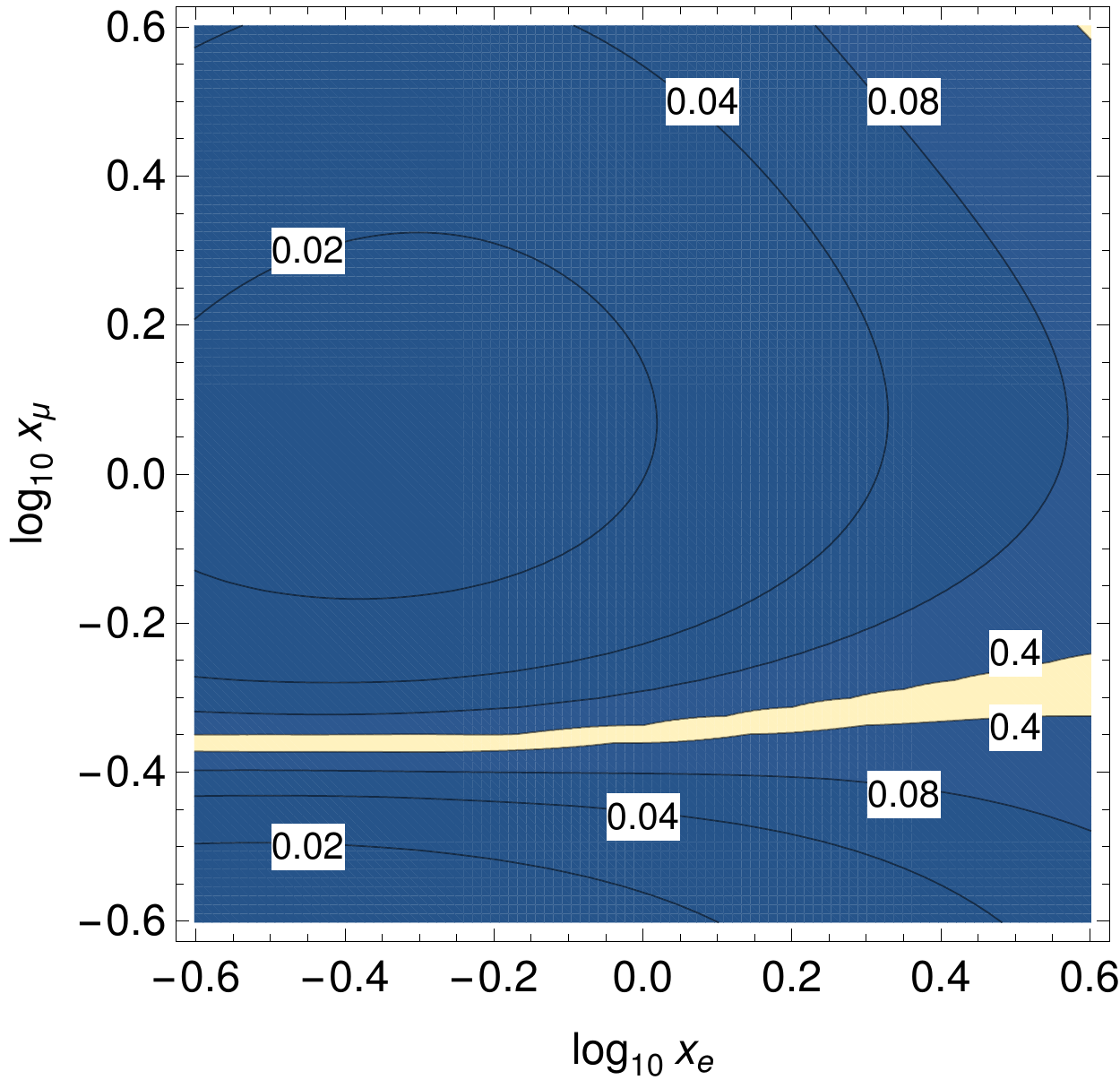}\\
    \end{tabular}
    \caption{\small Upper bounds on the flavour violating LL and RR
      parameters using the current experimental limit on $\Br(\mu\to e
      \gamma)$ for $\tan\beta=2$ and $M=800$ GeV as a function of
      splitting between the masses of gaugino and sleptons of various
      flavours. The normalised selectron and smuon masses, $x_{e(\mu)}
      = m_{\tilde e(\tilde\mu)}/M$ are plotted on the axes.
      \label{fig:split_ij}}
\end{center}
\end{figure}

Using the formulae collected in Appendix~\ref{app:miexp}, a similar
discussion can be, if necessary, performed to explain the features or
cancellation areas of other plots presented in this Section. However,
as the general analytical formulae in the MI approximation are rather
complicated, we illustrate here other scenarios with numerical plots
only.

Fig.~\ref{fig:split_ij} shows similar bounds assuming identical
left-and right-handed slepton masses which however differ among the
generations, so that we choose
\begin{center}
\begin{tabular}{cc}
$\tan\beta=2$\,, & $\mu=M_1=M_2\equiv M = 800$ GeV\,,\\[1mm]
$m_{\tilde e_L} =m_{\tilde e_R} = m_{\tilde e}$ & $m_{\tilde \mu_L} =
  m_{\tilde \mu_R} = m_{\tilde \mu}$\,,\\[1mm]
$A_{\ell}^{ee} = A_{\ell}^{'ee} = Y_e\; m_{\tilde e} $\,,
  &$A_{\ell}^{\mu\mu} = A_{\ell}^{'\mu\mu} = Y_\mu\; m_{\tilde
    \mu}$\,, \\
\end{tabular}
\end{center}
and plot the results in terms of $x_e = m_{\tilde e}/M$ and $x_\mu =
m_{\tilde \mu}/M$. Again, a cancellation only exists for the bounds on
$\Delta_{RR}^{12}$, for an almost constant ratio $m_{\tilde \mu}\sim
2.5 M$. In this case, the bounds on $\Delta_{LL}^{12}$ are strongest
for small splitting between slepton and SUSY fermion masses, while the
bounds on $\Delta_{RR}^{12}$ are, apart from the cancellation region,
stronger for $m_{\tilde\mu}\lsim M$.  It is obvious from the form of
$X_{\gamma N1}^{e\mu}$ in~\eq{eq:llgmineut1} that the bounds on the LR
parameters, both holomorphic and non-holomorphic, have an identical
behaviour as in the case of the $m_L-m_R$ splitting plotted in
Fig.~\ref{fig:split_lr}, with the replacements $x_L \leftrightarrow
x_e$, $x_R \leftrightarrow x_\mu$.

\begin{figure}[tbp]
  \begin{center}
    \begin{tabular}{cc}
      $\Delta_{LL}^{12}$ & $\Delta_{RR}^{12}$ \\[3mm]
      \includegraphics[width=0.47\textwidth]{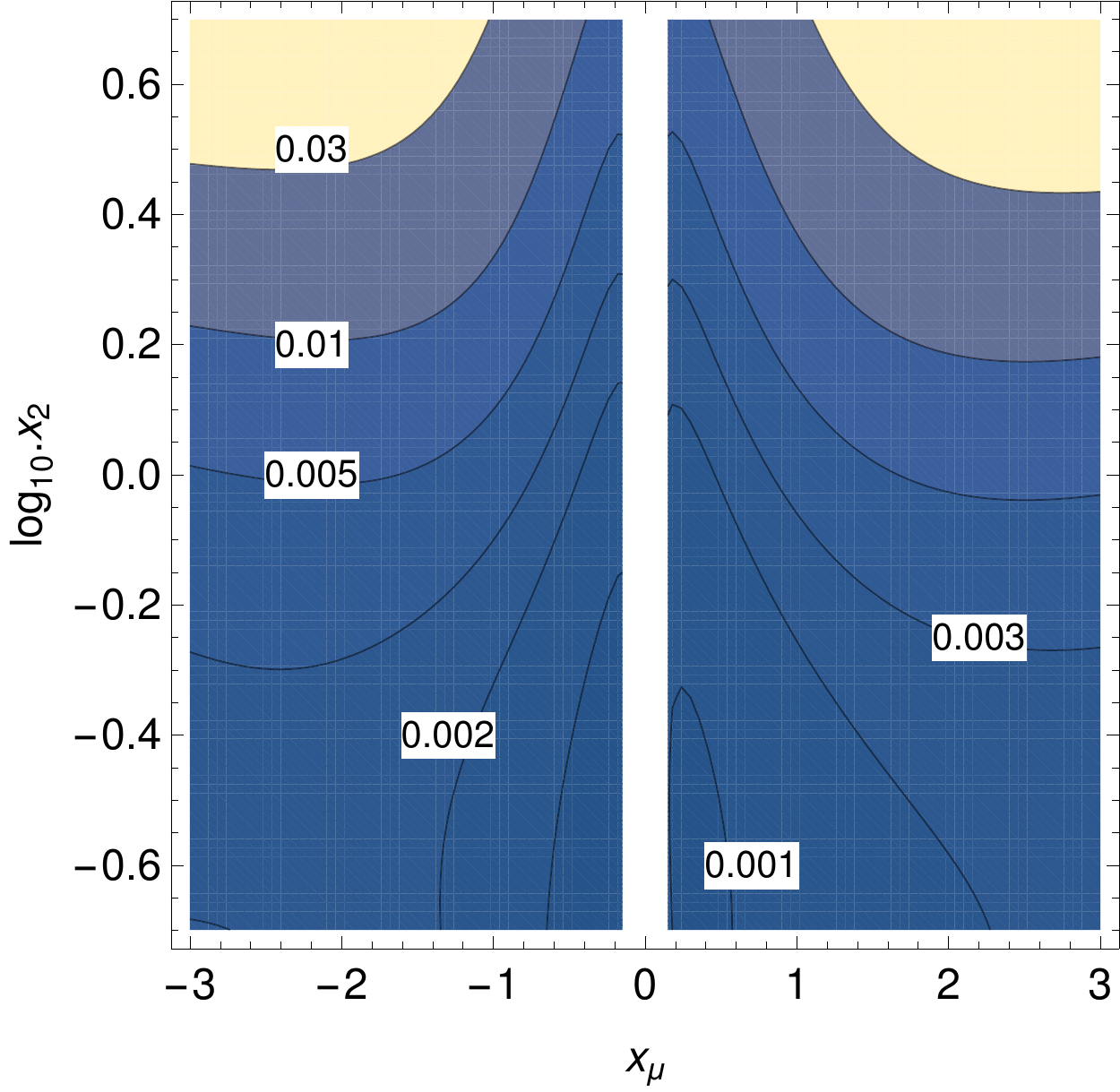}&
      \includegraphics[width=0.47\textwidth]{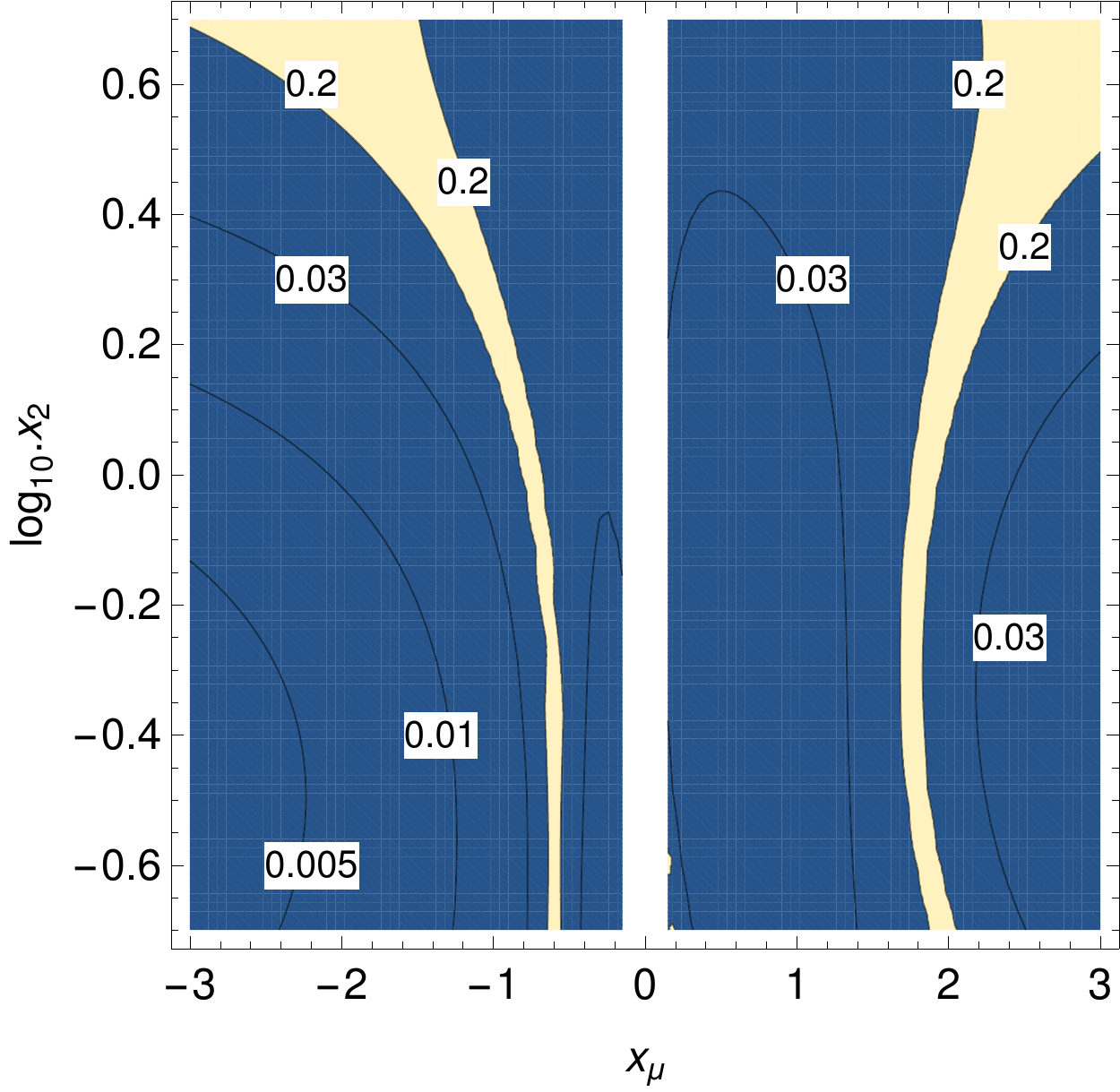}\\
      $\Delta_{LR}^{12},\Delta_{LR}^{21}$ & $\Delta_{LR}^{\prime
        12},\Delta_{LR}^{\prime 21}$ \\[3mm]
      \includegraphics[width=0.47\textwidth]{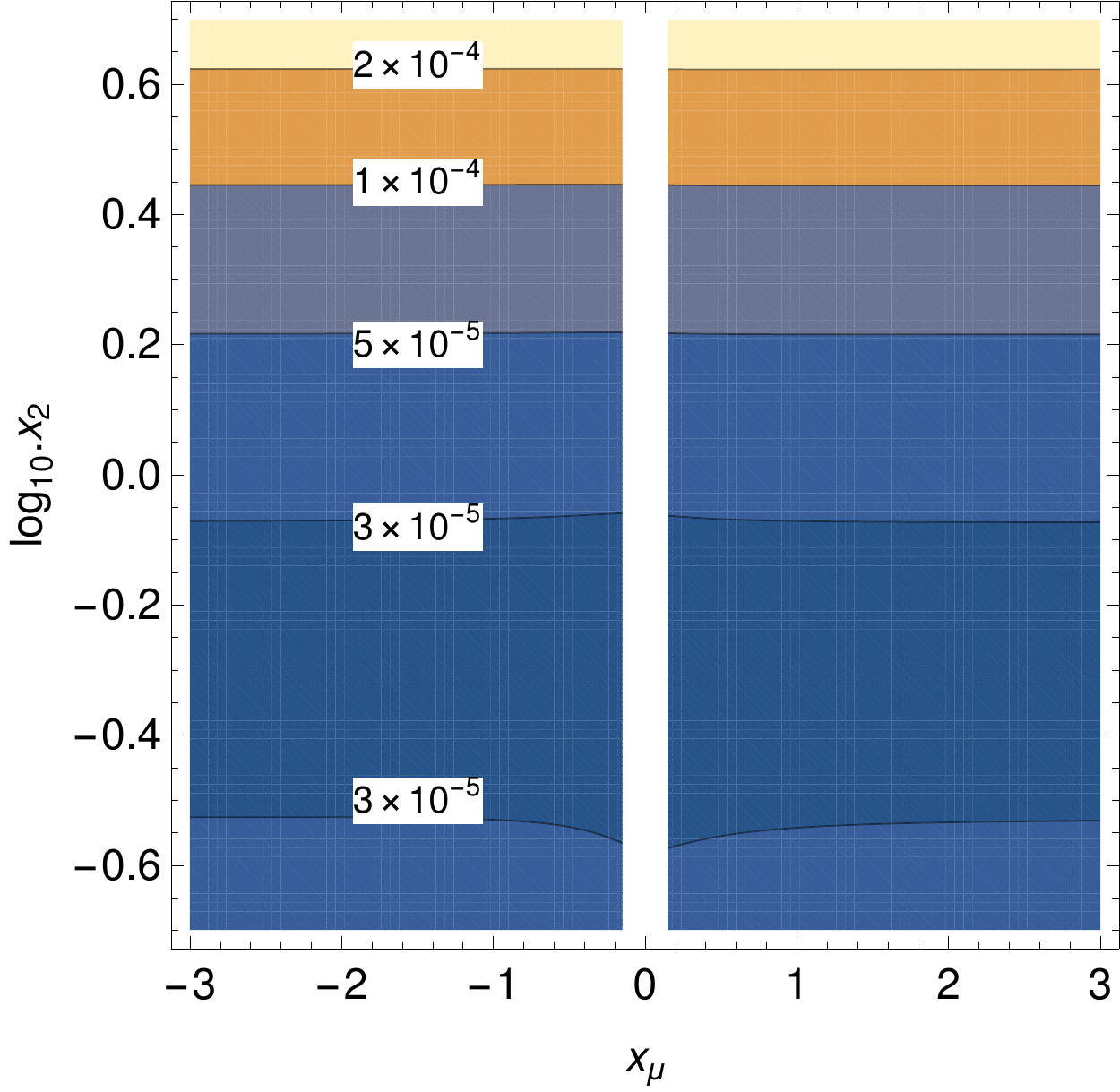}&
      \includegraphics[width=0.47\textwidth]{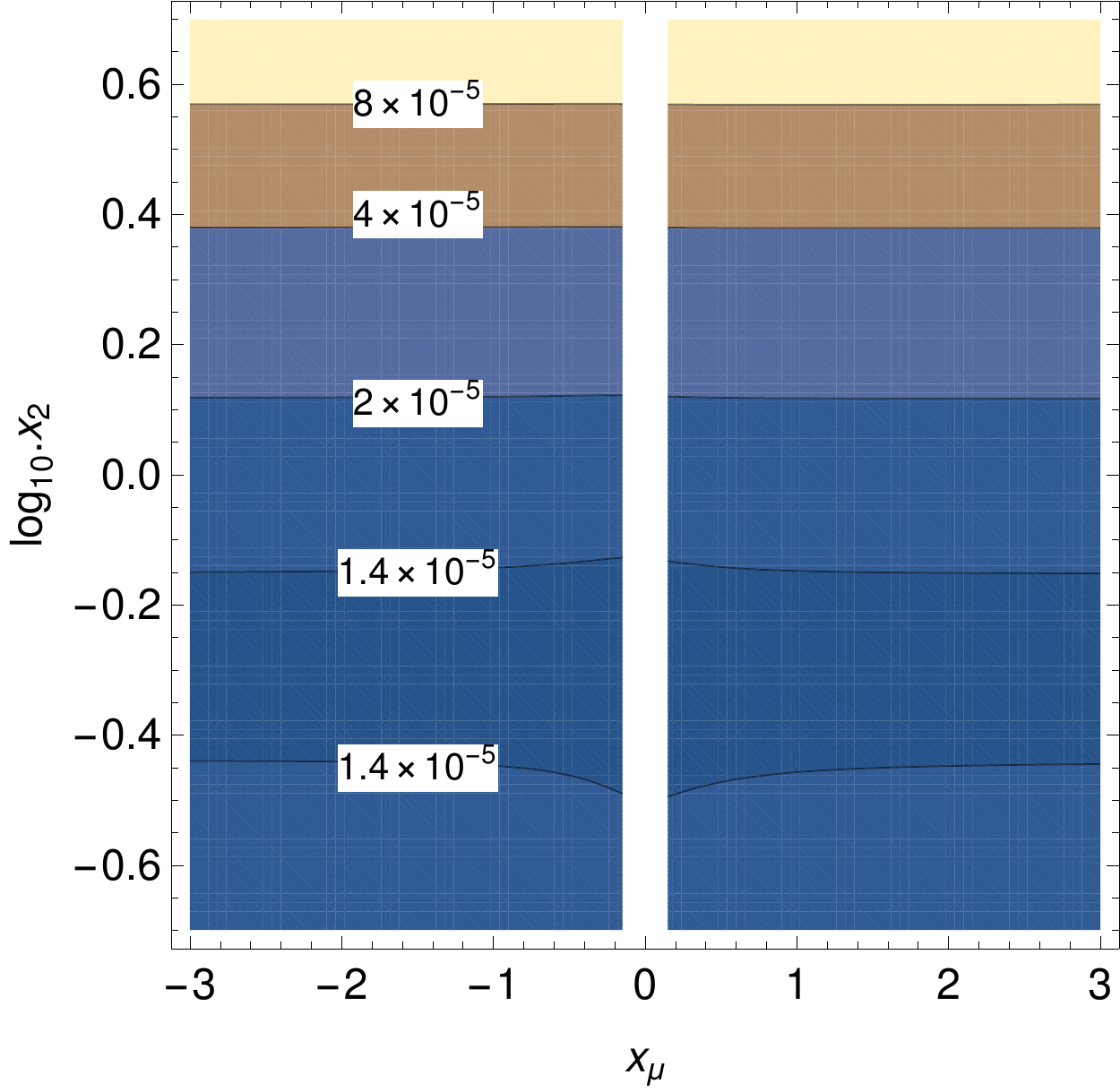}\\
\end{tabular}
    \caption{\small Upper bounds on the LFV parameters using the
      current experimental limit on the $\Br(\mu\to e \gamma)$ for
      degenerate slepton masses $M=800$ GeV as a function of mass
      splitting between the gaugino and the $\mu$ related parameters,
      $x_\mu = \mu/M$, $x_2 = M_1/M = M_2/M$.
      \label{fig:split_mu}}
  \end{center}
\end{figure}

Finally in Fig.~\ref{fig:split_mu} we assume an identical mass of
$m=400$ GeV for all sleptons but vary $M_1=M_2$ and $\mu$. The results
are displayed as a function of $x_2 = M_2/M$, $x_\mu = \mu/M$ (we do
not plot small values of $|\mu|<100$~GeV which are excluded by the
direct searches for charginos and neutralinos). The structure of
cancellation areas is more complicated, but again the ``blind spots'',
where the bounds on MI's disappear, exist only for $\Delta_{RR}^{12}$.
As expected from the form of $X_{\gamma N1}^{e\mu}$
in~\eq{eq:llgmineut1}, the bounds on $\Delta_{LR}^{12}$,
$\Delta_{LR}^{'12}$ are at leading order independent of the $\mu$
parameter.  They are also correlated with the bounds displayed in
lower plots of Fig.~\ref{fig:split_lr}, as for a fixed slepton mass
and varied $M_2$ the coefficient $X_{\gamma N1}^{e\mu}$ is now
proportional to
\bea
C_{12}(\{m,m\},M_2)=\left\{
\begin{array}{lp{1cm}l}
\phantom{-}\frac{1}{2m^2} && x \ll 1 \\[2mm]
-\frac{5}{2m^2 x_2^2} && x \gg 1 \\[2mm]
\end{array}
\right.
\label{eq:c12lra}
\eea
so that again the bounds saturate for small $x_2$ and fall like
$1/x_2^2$ in the opposite limit.

Similar plots constraining $13$ and $23$ mass insertions have almost
identical shape; the bounds are just rescaled by constant factors. The
bounds on $\Delta_{LL}^{13}$ and $\Delta_{RR}^{13}$
($\Delta_{LL}^{23}$ and $\Delta_{RR}^{23}$) are approximately 550
(650) times weaker than the bounds on $\Delta_{LL}^{12}$ and
$\Delta_{RR}^{12}$, respectively.  The bounds on
$\Delta_{LR}^{13(31)}$ and $\Delta_{LR}^{\prime 13(31)}$
($\Delta_{LR}^{23(32)}$ and $\Delta_{LR}^{\prime 23(32)}$) are
respectively 9000 (11000) times weaker than the bounds on
$\Delta_{LR}^{12(21)}$ and $\Delta_{LR}^{\prime 12(21)}$.

\subsection{Correlations between LFV processes}
\label{sec:corr}

The correlations between various leptonic decays, in particular
radiative and 3-body charged lepton decays, are important for
designing new experiments searching for the LFV phenomena. In the
photon penguin domination scenario the ratio of decay rates for both
processes is given by the simple formula:
\bea
\frac{\Br(\ell\to 3\ell')}{\Br(\ell\to\ell'\gamma)} \approx
\frac{\alpha_{em}}{3\pi}\left(\log\frac{m_\ell^2}{m_{\ell'}^2} -
\frac{11}{4}\right)\,.
\label{eq:phdomin}
\eea
In this case the decision which measurement is more promising depends
purely on experimental accuracy achievable for each of them.  However,
other type of contributions, like $Z$-penguin and box diagrams, can
modify the ratio~(\ref{eq:phdomin}). Such contributions may be
particularly important for a ``blind spot'' scenario, like the
weakened limit on $\Delta_{RR}$ for some ratios of slepton and gaugino
masses.

\begin{figure}[tb!]
  \begin{center}
    \begin{tabular}{cc}
      $\Delta_{LL}^{12}$ & $\Delta_{RR}^{12}$ \\[3mm]
      \includegraphics[width=0.47\textwidth]{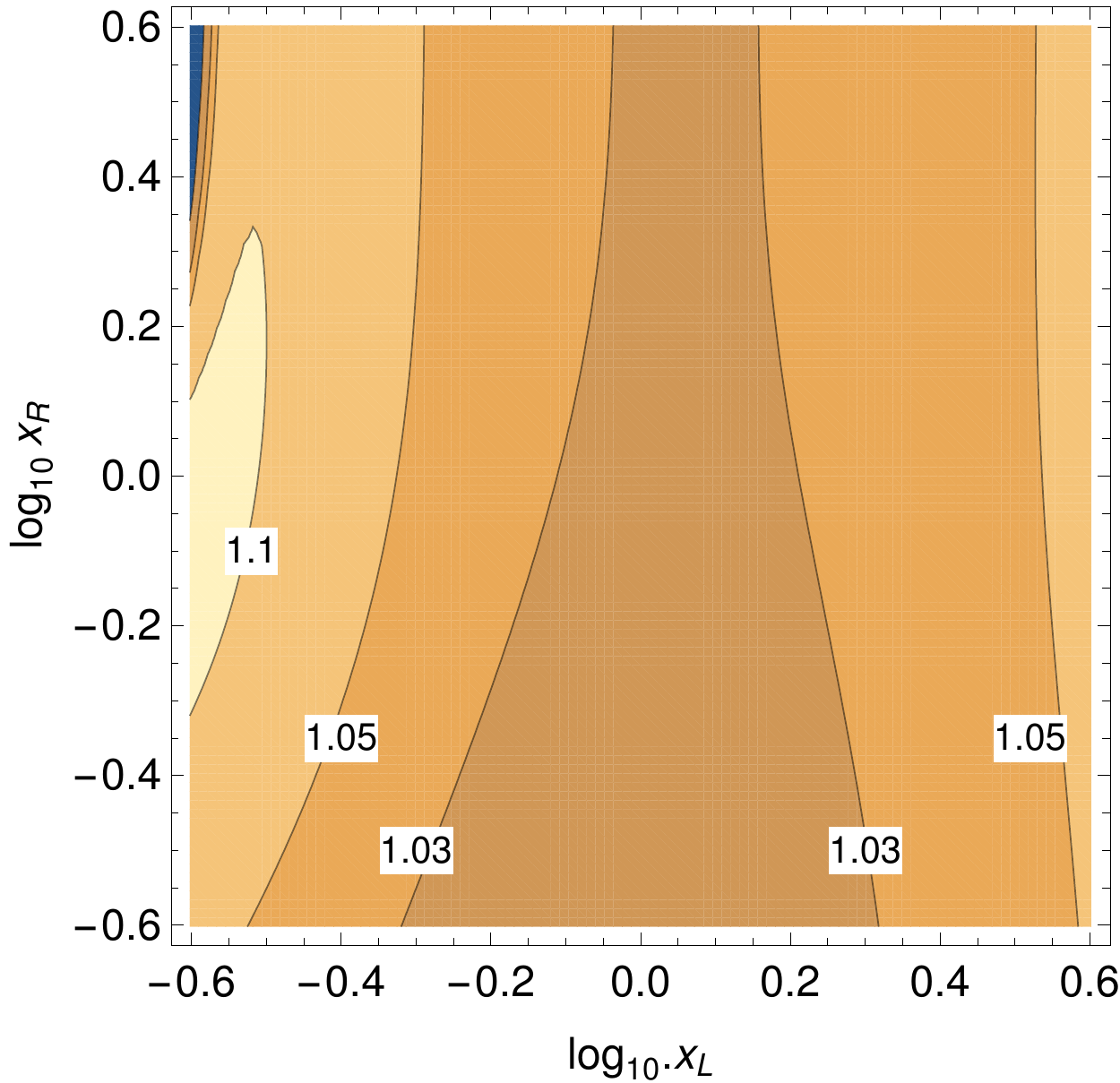}&
      \includegraphics[width=0.47\textwidth]{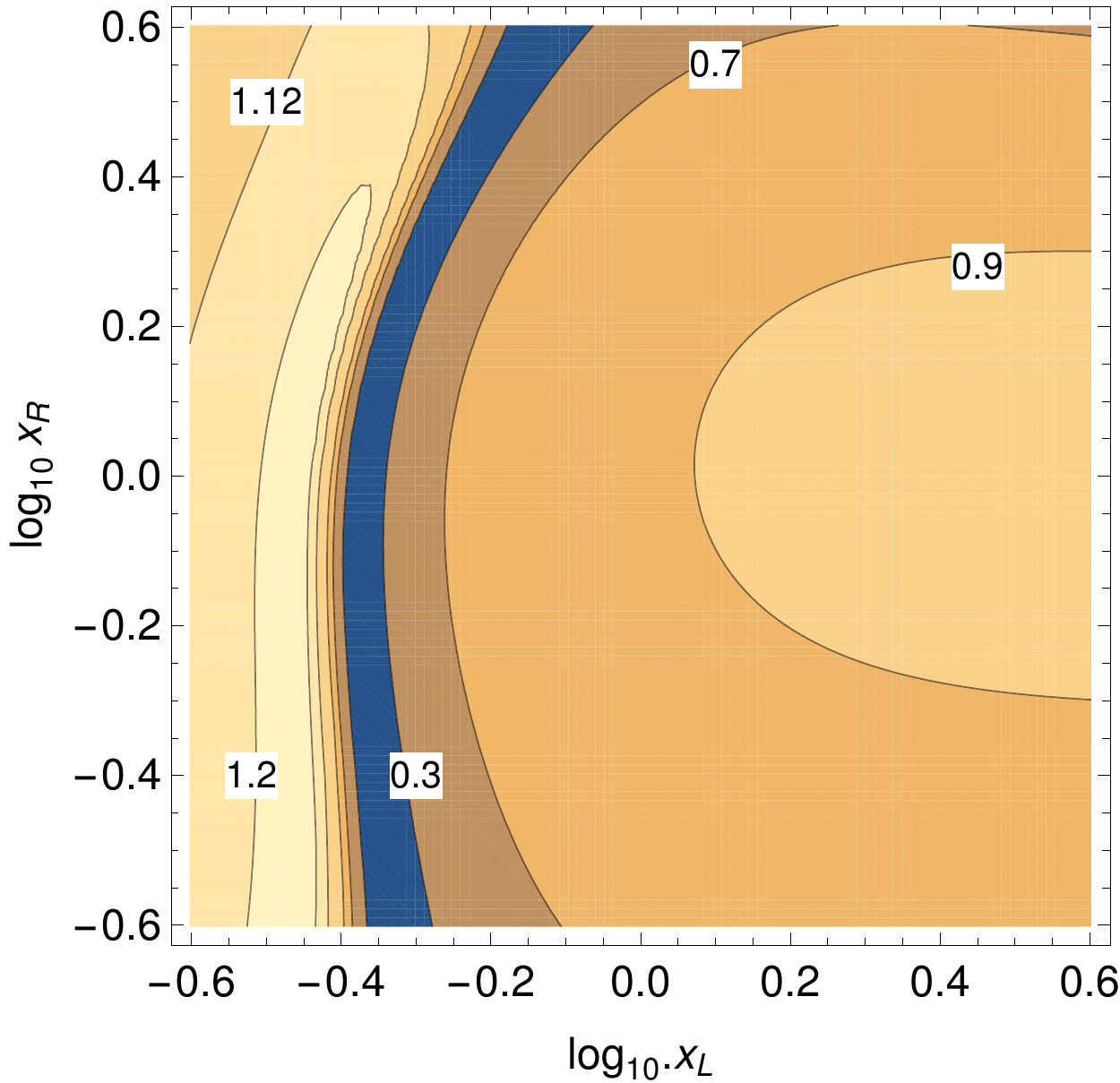}\\
      \includegraphics[width=0.47\textwidth]{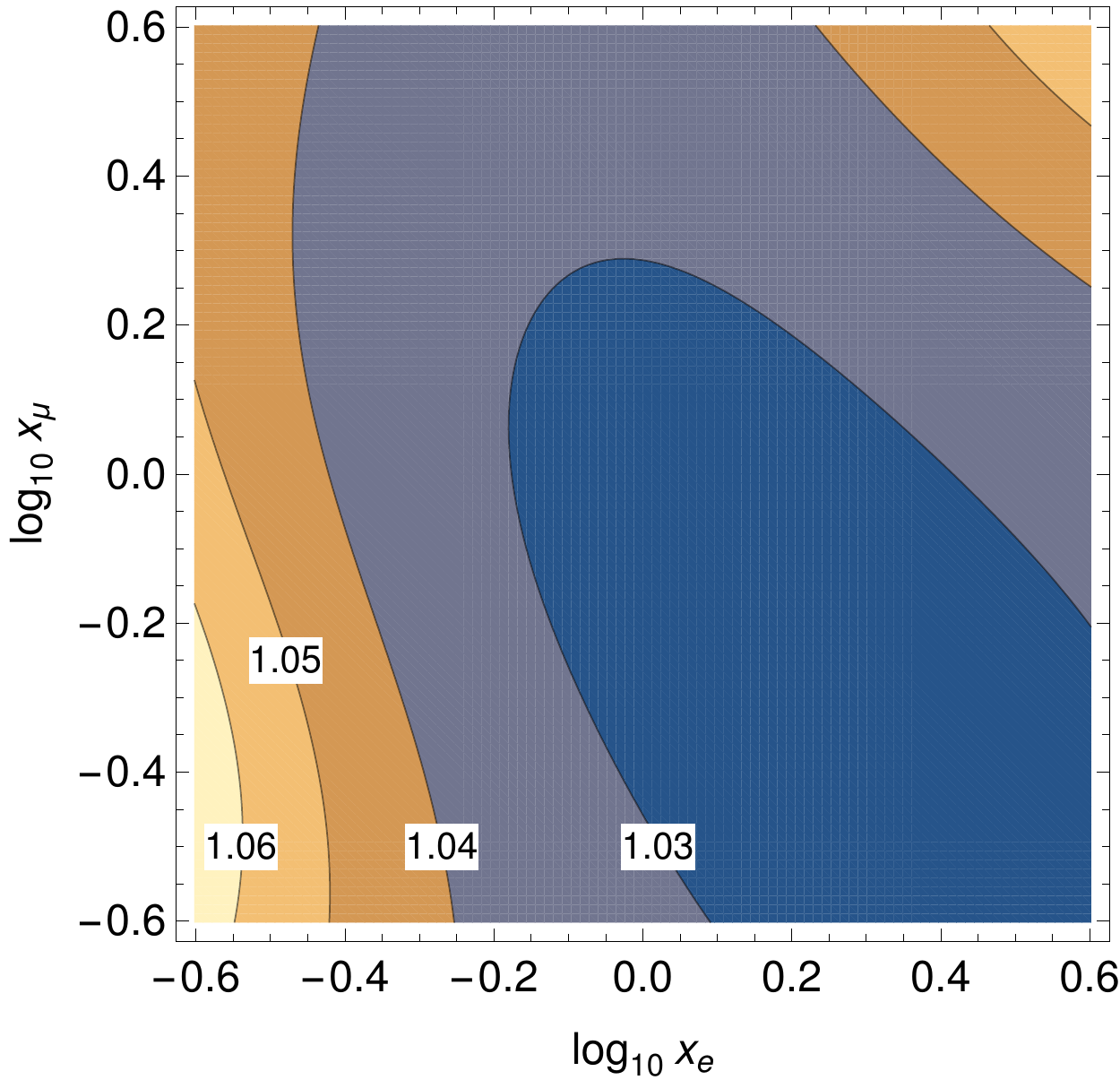}&
      \includegraphics[width=0.47\textwidth]{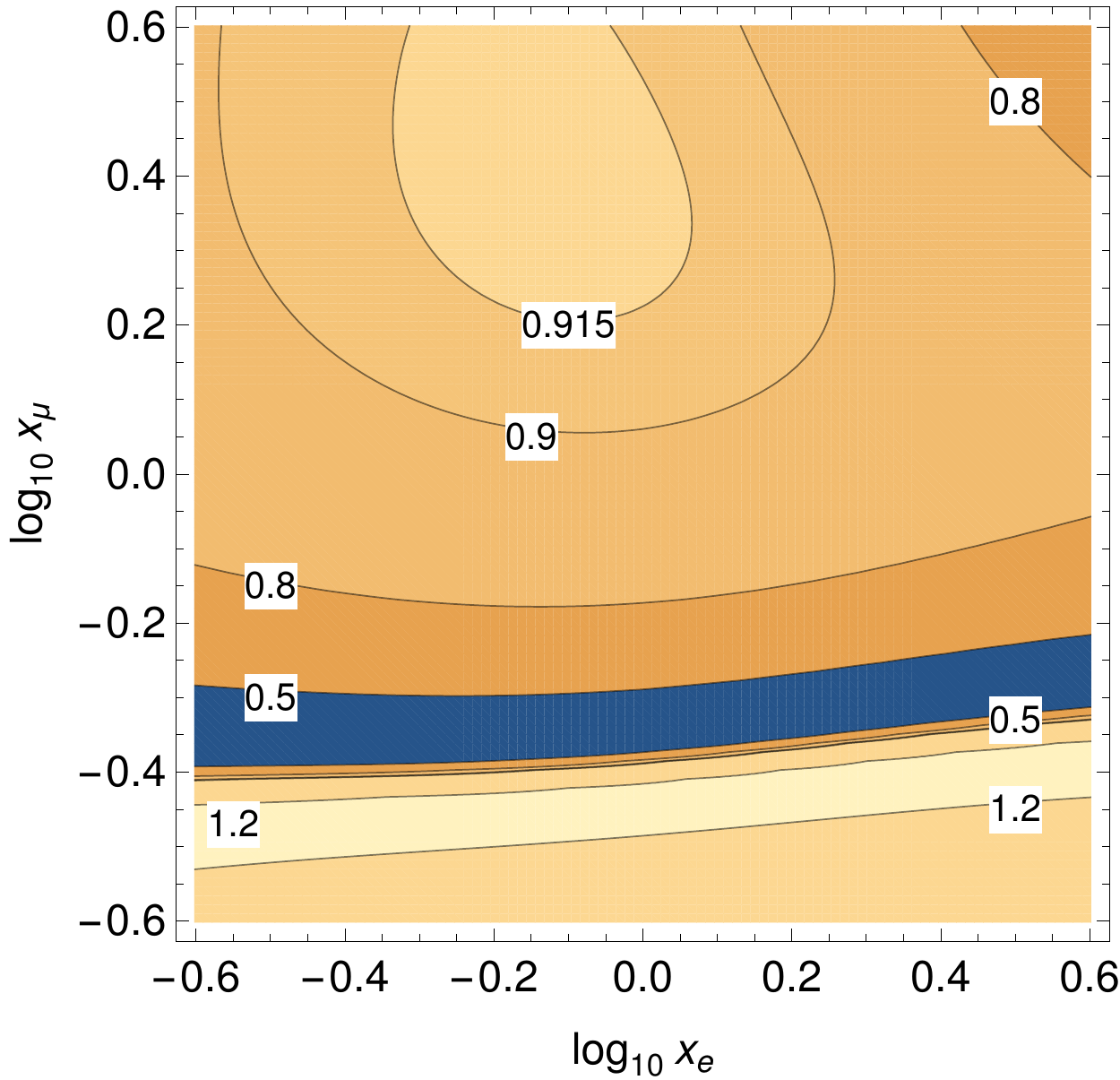}\\
\end{tabular}
\caption{\small The ratio $R_{\ell\ell'}$ as a function of the mass
  splitting between left and right slepton masses (upper row) and
  between the selectron and smuon masses (lower row) for $M=800$ GeV
  as a function of the normalised slepton masses $x_{L(R)}=m_{L(R)}/M$
  and $x_{\tilde e(\tilde\mu)} = m_{\tilde e(\tilde\mu)}/M$.
  \label{fig:correl}}
  \end{center}
\end{figure}

In Fig.~\ref{fig:correl} we plot the quantity $R_{\ell\ell'}$ defined as
\bea
R_{\ell\ell'} =
\frac{\alpha_{em}}{3\pi}\left(\log\frac{m_\ell^2}{m_{\ell'}^2} -
\frac{11}{4}\right)\frac{\Br(\ell\to\ell'\gamma)}{\Br(\ell\to 3\ell')}\,,
\label{eq:rll}
\eea
as a function of the SUSY mass splittings, in the same scenarios as
described in Fig.~\ref{fig:split_lr} and Fig.~\ref{fig:split_ij}. We
assume non-vanishing $\Delta_{LL}^{12}$ and $\Delta_{RR}^{12}$ terms.
For LR terms, both holomorphic and non-holomorphic, a photon
penguin dominated scenario is always realised and $R_{\ell\ell'}$ is
very close to 1.

As one can see from Fig.~\ref{fig:correl}, radiative and 3-body decays
are almost always closely correlated, with $R_{\ell\ell'}$ differing
from 1 by a few \% at most.  Exceptions are only possible for
parameter combinations for which $\Br(\ell\to\ell'\gamma)$ becomes
small due to cancellations or some other type of suppression, like in
scenarios with large mass splitting (compare Figs.~\ref{fig:split_lr}
and \ref{fig:split_ij}). Simultaneously, $\Br(\ell\to 3\ell')$ is
given by the more complicated expression~(\ref{eq:br4l}), which in the
limit of small photon penguin contribution becomes the sum of positive
terms and cannot vanish. Thus, although both decays are usually
strongly correlated and only relative experimental sensitivities
decide which of them has better chances to discover generic LFV
effects mediated by the slepton sector, for some particular ranges of
MSSM parameter searches for 3-body charged lepton decays are a safer
choice, allowing to avoid blind spots appearing for such setups due to
the suppression of $\ell\to\ell'\gamma$ decay rates.

\subsection{Non-decoupling effects in LFV Higgs decays}
\label{sec:hllndec}

LFV Higgs decays in the SM are absent at the tree level and strongly
suppressed also at the loop level. Examining LFV Higgs boson decays
within the MSSM is very interesting because, contrary to the other
processes discussed in this paper, some contributions to the Higgs
decay amplitudes proportional to the lepton Yukawa couplings or to the
non-holomorphic trilinear slepton soft terms do not decouple in the
limit of heavy SUSY masses and can be potentially large.

As can be seen from Tables~\ref{tab:hbounds}, for an average SUSY mass
scale of $M=400$ GeV and the parameter setup of~\eq{eq:mpatt} the
upper bounds on the flavour violating parameters from Higgs decays are
much weaker than from the other processes.  However, the bounds from
Higgs decays on the $\Delta_{LL}^{IJ}$, $\Delta_{RR}^{IJ}$ and on the
non-holomorphic LR terms $\Delta_{LR}^{'IJ}$ do not scale like
$1/M^2$.  Thus, comparing the limits on $\Delta_{LR}^{'13}$ and
$\Delta_{LR}^{'23}$ entries from $h\to \ell\ell'$ and $\ell\to\ell'
\gamma$ decays one can check that e.g. for $\tan\beta=20$ and
\bea
M_{SUSY}\gsim \frac{1.5}{\sqrt{|\cos(\alpha-\beta)|}}~\mathrm{TeV}\,,
\eea
the latter are becoming weaker. For $\Delta_{LR}^{'12}$ the same
occurs at much higher scale
\bea M_{SUSY}\gsim
\frac{220}{\sqrt{|\cos(\alpha-\beta)|}}~\mathrm{TeV}\,.
\eea
The bounds on $\Delta_{LL}^{IJ}$, $\Delta_{RR}^{IJ}$ are obtained
assuming that the flavour diagonal $A'_l$ terms vanish, so that all
non-decoupling LL and RR contributions are proportional to the Yukawa
couplings (see~\eq{eq:ly}). In this case the Higgs decays become most
constraining for slightly higher SUSY scales, again for $\tan\beta=20$
and $\alpha$ angle of~\eq{eq:alval} bounds on $\Delta_{LL}^{IJ}$ and
$\Delta_{RR}^{IJ}$ from Higgs decays become stronger than the bounds
from $\ell\to\ell' \gamma$ decays for $M_{SUSY}\gsim
2/\sqrt{|\cos(\alpha-\beta)|}$ TeV for $\tau\mu$ transitions,
$M_{SUSY}\gsim 3/\sqrt{|\cos(\alpha-\beta)|}$ TeV for $\tau e$
transitions and $M_{SUSY}\gsim 200/\sqrt{|\cos(\alpha-\beta)|}$ TeV
for $\mu e$ transitions.

The Higgs decays in supersymmetric extensions of the SM have been
already studied e.g. in~\cite{Arganda:2004bz, Azatov:2012wq,
  Petersson:2012nv, Bartl:2014bka, Arana-Catania:2013xma,
  Aloni:2015wvn, Arganda:2015uca, Barenboim:2015fya, Gomez:2017dhl}.
In this Section we analyse within the general MSSM the decays of the
lighter CP-even Higgs boson $h$.  The mass eigenstates formulae for
the MSSM contributions to the effective leptonic Yukawa couplings of
$h$ are given in Eqs.~(\ref{eq:yukeff} -- \ref{eq:hnn}) while the
relevant MI expressions are collected in Appendix~\ref{app:lhmi}.  The
potentially largest contributions to $h\to \ell\ell'$ decays come from
the effects non decoupling in the limit of large SUSY masses and
proportional to non-holomorphic trilinear terms (see~\eq{eq:lnh}).
Assuming that flavour violating $A_l^{'IJ}$ terms are the only source
of LFV and using Eqs.~(\ref{eq:brhll}), (\ref{eq:lnh}) and
(\ref{eq:zrdef}), one can write
\bea
Br(h\to \ell^I \bar\ell^J) &\approx& \frac{e^4 M_h}{8192 \pi^5 c_W^4
  \Gamma_h} \frac{\cos^2(\alpha-\beta)}{\cos^2\beta} \left(
g(x_{\tilde e_{LI}}, x_{\tilde e_{RJ}})^2
\left|\Delta_{LR}^{'IJ}\right|^2 \right. \nn
&+& \left. g(x_{\tilde e_{LJ}}, x_{\tilde e_{RI}})^2
\left|\Delta_{LR}^{'JI}\right|^2 \right)\,,
\eea
where $\alpha,\beta$ are the mixing angles in the Higgs sector (see
Appendix~\ref{app:lagr}), the dimensionless mass ratios are
\bea
x_{\tilde e_{L(R)I}} = \frac{m_{\tilde e_{L(R)I}}}{|M_1|}\,,
\eea
and we defined
\bea
g(x,y) = -\sqrt{x y}\; C_0(x,y,1) \;.
\label{eq:gratdef}
\eea

\begin{figure}[tbp]
  \begin{center}
    \includegraphics[width=0.5\textwidth]{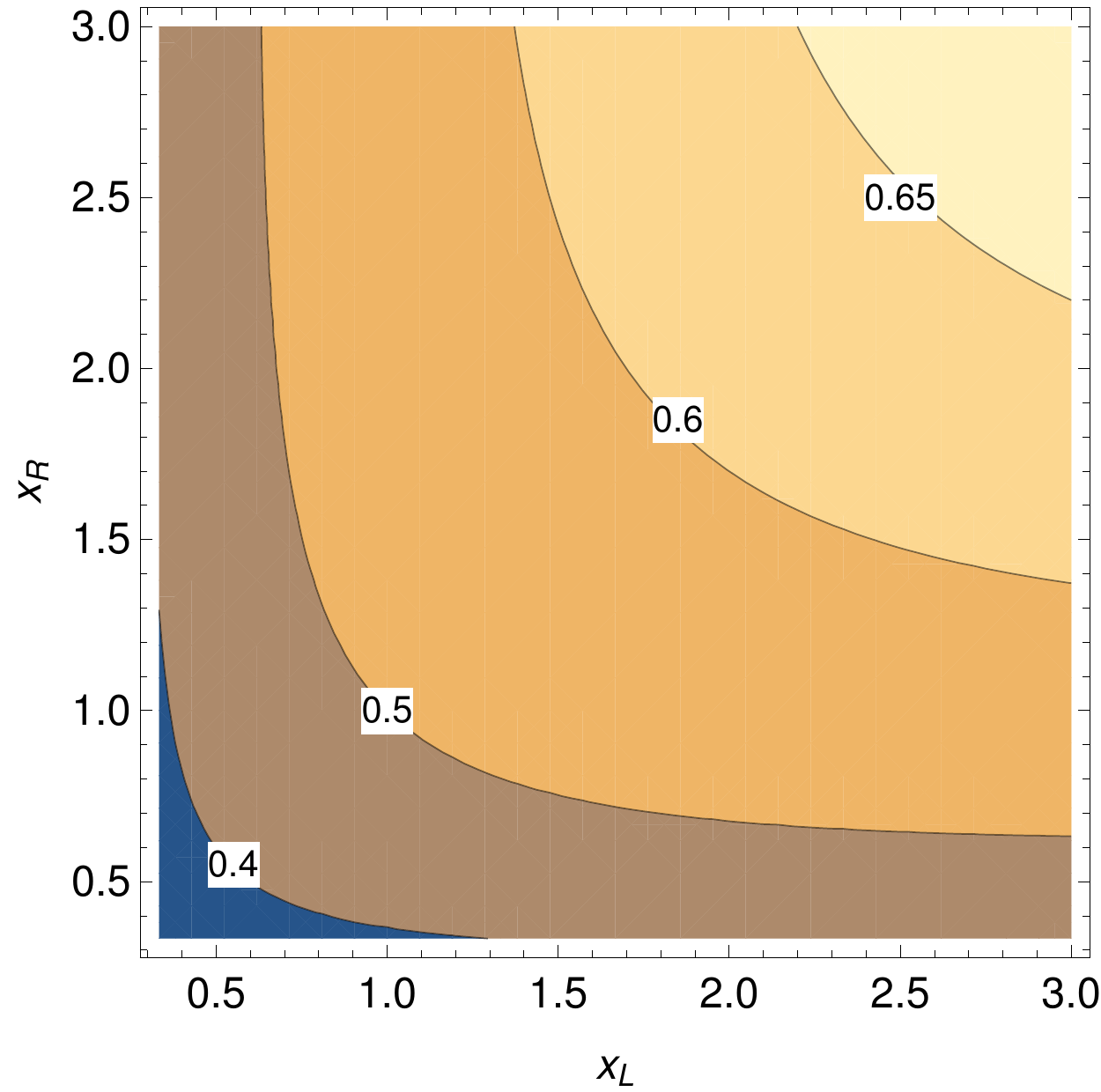}
\end{center}
\caption{\small Dependence of function $g(x_L, x_R)$
  of~\eq{eq:gratdef} on the splitting between the slepton and bino
  masses.\label{fig:cprat}}
\end{figure}

As can be seen from of Fig.~\ref{fig:cprat}, for reasonable mass
splittings $g(x,y)\sim{\cal O}(1)$ and, inserting the numerical values
of known quantities, one has
\bea
Br(h\to \ell^I \bar\ell^J) \sim 2\cdot 10^{-4} \;
\frac{\cos^2(\alpha-\beta)}{\cos^2\beta}\left
|\Delta_{LR}^{'IJ(JI)}\right|^2\,.
\label{eq:hlllim}
\eea
Even if for large SUSY mass scale $\Delta_{LR}^{'IJ}$ insertions are
not constrained experimentally by other LFV measurements, $Br(h\to
\ell^I \bar\ell^J)$ cannot be arbitrarily large in the MSSM because
$\Delta_{LR}^{'IJ}$ are constrained to ${\cal O}(1)$ by the vacuum
stability conditions and the requirement of the absence of charge and
colour breaking (CCB) minima of the scalar potential (see
e.g. discussion in~\cite{Dedes:2014asa}).

The Higgs mixing angle $\alpha$ is subject to strong radiative
corrections from the squark sector and thus from the point of view of
pure leptonic sector can be treated as a free parameter.  However, the
allowed values of the Higgs mixing angles $\alpha, \beta$ are limited
by the existing experimental constraints (see e.g. Fig.~6 in Appendix
B of ref.~\cite{Aloni:2015wvn}), thus also the overall pre-factor
$\frac{\cos^2(\alpha-\beta)}{\cos^2\beta}$ in~\eq{eq:hlllim} can be at
most ${\cal O}(1)$. Summarising, the maximal $Br(h\to \ell^I
\bar\ell^J)$ which can be generated with the non-holomorphic trilinear
terms is ${\cal O}(10^{-4})$, not much below the current experimental
sensitivities collected in Table~\ref{tab:leplim} (including of
decoupling contributions does not change this conclusion even for a
light SUSY spectrum~\cite{Arana-Catania:2013xma, Aloni:2015wvn}).
Further searches may therefore find the effects of non-holomorphic
trilinear terms or provide stricter bounds on them.

Similar analysis could be done for non-decoupling contributions
proportional to $\Delta_{LL}^{IJ}$ and $\Delta_{RR}^{IJ}$ parameters.
However, in this case non-decoupling terms are proportional also
either to the diagonal $A_l'$ soft terms or to lepton Yukawa
couplings, so the formulae become complicated and a more involved
numerical analysis is required. Terms proportional to
$\Delta_{LL}^{IJ}$ and $\Delta_{RR}^{IJ}$ multiplied by diagonal
$A_l'$ terms can generate similar LFV Higgs decay rates as the flavour
off-diagonal non-holomorphic $A_l^\prime$-terms. However, assuming
that all non-holomorphic terms vanish, and including only the Yukawa
suppressed contributions one has a much stricter bound $Br(h\to \ell^I
\bar\ell^J) \lsim 10^{-4} (Y_l^I)^2$ in the MSSM.

For a complete phenomenological analysis of LFV Higgs decays in the
MSSM one would need to go beyond the one-loop analysis of this
article.  First, one would need to perform the matching of the MSSM on
the 2HDM with generic Yukawa couplings including the resummation of
the higher order chirally enhanced effects (see for
example~\cite{Crivellin:2010er, Crivellin:2011jt, Crivellin:2012zz}).
Then, one has to calculate the loop effects for flavour observables
within this generic 2HDM~\cite{Crivellin:2013wna}.


\section{Conclusions}
\label{sec:conclusion}

New precision data in the lepton flavour sector are expected to come
in the foreseeable future.  In the search for beyond the SM effects,
they will require precision and efficient calculations in various BSM
models. In this article lepton flavour violating processes within MSSM
have been calculated using the Flavour Expansion Theorem, a recently
developed new technique of a purely algebraic mass-insertion expansion
of the amplitudes~\cite{Dedes:2015twa}. Both flavour-violating
off-diagonal terms and flavour-conserving mass-insertions are
considered.  The expansion in the flavour conserving off-diagonal mass
terms leads to a transparent qualitative understanding of the
coefficients in front of the flavour violating mass insertions
(see~\eq{eq:miform}) in various decoupling limits. Most flavour
violating one-loop amplitudes decouple as $v^2/M^2$ where $M$ is one
of the soft SUSY breaking mass parameters. The exception are the Higgs
flavour violating decays where the amplitudes decouple as $v^2/M^2_A$.
We find that our full MI approximation, both in flavour violating and
flavour conserving off-diagonal mass terms is an excellent
approximation to the calculations in the mass eigenstates basis for a
very broad pattern of supersymmetric spectra, in particular for highly
non-degenerate spectra.  This is useful because in the MI
approximation we work directly with the Lagrangian parameters and can
constrain them with experimental limits.

On the physics side, the considered processes are:
$\ell \to \ell' \gamma$, $\ell \to 3 \ell'$, $\ell \to 2\ell'\ell''$,
$h \to \ell\ell'$ as well as $\mu \to e$ conversion in nuclei.  The
bounds on the flavour changing parameters of the MSSM have been
updated and their sensitivity to the forthcoming experimental results
in different channels has been discussed.  We have emphasised that,
given the foreseen experimental progress, precision measurements of
different processes have very different potential for the discovery of
supersymmetric effects. The radiative and leptonic muon decays are
likely to remain the most important source of information on
supersymmetric LFV. The leptonic decays play a complementary role to
the radiative ones in eliminating some "blind spots" of weakly
constrained by the latter LFV mass insertions.  This is illustrated in
Sections~\ref{sec:msplit} and~\ref{sec:corr}. Our complete analytical
MI expansion facilitates the investigation of the LFV processes when
the SUSY spectra are non-degenerate and finding such "blind spots"
with suppressed branching ratios and regions of correlations between
various processes. This is illustrated in Sections~\ref{sec:msplit}
and~\ref{sec:corr}. The LFV Higgs decays are discussed in some detail.
For the supersymmetric spectrum of order of 1 TeV, the current
experimental limits on the LFV Higgs decays give several orders of
magnitude weaker bounds on lepton violating MI than the radiative
lepton decays.  However, for the superpartner masses of several TeV
Higgs decays provide stronger bounds than the latter because the
bounds from Higgs decays do not scale with superpartner masses.  We
have also analysed the role of the so-called non-holomorphic $A$-terms
in the flavour-violating Higgs boson decays, which can give branching
ratios not much below the present experimental sensitivity.

\section*{Acknowledgements}

ZF, WM, SP and JR are supported in part by the National Science
Centre, Poland, under research grants
DEC-2015/19/B/ST2/02848,
DEC-2015/18/M/ST2/00054
and DEC-2014/15/B/ST2/02157.  
U.N. is supported by BMBF under grant no.~05H15VKKB1. The work of
A.C. is supported by an Ambizione Grant of the Swiss National Science
Foundation (PZ00P2\_154834). JR would also like to thank CERN for
hospitality during his visits there. 

\newpage

\appendix
\def\theequation{\thesection.\arabic{equation}}

\section{MSSM Lagrangian and vertices}
\label{app:lagr}

Throughout this article we use the notation of
Refs.~\cite{Rosiek:1989rs, Rosiek:1995kg} which is very similar to
SLHA2 conventions~\cite{Allanach:2008qq}, up to minor differences
listed in Table~\ref{tab:slha2}.

For completeness, we collect here the definitions of the mass and
  mixing matrices for the supersymmetric particles and the relevant
  MSSM Feynman rules.  The slepton and sneutrino mass and mixing
matrices are defined as:
\bea
Z_{\nu}^{\dagger} \left(M_{LL}^2 + \frac{M_Z^2\cos2\beta}{2} \hat
1\right) Z_{\nu} &=& {\rm diag}\left( m_{\nu_1}^2 \ldots m_{\nu_3}^2
\right)\,,
\eea
\bea
Z_L^{\dagger} \left(
\begin{array}{cc}
\left(
{\cal M}_L^2\right)_{LL} & 
\left({\cal M}_L^2\right)_{LR} \\
\left({\cal M}_L^2\right)_{LR}^{\dagger} & 
\left({\cal M}_L^2\right)_{RR} 
\end{array}
\right)Z_L = {\rm diag}\left( m_{L_1}^2 \ldots m_{L_6}^2 \right)\,,
\eea
\bea
\left({\cal M}_L^2\right)_{LL} &=& (M_{LL}^2)^T +
\frac{M_Z^2\cos2\beta}{2} (1-2c_W^2) {\hat 1} + \frac{v_1^2 Y_l^2}{2}\,,
\\
\left({\cal M}_L^2\right)_{RR} &=& M_{RR}^2 -
\frac{M_Z^2\cos2\beta}{2} s_W^2 {\hat 1} + \frac{v_1^2 Y_l^2}{2}\,, \\
\left({\cal M}_L^2\right)_{LR} &=& \frac{1}{\sqrt{2}} \left(v_2 (Y_l
\mu^{\star} - A_l^{'}) + v_1 A_l \right)\,,
\eea
where, as usual, we use $\tan\beta=v_2/v_1$ and $M_{LL}^2$,
$M_{RR}^2$, $A_l$, $A_l^{'}$, $Y_l=-\sqrt{2}m_l/v_1$ are $3\times 3$
matrices in flavour space.

The neutralino and chargino mass and mixing matrices can be written
down as:
\bea
Z_N^T
\left(
\begin{array}{cccc}
  M_1 & 0 & -\frac{ev_1}{2c_W} & \frac{ev_2}{2c_W} \\ 
  0 & M_2 & \frac{ev_1}{2s_W} & -\frac{ev_2}{2s_W} \\ 
  -\frac{ev_1}{2c_W} & \frac{ev_1}{2s_W} & 0 & -\mu \\ 
  \frac{ev_2}{2c_W} & -\frac{ev_2}{2s_W} &-\mu
  & 0
\end{array}
\right) Z_N &=& {\rm diag}\left( m_{\chi^0_1} \ldots m_{\chi^0_4}
\right)\,,
\eea
\bea
(Z_-)^T 
\left(
\begin{array}{cc}
 M_2 & \frac{ev_2}{\sqrt{2}s_W}
\\  \frac{ev_1}{\sqrt{2}s_W} & \mu
\end{array}
\right) Z_+ = {\rm diag} \left( m_{\chi_1}, m_{\chi_2} \right)\,. %
\eea
We also use the following abbreviation for the matrix $Z_R$
parametrizing the mixing in the CP-even Higgs sector:
\bea
Z_R =  \left(
\begin{array}{cc}
    {\cos}{\alpha}&-  {\sin}{\alpha}\\
    {\sin}{\alpha}&  {\cos}{\alpha}
\end{array}
\right)\,.
\label{eq:zrdef}
\eea

\begin{table}[tb]
\begin{center}
\begin{tabular}{|c|c|}
\hline
SLHA2~\cite{Allanach:2008qq} & Ref.~\cite{Rosiek:1989rs,Rosiek:1995kg}\\
\hline 
& \\[-4mm]
$\hat T_U$, $\hat T_D$, $\hat T_E$ & $-A_u^T$, $+A_d^T$, $+A_l^T$\\
$\hat m_{\tilde Q}^2$, $\hat m_{\tilde L}^2$ & $m_Q^2$, $m_L^2$ \\
$\hat m_{\tilde u}^2$, $\hat m_{\tilde d}^2$, $\hat m_{\tilde l}^2$ &
$(m_U^2)^T$, $(m_D^2)^T$, $(m_E^2)^T$ \\
${\cal M}_{\tilde u}^2$, ${\cal M}_{\tilde d}^2$ & $({\cal M}_U^2)^T$,
$({\cal M}_D^2)^T$ \\[1mm]
\hline
\end{tabular}
\end{center}
\caption{\small Comparison of~SLHA2~\cite{Allanach:2008qq} and
  ref.~\cite{Rosiek:1989rs,Rosiek:1995kg} conventions.}
\label{tab:slha2}
\end{table}

Below we list the vertices used in calculations of the LFV processes
expressed in terms of the mixing matrices defined above.

\noindent 1) Lepton-slepton-neutralino and lepton-sneutrino-chargino
vertices (for an {\em incoming} charged lepton of flavour $I$):
\bea
V_{\ell \tilde L N,L}^{Iij} &=& {e \over \sqrt{2}s_Wc_W} Z_L^{Ii} (Z_N^{1j} s_W
+ Z_N^{2j} c_W) + Y_l^I Z_L^{(I+3)i} Z_N^{3j}\,, \nn
V_{\ell \tilde L N,R}^{Iij} &=& {-e\sqrt{2} \over c_W} Z_L^{(I+3)i}
Z_N^{1j\star} + Y_l^I Z_{L}^{Ii} Z_{N}^{3j\star}\,,\nn
V_{\ell\tilde\nu C,L}^{IKj} &=& - {e \over s_W} Z_+^{1j}\,,
Z_{\nu}^{IK\star} \nn
V_{\ell\tilde\nu C,R}^{IKj} &=& - Y_l^I Z_-^{2j\star} Z_{\nu}^{IK\star}\,.
\label{eq:vlsl}
\eea
2) $Z$-chargino and $Z$-neutralino vertices:
\bea
V_{CCZ,L}^{ij} &=& - \frac{e}{2 s_W c_W} \left(Z_+^{1i*}Z_+^{1j} +
\delta^{ij} (c_W^2 - s_W^2)\right)\,, \nn
V_{CCZ,R}^{ij} &=& - \frac{e}{2 s_W c_W} \left(Z_-^{1i}Z_-^{1j*} +
\delta^{ij} (c_W^2 - s_W^2)\right)\,, \nn
V_{NNZ,L}^{ij} &=& - V_{NNZ,R}^{ji} = \frac{e}{2 s_W
  c_W}\left(Z_N^{4i*} Z_N^{4j} - Z_N^{3i*} Z_N^{3j}\right)\,.
\label{eq:vzcn}
\eea
3) CP-even-Higgs-slepton and  CP-even-Higgs-sneutrino vertices:
\bea
V_{HLL}^{Kil} &=& \sum_{C=1}^3 \left( {e^2 \over 2c_W^2} \left( v_1
Z_R^{1K} - v_2 Z_R^{2K} \right) \left(\delta^{il} + {1-4s_W^2 \over
2s_W^2} Z_L^{Ci\star} Z_L^{Cl}\right) \right.  \nonumber\\
& - &  (Y_l^C)^2 v_1 Z_R^{1K} (Z_L^{Ci\star} Z_L^{Cl} +
 Z_L^{(C+3)i\star} Z_L^{(C+3)l}) \nonumber\\
&-& \left. {Z_R^{2K}\over \sqrt{2}}   (Y_l^{C*}\mu^* Z_L^{Ci\star}
Z_L^{(C+3)l} + Y_l^{C} \mu Z_L^{Cl} Z_L^{(C+3)i\star})\right)\nonumber\\
&-& {1\over \sqrt{2}} \sum_{C,D=1}^3 \left( Z_R^{1K} (A_l^{CD\star} Z_L^{Cl}
Z_L^{(D+3)i\star} + A_l^{CD} Z_L^{Ci\star} Z_L^{(D+3)l} ) \right. \nonumber\\
& -& \left. Z_R^{2K} (A_l^{'CD\star} Z_L^{Cl} Z_L^{(D+3)i\star} +
A_l^{'CD} Z_L^{Ci\star} Z_L^{(D+3)l}) \right)\,,\nn
V_{H\tilde\nu\tilde\nu}^{KLI} &=& -\frac{e^2}{4s_W^2c_W^2}(v_1
Z_R^{1K} - v_2 Z_R^{2K})\delta_{LI} \,.
\label{eq:vhsl}
\eea
4) CP-odd-Higgs-slepton and CP-odd-Higgs-sneutrino vertices:
\bea
V_{ALL}^{1il} &=& {i\cos\beta\over \sqrt{2}} \sum_{C,D=1}^3 \left( (
A_l^{CD*}\tan\beta + A_l^{'CD*} - Y_l^{C}\mu\delta^{CD}) Z_L^{Cj}
Z_L^{(D+3)i\star} \right.\nn
&-&\left. (A_l^{CD}\tan\beta + A_l^{'CD} - Y_l^{C} \mu^{*}\delta^{CD})
Z_L^{Ci\star} Z_L^{(D+3)j} \right) \,,\nn
V_{A\tilde\nu\tilde\nu}^{1LI} &=& 0 \,.
\label{eq:ahsl}
\eea
5) CP-even-Higgs-neutralino and  CP-even-Higgs-chargino vertices:
\bea
V_{NHN,L}^{iKl} &=& V_{NHN,R}^{iKl\,*} = {e \over 2 s_W c_W} \left(
(Z_R^{1K} Z_N^{3l} - Z_R^{2K} Z_N^{4l}) (Z_N^{1i} s_W - Z_N^{2i}
c_W)\right.\nonumber\\
&+&\left.  (Z_R^{1K} Z_N^{3i} - Z_R^{2K} Z_N^{4i}) (Z_N^{1l} s_W -
Z_N^{2l} c_W) \right) \,,\nn
V_{CHC,L}^{iKl} &=& V_{CHC,R}^{iKl\,*} = -{e \over \sqrt{2} s_W }
\left( Z_R^{1K} Z_-^{2i} Z_+^{1l} + Z_R^{2K} Z_-^{1i} Z_+^{2l} \right)\,.
\label{eq:vhcn}
\eea
6) CP-odd-Higgs-neutralino and  CP-odd-Higgs-chargino vertices:
\bea
V_{NAN,L}^{i1l} &=& V_{NAN,R}^{i1l\,*} = {-ie^2 \over 4 s_W^2 c_W^2
  M_Z} \left( (v_2 Z_N^{3j} - v_1 Z_N^{4j}) (Z_N^{1i} s_W - Z_N^{2i}
c_W)\right.\nn
&+&\left. (v_2 Z_N^{3i} - v_1 Z_N^{4i} ) (Z_N^{1j} s_W - Z_N^{2j}
c_W)\right) \,,\nn
V_{CAC,L}^{i1l} &=& V_{CAC,R}^{i1l\,*} = {ie^2 \over 2 \sqrt{2}s_W^2
  M_W} (v_2 Z_-^{2i} Z_+^{1j} + v_1 Z_-^{1i} Z_+^{2j}) \,.
\label{eq:vacn}
\eea
7) $Z$-slepton vertex:
\bea
V_{LLZ}^{ij} &=& {e \over 2 s_W c_W} \left(Z_L^{ki*} Z_L^{kj} - 2
s_W^2 \delta^{ij} \right)\,.
\label{eq:vzll}
\eea

\section{Loop integrals}
\label{app:loop}

We define the following loop integrals for 2-point and 3-point
functions with non-vanishing external momenta $p$ and $q$:
\bea
\frac{i}{(4\pi)^2}B_0(p,m_1,m_2) &=& \int \frac{d^4k}{(2\pi)^4}
\frac{1}{(k^2 -m_1^2)((k-p)^2 -m_2^2)} \,,\nn
\frac{i}{(4\pi)^2}p_{\mu} B_1(p,m_1,m_2) &=& \int
\frac{d^4k}{(2\pi)^4} \frac{k_{\mu}}{(k^2 -m_1^2)((k-p)^2 -m_2^2)} \,,\\
\frac{i}{(4\pi)^2}C_{2n}(p,q,m_1,m_2,m_3) &=& \int \frac{d^4k}{(2\pi)^4}
\frac{(k^2)^n}{(k^2 -m_1^2)((k+p)^2 -m_2^2)((k+p+q)^2 -m_3^2)} \,,\nn
\frac{i}{(4\pi)^2}\left(p_{\mu} C_{11}(p,q,m_1,m_2,m_3)\right. &+&
\left.  q_{\mu} C_{12}(p,q,m_1,m_2,m_3)\right)\nn
&=& \int \frac{d^4k}{(2\pi)^4} \frac{k_{\mu}}{(k^2 -m_1^2)((k+p)^2
  -m_2^2)((k+p+q)^2 -m_3^2)}\,.\nonumber
\eea

In our expanded results we need only the integrals above, their
derivatives and higher point 1-loop integrals calculated at vanishing
external momenta. Let us define
\bea
\frac{i}{(4\pi)^2}L_i^{2n}(m_1,\ldots,m_i) &=& \int
\frac{d^4k}{(2\pi)^4} \frac{(k^2)^{n}}{\prod\limits_{j=1}^i (k^2 -
  m_j^2)}\,.
\eea
In common notation $L_3^{2n}=C_{2n}, L_4^{2n}=D_{2n}, L_5^{2n}=E_{2n}$
etc.

For $i\geq 3$ one has:
\bea
L_i^0(m_1,\ldots,m_i) &=& - \sum_{j=2}^i \frac{m_j^2 \log
  \frac{m_j^2}{m_1^2}}{\prod\limits_{k=1,k\neq j}^i (m_j^2 -
  m_k^2)}\,,\nn
L_i^2(m_1,\ldots,m_i) &=& \sum_{j=2}^i \frac{m_j^4 \log
  \frac{m_j^2}{m_1^2}}{\prod\limits_{k=1,k\neq j}^i (m_j^2 -
  m_k^2)}\,,
\label{eq:l01def}
\eea
(with the exception of $L_3^2\equiv C_2$ having also an infinite part,
which however is always cancelled out in flavour violating processes
and is thus not given here explicitly).

To simplify our formulae, we use the relation
\bea
2 L_i^0(m_1,m_2,\ldots,m_i) &=& L_{i+1}^{2}(m_1,m_1,m_2,\ldots, m_i) +
L_{i+1}^{2}(m_1,m_2,m_2,\ldots, m_i) \nn
&+& \ldots + L_{i+1}^{2}(m_1,\ldots, m_{i-1}, m_i, m_i)\,,
\label{eq:lrel}
\eea
which can be obtained by differentiating with respect to $\lambda$ the
integral form of the homogeneity property
\bea
L_i^{0}(\lambda m_1,\ldots,\lambda m_i) &=& \lambda^{4-2i}
L_i^{0}( m_1,\ldots, m_i)\, ,
\eea
and using the relation ($k=1,\ldots,i$)
\bea
m_k^2 L_{i+1}^{0}(m_1,\ldots, m_k, m_k, \ldots,m_i) &=&
L_{i+1}^{2}(m_1, \ldots, m_k, m_k, \ldots, m_i) \nn
&-& L_i^{0}( m_1,\ldots,m_k,\ldots, m_i)\, .
\eea

In addition, we define the following integrals:
\bea
C_0'(m_1,m_2,m_3) &=& \left.\frac{\partial
  C_0(p,q,m_1,m_2,m_3)}{\partial q^2}\right|_{p=q=0} \nn
&=& \frac{ 2 m_2^2
  m_3^2 - m_1^2 (m_2^2 + m_3^2) }{2 (m_1^2 - m_2^2) (m_1^2 - m_3^2)
  (m_2^2 - m_3^2)^2}
\label{eq:c0diff}
\\
&+& \frac{m_1^4 \log\frac{m_1^2}{m_2^2}}{2 (m_1^2 - m_2^2)^2 (m_1^2 -
  m_3^2)^2}
+\frac{m_3^2 (m_3^4 -2 m_1^2 m_2^2 + m_2^2 m_3^2 )
  \log\frac{m_3^2}{m_2^2}}{ 2 (m_1^2 - m_3^2)^2 (m_2^2 -
  m_3^2)^3}\,,\nonumber
\eea
\bea
C_{11}(m_1,m_2) &=& - \frac{m_1^2 - 3 m_2^2}{4(m_1^2-m_2^2)^2} +
\frac{ m_2^4}{2(m_1^2-m_2^2)^3} \log\frac{m_2^2}{m_1^2} \,,\\[2mm]
C_{12}(m_1,m_2) &=& - \frac{m_1^2 + m_2^2}{2 (m_1^2-m_2^2)^2} -
\frac{ m_1^2 m_2^2}{(m_1^2-m_2^2)^3} \log\frac{m_2^2}{m_1^2} \,,\\[2mm]
C_{23}(m_1,m_2) &=& - \frac{m_1^4 - 5 m_1^2 m_2^2 - 2 m_2^4}{12
  (m_1^2-m_2^2)^3} + \frac{m_1^2 m_2^4}{2 (m_1^2-m_2^2)^4}
\log\frac{m_2^2}{m_1^2}\,,\\[2mm]
C_{01}(m_1,m_2) &=& \frac{7 m_1^4 - 29 m_1^2 m_2^2 + 16 m_2^4}{36
  (m_1^2 - m_2^2)^3 } + \frac{ m_2^4 (-3 m_1^2 + 2 m_2^2) }{6 (m_1^2 -
  m_2^2)^4 }\log\frac{m_2^2}{m_1^2}\,,\\[2mm]
C_{02}(m_1,m_2) &=& \frac{11 m_1^4 - 7 m_1^2 m_2^2 + 2 m_2^4}{36
  (m_1^2 - m_2^2)^3 }+ \frac{m_1^6}{6 (m_1^2 - m_2^2)^4 }
\log\frac{m_2^2}{m_1^2}\,.
\label{eq:caux}
\eea

\section{Divided differences}
\label{app:ddiff}

The expansion of the amplitudes given in the mass eigenbasis in terms
of mass insertions can be naturally expressed~\cite{Dedes:2015twa} by
the so-called divided differences of the loop functions.

In case of a function of a single argument, $f(x)$, divided
differences are defined recursively as:
\bea
f^{[0]}(x) & =&  f(x)\,, \nn
f^{[1]}(x,y) & =& {f^{[0]}(x) - f^{[0]}(y) \over x -y}\,, \nn
f^{[2]}(x,y,z) & =& {f^{[1]}(x,y) - f^{[1]}(x,z) \over y -z}\,, \nn
&\ldots&\,.
\eea
As can be easily checked, a divided difference of order $n$ is
symmetric under permutation of any subset of its arguments. It also
has a smooth limit for degenerate arguments:
\bea
\lim_{\{ x_0,\dots ,x_m\} \to \{ \xi,\dots,\xi \}}
f^{[k]}(x_0,\dots,x_k)=\frac{1}{m!}\frac{\partial^{m} }{\partial
  \xi^{m}}f^{[k-m]}(\xi,x_{m+1}\dots,x_k)\;.
\label{eq:ddlimit}
\eea
To compactify the formulae for functions of many arguments, we use the
notation
\bea
f^{[k]}(x_0,\dots,x_k) & \equiv& f(\{x_0,\dots,x_k\})\,,
\eea
where the order of the divided difference is defined by the number of
arguments inside curly brackets. Then, for example a divided
difference of the 1st order in the 1st argument and of the 3rd order
in the 2nd argument for the function of 3 variables, $g(x,y,z)$, can
be written down as:
\bea
g(\{x_1,x_2\},\{y_1,y_2,y_3,y_4\},z)\,.
\eea
For the loop functions defined in Appendix~\ref{app:loop} one should
note that their natural arguments are squares of masses. However, we
use $m_i$'s instead of $m_i^2$'s to compactify the notation. Thus, for
loop functions we write divided differences as
\bea
L(m_1,\ldots,\{m_i,m_i'\},\ldots,m_n) =
\frac{L(m_1,\ldots,m_i,\ldots,m_n) -
  L(m_1,\ldots,m_i',\ldots,m_n)}{m_i^2 - m_i^{'2}} \,,
\eea
with squared masses in the denominator.

The FET expansion works for any transition amplitude, also in the
  case of non-vanishing external momenta or for multi-loop
  calculations.  However, it is particularly effective for 1-loop
  functions with vanishing external momenta, due to the fact that the
  notion of the divided differences is naturally encoded in the
  structure of such functions: a divided difference of a $n$-point
  scalar 1-loop function is a $(n+1)$-point function (see eq.~3.13 in
  ref.~\cite{Dedes:2015twa} for generalisation to the case of
  non-vanishing external momenta). Thus, for example one has
\bea
B_0(m_1,\{m_2,m_3\}) &=& B_0(\{m_1,m_2\},m_3) = C_0(m_1,m_2,m_3)\nn
B_0(m_1,\{m_2,m_3,m_4\}) &=& C_0(m_1,m_2,\{m_3,m_4\}) =
D_0(m_1,m_2,m_3,m_4)\\
&\ldots& \nonumber
\eea
We use such relations extensively to find cancellations between
various terms and to identify the lowest non-vanishing order of mass
insertion expansion for a given process.

\section{Box diagrams in the mass eigenstates basis}
\label{app:fullbox}

\begin{figure}[tb]
\begin{center}
\includegraphics[width=0.8\textwidth]{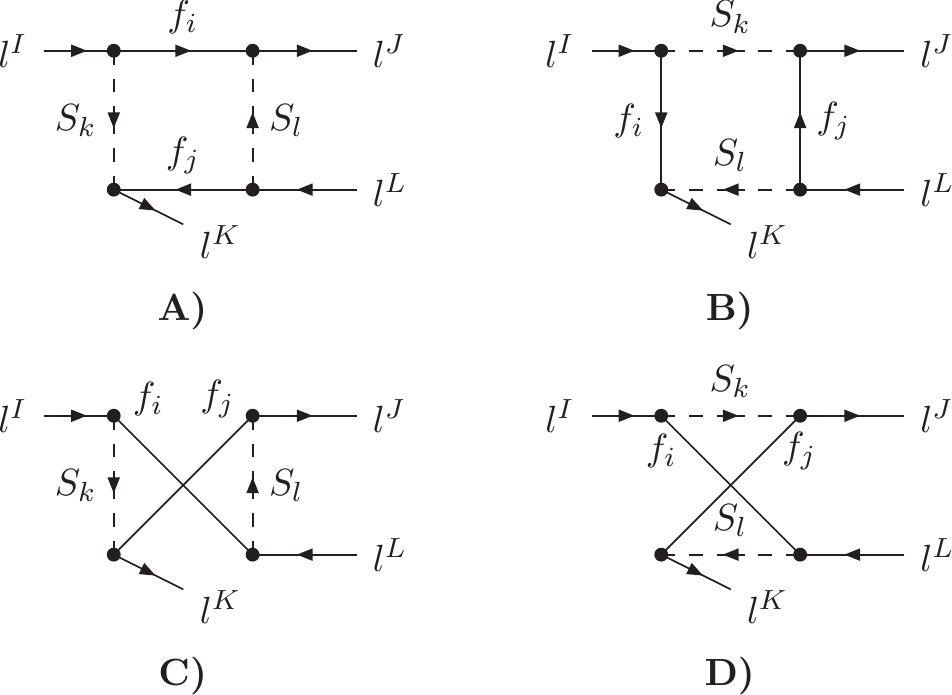}
\caption{\small MSSM box diagrams with 4 external charged
  leptons\label{fig:lbox}}
\end{center}
\end{figure}

There are four types of box diagrams with four external leptons
involving slepton (sneutrinos) and neutralinos (charginos) in the
loop, displayed in Fig.~\ref{fig:lbox}. Both chargino-sneutrino and
neutralino-slepton pairs contribute to diagrams A) and B), while only
neutralinos (Majorana fermions) can be exchanged in the ``crossed''
diagrams C) and D).

Using whenever necessary Fierz identities, the amplitudes describing
  each of the diagrams $N=A,B,C,D$ can be brought into the form
\bea
i A_{N}^{JIKL} = i \sum_{Q=V,S,T} B_{N \;Q XY}^{JIKL} [\bar u(p_J)
  \Gamma_Q P_X u(p_I)] [\bar u(p_K) \Gamma_Q P_Y v(p_L)]
\eea
with $\Gamma_V=\gamma^\mu$, $\Gamma_S=1$ and
$\Gamma_T=\sigma_{\mu\nu}$.  Note that for $\Gamma_T$ only the case
$X=Y$ is non vanishing.
Assuming that the generic couplings for an incoming lepton $\ell^I$ -
an incoming scalar particle $S_k$ and an outgoing fermion $f_i$ takes
the form
\bea
i V_{\ell Sf}^{Iki} = i(A_{\ell Sf}^{Iki} P_L + B_{\ell Sf}^{Iki}
P_R)\, ,
\eea
the contribution from diagram A) in Fig.~\ref{fig:lbox}) to the Wilson
coefficients $B_{QXY}$ can be written down as:
\bea
(4\pi)^2 B_{A\;VLL}^{JIKL} &=& \frac{1}{4} A_{\ell Sf}^{Iki} A_{\ell
  Sf}^{Jli*} A_{\ell Sf}^{Kkj*} A_{\ell Sf}^{Llj} D_2\,, \nn
(4\pi)^2 B_{A\;VRR}^{JIKL} &=& \frac{1}{4} B_{\ell Sf}^{Iki} B_{\ell
  Sf}^{Jli*} B_{\ell Sf}^{Kkj*} B_{\ell Sf}^{Llj} D_2\,, \nn
(4\pi)^2 B_{A\;VLR}^{JIKL} &=& \frac{1}{4} A_{\ell Sf}^{Iki} A_{\ell
  Sf}^{Jli*} B_{\ell Sf}^{Kkj*} B_{\ell Sf}^{Llj} D_2\,, \nn
(4\pi)^2 B_{A\;VRL}^{JIKL} &=& \frac{1}{4} B_{\ell Sf}^{Iki} B_{\ell
  Sf}^{Jli*} A_{\ell Sf}^{Kkj*} A_{\ell Sf}^{Llj} D_2\,, \nn
(4\pi)^2 B_{A\;SLL}^{JIKL} &=& A_{\ell Sf}^{Iki} B_{\ell Sf}^{Jli*}
B_{\ell Sf}^{Kkj*} A_{\ell Sf}^{Llj} m_{f_i} m_{f_j} D_0\,, \nn
(4\pi)^2 B_{A\;SRR}^{JIKL} &=& B_{\ell Sf}^{Iki} A_{\ell Sf}^{Jli*}
A_{\ell Sf}^{Kkj*} B_{\ell Sf}^{Llj} m_{f_i} m_{f_j} D_0\,, \nn
(4\pi)^2 B_{A\;SLR}^{JIKL} &=& A_{\ell Sf}^{Iki} B_{\ell Sf}^{Jli*}
A_{\ell Sf}^{Kkj*} B_{\ell Sf}^{Llj} m_{f_i} m_{f_j} D_0\,, \nn
(4\pi)^2 B_{A\;SRL}^{JIKL} &=& B_{\ell Sf}^{Iki} A_{\ell Sf}^{Jli*}
B_{\ell Sf}^{Kkj*} A_{\ell Sf}^{Llj} m_{f_i} m_{f_j} D_0\,, \nn
(4\pi)^2 B_{A\;TL}^{JIKL} &=& 0\,, \nn
(4\pi)^2 B_{A\;TR}^{JIKL} &=& 0\,.
\label{eq:lboxa}
\eea
where $D_0,D_2$ above are the abbreviations for 4-point loop functions
with respective mass arguments, $D_0 =
D_0(m_{f_i},m_{f_j},m_{S_k},m_{S_l}), D_2 =
D_2(m_{f_i},m_{f_j},m_{S_k},m_{S_l})$ (see Appendix~\ref{app:loop}).

Using the same notation, the contributions from diagram B), C), D) are:
\bea
(4\pi)^2 B_{B\;VLL}^{JIKL} &=& \frac{1}{4} A_{\ell Sf}^{Iki} A_{\ell
  Sf}^{Jkj*} A_{\ell Sf}^{Kli*} A_{\ell Sf}^{Llj} D_2 \,,\nn
(4\pi)^2 B_{B\;VRR}^{JIKL} &=& \frac{1}{4} B_{\ell Sf}^{Iki} B_{\ell
  Sf}^{Jkj*} B_{\ell Sf}^{Kli*} B_{\ell Sf}^{Llj} D_2\,, \nn
(4\pi)^2 B_{B\;VLR}^{JIKL} &=& - \frac{1}{2} A_{\ell Sf}^{Iki} A_{\ell
  Sf}^{Jkj*} B_{\ell Sf}^{Kli*} B_{\ell Sf}^{Llj} m_{f_i} m_{f_j} D_0\,,\nn
(4\pi)^2 B_{B\;VRL}^{JIKL} &=& - \frac{1}{2} B_{\ell Sf}^{Iki} B_{\ell
  Sf}^{Jkj*} A_{\ell Sf}^{Kli*} A_{\ell Sf}^{Llj} m_{f_i} m_{f_j} D_0\,,\nn
(4\pi)^2 B_{B\;SLL}^{JIKL} &=& - \frac{1}{2} A_{\ell Sf}^{Iki} B_{\ell
  Sf}^{Jkj*} B_{\ell Sf}^{Kli*} A_{\ell Sf}^{Llj} m_{f_i} m_{f_j} D_0\,,
\nn
(4\pi)^2 B_{B\;SRR}^{JIKL} &=& - \frac{1}{2} B_{\ell Sf}^{Iki} A_{\ell
  Sf}^{Jkj*} A_{\ell Sf}^{Kli*} B_{\ell Sf}^{Llj} m_{f_i} m_{f_j} D_0\,,
\nn
(4\pi)^2 B_{B\;SLR}^{JIKL} &=& - \frac{1}{2} A_{\ell Sf}^{Iki} B_{\ell
  Sf}^{Jkj*} A_{\ell Sf}^{Kli*} B_{\ell Sf}^{Llj} D_2\,, \nn
(4\pi)^2 B_{B\;SRL}^{JIKL} &=& - \frac{1}{2} B_{\ell Sf}^{Iki} A_{\ell
  Sf}^{Jkj*} B_{\ell Sf}^{Kli*} A_{\ell Sf}^{Llj} D_2 \,,\nn
(4\pi)^2 B_{B\;TL}^{JIKL} &=& -\frac{1}{8} A_{\ell Sf}^{Iki} B_{\ell
  Sf}^{Jkj*} B_{\ell Sf}^{Kli*} A_{\ell Sf}^{Llj} m_{f_i} m_{f_j} D_0\,,
\nn
(4\pi)^2 B_{B\;TR}^{JIKL} &=& -\frac{1}{8} B_{\ell Sf}^{Iki} A_{\ell
  Sf}^{Jkj*} A_{\ell Sf}^{Kli*} B_{\ell Sf}^{Llj} m_{f_i} m_{f_j} D_0\,,
\label{eq:lboxb}
\eea
\bea
(4\pi)^2 B_{C\;VLL}^{JIKL} &=& \frac{1}{2} A_{\ell Sf}^{Iki} A_{\ell
  Sf}^{Lli} A_{\ell Sf}^{Jlj*} A_{\ell Sf}^{Kkj*} m_{f_i} m_{f_j} D_0\,,
\nn
(4\pi)^2 B_{C\;VRR}^{JIKL} &=& \frac{1}{2} B_{\ell Sf}^{Iki}B_{\ell
  Sf}^{Lli} B_{\ell Sf}^{Jlj*} B_{\ell Sf}^{Kkj*} m_{f_i} m_{f_j} D_0\,,
\nn
(4\pi)^2 B_{C\;VLR}^{JIKL} &=& \frac{1}{4} B_{\ell Sf}^{Iki}A_{\ell
  Sf}^{Lli} B_{\ell Sf}^{Jlj*} A_{\ell Sf}^{Kkj*} D_2\,,\nn
(4\pi)^2 B_{C\;VRL}^{JIKL} &=& \frac{1}{4} A_{\ell Sf}^{Iki} B_{\ell
  Sf}^{Lli} A_{\ell Sf}^{Jlj*} B_{\ell Sf}^{Kkj*} D_2\,,\nn
(4\pi)^2 B_{C\;SLL}^{JIKL} &=& - \frac{1}{2} A_{\ell Sf}^{Iki} A_{\ell
  Sf}^{Lli} B_{\ell Sf}^{Jlj*} B_{\ell Sf}^{Kkj*} m_{f_i} m_{f_j} D_0\,,
\nn
(4\pi)^2 B_{C\;SRR}^{JIKL} &=& - \frac{1}{2} B_{\ell Sf}^{Iki}B_{\ell
  Sf}^{Lli} A_{\ell Sf}^{Jlj*} A_{\ell Sf}^{Kkj*} m_{f_i} m_{f_j} D_0\,,
\nn
(4\pi)^2 B_{C\;SLR}^{JIKL} &=& \frac{1}{2} B_{\ell Sf}^{Iki}A_{\ell
  Sf}^{Lli} A_{\ell Sf}^{Jlj*} B_{\ell Sf}^{Kkj*} D_2 \,,\nn
(4\pi)^2 B_{C\;SRL}^{JIKL} &=& \frac{1}{2} A_{\ell Sf}^{Iki} B_{\ell
  Sf}^{Lli} B_{\ell Sf}^{Jlj*} A_{\ell Sf}^{Kkj*} D_2 \,,\nn
(4\pi)^2 B_{C\;TL}^{JIKL} &=& \frac{1}{8} A_{\ell Sf}^{Iki} A_{\ell
  Sf}^{Lli} B_{\ell Sf}^{Jlj*} B_{\ell Sf}^{Kkj*} m_{f_i} m_{f_j} D_0\,,
\nn
(4\pi)^2 B_{C\;TR}^{JIKL} &=& \frac{1}{8} B_{\ell Sf}^{Iki} B_{\ell
  Sf}^{Lli} A_{\ell Sf}^{Jlj*} A_{\ell Sf}^{Kkj*} m_{f_i} m_{f_j} D_0\,,
\label{eq:lboxc}
\eea
\bea
(4\pi)^2 B_{D\;VLL}^{JIKL} &=& \frac{1}{2} A_{\ell Sf}^{Iki} A_{\ell
  Sf}^{Lli} A_{\ell Sf}^{Jkj*} A_{\ell Sf}^{Klj*} m_{f_i} m_{f_j} D_0\,,\nn
(4\pi)^2 B_{D\;VRR}^{JIKL} &=& \frac{1}{2} B_{\ell Sf}^{Iki}B_{\ell
  Sf}^{Lli} B_{\ell Sf}^{Jkj*} B_{\ell Sf}^{Klj*} m_{f_i} m_{f_j} D_0\,,\nn
(4\pi)^2 B_{D\;VLR}^{JIKL} &=& - \frac{1}{4} B_{\ell Sf}^{Iki}A_{\ell
  Sf}^{Lli} B_{\ell Sf}^{Jkj*} A_{\ell Sf}^{Klj*} D_2\,,\nn
(4\pi)^2 B_{D\;VRL}^{JIKL} &=& - \frac{1}{4} A_{\ell Sf}^{Iki} B_{\ell
  Sf}^{Lli} A_{\ell Sf}^{Jkj*} B_{\ell Sf}^{Klj*} D_2\,,\nn
(4\pi)^2 B_{D\;SLL}^{JIKL} &=& - \frac{1}{2} A_{\ell Sf}^{Iki} A_{\ell
  Sf}^{Lli} B_{\ell Sf}^{Jkj*} B_{\ell Sf}^{Klj*} m_{f_i} m_{f_j} D_0\,,\nn
(4\pi)^2 B_{D\;SRR}^{JIKL} &=& - \frac{1}{2} B_{\ell Sf}^{Iki}B_{\ell
  Sf}^{Lli} A_{\ell Sf}^{Jkj*} A_{\ell Sf}^{Klj*} m_{f_i} m_{f_j} D_0\,,\nn
(4\pi)^2 B_{D\;SLR}^{JIKL} &=& - \frac{1}{2} B_{\ell Sf}^{Iki}A_{\ell
  Sf}^{Lli} A_{\ell Sf}^{Jkj*} B_{\ell Sf}^{Klj*} D_2\,, \nn
(4\pi)^2 B_{D\;SRL}^{JIKL} &=& - \frac{1}{2} A_{\ell Sf}^{Iki} B_{\ell
  Sf}^{Lli} B_{\ell Sf}^{Jkj*} A_{\ell Sf}^{Klj*} D_2 \,,\nn
(4\pi)^2 B_{D\;TL}^{JIKL} &=& \frac{1}{8} A_{\ell Sf}^{Iki} A_{\ell
  Sf}^{Lli} B_{\ell Sf}^{Jkj*} B_{\ell Sf}^{Klj*} m_{f_i} m_{f_j} D_0\,,\nn
(4\pi)^2 B_{D\;TR}^{JIKL} &=& \frac{1}{8} B_{\ell Sf}^{Iki} B_{\ell
  Sf}^{Lli} A_{\ell Sf}^{Jkj*} A_{\ell Sf}^{Klj*} m_{f_i} m_{f_j} D_0\,,
\label{eq:lboxd}
\eea

To obtain the actual MSSM contributions to the 4-lepton operators, one
should add terms from eqs.~(\ref{eq:lboxa},\ref{eq:lboxb}) with
replacements $f\to C, S\to\tilde\nu, A_{\ell Sf}\to V_{\ell\tilde\nu
  C,L}, B_{\ell Sf}\to V_{\ell\tilde\nu C,R}$ and $f\to N, S\to \tilde
L, A_{\ell Sf}\to V_{\ell\tilde L N,L}, B_{\ell Sf}\to V_{\ell\tilde L
  N,R}$ (summing over repeated indices of loop particles) and terms
from eqs.~(\ref{eq:lboxc},\ref{eq:lboxd}), substituting there only
$f\to N, S\to \tilde L, A_{\ell Sf}\to V_{\ell\tilde L N,L}, B_{\ell
  Sf}\to V_{\ell\tilde L N,R}$.

The contributions to 2-quark 2-lepton operators can be obtained from
diagrams A) and C) by replacing $\ell_K$ and $\ell_L$ with $q_K$ and
$q_L$ as defined in~\eq{eq:qboxbasis}.  Therefore, the expressions for
$B_{q\;QXY}$ can be obtained replacing vertices of leptons $\ell^K$
and $\ell^L$ by the relevant quark-squark vertices. Such vertices are
not listed in Appendix~\ref{app:lagr} but can be found in
Refs.~\cite{Rosiek:1989rs, Rosiek:1995kg}.  The explicit form of
$\ell\ell d d$ box amplitudes can be also found in Appendix A.3 of
Ref.~\cite{Dedes:2008iw}.

\section{Effective lepton couplings in the leading MI order}
\label{app:miexp}

We list below the MI expanded expressions for the leptonic penguin and
box diagram amplitudes. For penguins we follow the decomposition
of~\eq{eq:miform}, with $F_{XY}$ denoting functions of flavour diagonal
SUSY parameters multiplying the respective slepton mass insertions:
\bea
F_X^{IJ} &=& \frac{1}{(4\pi)^2}\left(F_{X~LL}^{IJ}\; \Delta_{LL}^{IJ} +
F_{X~RR}^{IJ}\; \Delta_{RR}^{JI} \right. \nn
&+&\left. F_{X~ALR}^{IJ}\; \Delta_{LR}^{JI} + F_{X~BLR}^{IJ} \;
\Delta_{LR}^{IJ*} + F_{X~ALR}^{'IJ}\; \Delta_{LR}^{'JI} +
F_{X~BLR}^{'IJ}\; \Delta_{LR}^{'IJ*}\right)\;.
%
%
\eea
To compactify the notation, we also introduce the abbreviation
\bea
\bar M_{XY}^{IJ} = \sqrt{ \left(M_{XX}^2\right)^{II}
  \left(M_{YY}^2\right)^{JJ} }
\eea
where $X,Y = L$ or $R$.

\subsection{Lepton-photon vertex}
\label{app:llgmi}

\subsubsection{Tensor (magnetic) couplings}

After performing MI expansion, one can see that terms coming from
$F_{\gamma A}$ in~\eq{eq:fgamab} are always suppressed by the powers
of lepton Yukawa couplings or lepton masses, and may add to or cancel
terms generated from $F_{\gamma L B}, F_{\gamma R B}$. Thus, in the
expressions below we give the sum of both types of contributions.

The chargino contributions contain only terms proportional to LL
slepton mass insertions (see Appendix~\ref{app:ddiff} for the notation
of divided differences and curly brackets around the function
arguments)
\bea
\left(F_{\gamma~LL}\right)^{JI}_C &=& \frac{e^2 v_1 Y_L^J}{2\sqrt{2}
  s_W^2} \bar M_{LL}^{IJ} \left(
C_{11}(\left|M_2\right|,\{m_{\tilde\nu_I},m_{\tilde\nu_J}\})
\right. \nn
&+& C_{11}(\left|\mu\right|,\{m_{\tilde\nu_I},m_{\tilde\nu_J}\})
- C_{23}(\left|M_2\right|,\{m_{\tilde\nu_I},m_{\tilde\nu_J}\})
\nn
&+& \left. \left(\left|\mu\right|^2 + \left|M_2\right|^2 + 2 \mu^*
M_2^* \tan\beta\right) C_{11}(\{\left|\mu\right|,
\left|M_2\right|\},\{m_{\tilde\nu_I},m_{\tilde\nu_J}\}) \right)
\eea
The non-vanishing neutralino contributions are:
\bea
\left(F_{\gamma~LL}\right)^{JI}_N &= & \frac{e^2}{2 c_W^2} \bar
M_{LL}^{IJ} \left( M_1^{\star} C_{12}(\{m_{\tilde
  e_{LI}},m_{\tilde e_{LJ}},m_{\tilde e_{RJ}}\},|M_1|)
\left(M_{LR}^2\right)_{JJ} \right. \nn
&-& \frac{v_1}{2\sqrt{2}} Y_L^J \left(\frac{c_W^2}{s_W^2}
(C_{12}(\{m_{\tilde e_{LI}},m_{\tilde e_{LJ}}\},|\mu|) -
C_{23}(\{m_{\tilde e_{LI}},m_{\tilde e_{LJ}}\},|M_2|))
\right. \nn
&-& C_{12}(\{m_{\tilde e_{LI}},m_{\tilde e_{LJ}}\},|\mu|) -
C_{23}(\{m_{\tilde e_{LI}},m_{\tilde e_{LJ}}\},|M_1|) \nn
&+& \left(|M_2|^2 + \mu^{\star} M_2^{\star}\tan\beta \right)
\frac{c_W^2}{s_W^2}C_{12}(\{m_{\tilde e_{LI}},m_{\tilde
  e_{LJ}}\},\{|\mu|,|M_2|\}) \nn
\label{eq:llgmineut1}
&-& \left.\left. \left(|M_1|^2 + \mu^{\star} M_1^{\star} \tan\beta
\right) C_{12}(\{m_{\tilde e_{LJ}},m_{\tilde
  e_{LI}}\},\{|\mu|,|M_1|\}) \right)\right)\\[2mm]
\left(F_{\gamma~RR}\right)^{JI}_N &= & \frac{e^2}{2 c_W^2} \bar
M_{RR}^{IJ} \left( M_1^{\star} C_{12} \left( \{m_{\tilde
  e_{LI}},m_{\tilde e_{RI}},m_{\tilde e_{RJ}}\},|M_1|\right)
\left(M_{LR}^2\right)_{II} \right. \nn
& -& \frac{v_1}{\sqrt{2}} Y_L^{I} \left( C_{12} \left(\{m_{\tilde
  e_{RI}},m_{\tilde e_{RJ}}\},|\mu|\right) - 2 C_{23}(\{m_{\tilde
  e_{RI}},m_{\tilde e_{RJ}}\},|M_1|) \right. \nn
& +&\left.\left. \left(|M_1|^2 + \mu^{\star} M_1^{\star}\tan\beta
\right) C_{12}(\{m_{\tilde e_{RI}},m_{\tilde
  e_{RJ}}\},\{\mu,|M_1|\})\right)\right)\nonumber\\[2mm]
\left(F_{\gamma~ALR}\right)_N^{JI} &=& -\frac{v_1}{v_2}
\left(F_{\gamma~ALR}'\right)_N^{JI} = \frac{e^2 v_1}{2\sqrt{2} c_W^2}
\sqrt{ \bar M_{LR}^{IJ}} \; M_1^{\star} C_{12}(\{m_{\tilde
  e_{LI}},m_{\tilde e_{RJ}}\},|M_1|) \nonumber
\eea

\subsubsection{Vector couplings}

Loop functions $C_{01}$ and $C_{02}$ appearing in~\eq{eq:llgv} scale
with the inverse of the squared SUSY scale $M^2$. Thus, only LL and RR
terms contribute to the MI expanded expressions at the $v^2/M^2$
order, as LR mass insertions always come with additional $v/M$
powers. The non-vanishing chargino and neutralino contributions are:
\bea
\left(V_{\gamma L~LL}\right)^{JI}_C &=& \frac{e^2}{s_W^2}\bar
M_{LL}^{IJ}\; C_{01}(|M_2|,\{m_{\tilde \nu_I},m_{\tilde \nu_J}\})
\eea
\bea
\left(V_{\gamma L~LL}\right)^{JI}_N & = & - \frac{e^2}{2 s_W^2 c_W^2}
\bar M_{LL}^{IJ}\; (c_W^2 C_{02}(|M_2|,\{m_{\tilde e_{LI}},m_{\tilde
  e_{LJ}}\}) + s_W^2 C_{02}(|M_1|,\{m_{\tilde e_{LI}},m_{\tilde
  e_{LJ}}\})\nn
\left(V_{\gamma R~RR}\right)^{JI}_N & = &- \frac{2 e^2}{c_W^2} \bar
M_{RR}^{IJ}\; C_{02}(|M_1|,\{m_{\tilde e_{RI}},m_{\tilde e_{RJ}}\})
\eea

\subsection{Lepton-$Z^0$ vertex}
\label{app:llzmi}

The leading $v^2/M_{SUSY}^2$ terms in the effective
$Z\bar\ell^{I}\ell^{J}$ vertex defined in~\eq{eq:lzdef}, expanded to
the 1st order in LFV mass insertions, depend on divided differences of
scalar $C_0$ and $C_2$ 3-point functions.  They can be expressed as
higher point 1-loop functions (see Appendices~\ref{app:loop} and
\ref{app:ddiff}). We give here the expressions using explicitly scalar
4-, 5- and 6-point functions $D$, $E$ and $F$.

The only non-negligible chargino contribution to $Z\ell\ell'$ vertex
read:
\bea
\left(F_{ZL~LL}\right)^{JI}_C &= & - \frac{e^5}{4 s_W^5 c_W} \bar
M_{LL}^{IJ} \left(v_2^2 \,
D_0(|M_2|,|\mu|,m_{\tilde\nu^I},m_{\tilde\nu^J})\right. \nn
&+& (v_1^2 - v_2^2)
E_2(|M_2|,|M_2|,|\mu|,m_{\tilde\nu^I},m_{\tilde\nu^J})  \nn
&+&\left.  \frac{1}{2} \left|v_2 M_2 + v_1 \mu^* \right|^2
F_2(|M_2|,|M_2|,|\mu|,|\mu|,m_{\tilde\nu^I},m_{\tilde\nu^J})\right)
\eea
Neutralino contributions have a more complicated form. They can be
written down as:
\bea
(F_{ZL~LL})_N^{JI} &=& \frac{e^3\sqrt{2}}{16 s_W^3 c_W^3} \bar
M_{LL}^{IJ}\; (X_{ZNL4}^{JI} + X_{ZNL5}^{JI} + X_{ZNL5}^{IJ*}) \nn
(F_{ZR~LL})_N^{JI} & =& \frac{e^3\sqrt{2}}{8 s_W c_W^3} \bar
M_{LL}^{IJ}\; (X_{ZNR4}^{IJ} + X_{ZNR5}^{JI} + X_{ZNR5}^{IJ*})
\eea
\bea
(F_{ZL~RR})_N^{JI} &=& \frac{e^3\sqrt{2}}{16 s_W^3 c_W^3} \bar
M_{RR}^{IJ}\; (X_{ZNL2}^{JI} + X_{ZNL3}^{JI} + X_{ZNL3}^{IJ*}) \nn
(F_{ZR~RR})_N^{JI} & =& \frac{e^3\sqrt{2}}{8 s_W c_W^3} \bar
M_{RR}^{IJ}\; (X_{ZNR2}^{JI} + X_{ZNR3}^{JI} + X_{ZNR3}^{IJ*})
\eea
\bea
(F_{ZL~ALR})_N^{JI} &=& (F_{ZL~BLR})_N^{IJ*} = - \frac{v_1}{v_2}
(F_{ZL~ALR}')_N^{JI} = - \frac{v_1}{v_2} (F_{ZL~BLR})_N^{IJ*}\nn
&=& \frac{e^3 v_1}{16 s_W^3 c_W^3} \sqrt{\bar M_{LR}^{IJ}}\;
X_{ZNL1}^{JI} \nn
(F_{ZR~ALR})_N^{JI} &=& (F_{ZR~BLR})_N^{IJ*} = - \frac{v_1}{v_2}
(F_{ZR~ALR}')_N^{JI} = - \frac{v_1}{v_2} (F_{ZR~BLR})_N^{IJ*} \nn
&=& \frac{e^3 v_1}{4 s_W c_W^3} \sqrt{\bar M_{LR}^{IJ}}\;
X_{ZNR1}^{JI}
\eea
%
%
%
%
where we defined
\bea
X_{ZNL1}^{JI} &=& \sqrt{2} (s_W^2 E_2(|M_1|,m_{\tilde e_{LJ}},m_{\tilde
  e_{LI}},m_{\tilde e_{RJ}},m_{\tilde e_{RJ}}) \nn
&+& c_W^2 E_2(|M_2|,m_{\tilde e_{LJ}},m_{\tilde e_{LI}},m_{\tilde
  e_{RJ}},m_{\tilde e_{RJ}}))(M_{LR}^2)_{JJ}^* \nn
&+ & Y_l^J\left( 2 v_1 (M_1^* s_W^2 D_0(|M_1|,|\mu|,m_{\tilde
  e_{LI}},m_{\tilde e_{RJ}}) - c_W^2 M_2^* D_0(|M_2|,|\mu|,m_{\tilde
  e_{LI}},m_{\tilde e_{RJ}}))\right.\nn
&-& s_W^2 (v_1 M_1^* + v_2 \mu) (E_2(|M_1|,|\mu|,m_{\tilde e_{LI}},m_{\tilde
  e_{RJ}},m_{\tilde e_{RJ}}) \nn
&+& E_2(|M_1|,|\mu|,|\mu|,m_{\tilde e_{LI}},m_{\tilde e_{RJ}}))\nn
&+&  c_W^2 (v_1 M_2^* + v_2 \mu) (E_2(|M_2|,|\mu|,m_{\tilde
  e_{LI}},m_{\tilde e_{RJ}},m_{\tilde e_{RJ}}) \nn
&+& \left. E_2(|M_2|,|\mu|,|\mu|,m_{\tilde e_{LI}},m_{\tilde
  e_{RJ}}))\right)
\eea
\bea
X_{ZNL2}^{JI} &=& \sqrt{2} (M_{LR}^2)_{JJ}^* (M_{LR}^2)_{II} (s_W^2
(F_2(|M_1|,m_{\tilde e_{LJ}},m_{\tilde e_{LI}},m_{\tilde
  e_{RJ}},m_{\tilde e_{RJ}},m_{\tilde e_{RI}}) \nn
&+& F_2(|M_1|,m_{\tilde e_{LJ}},m_{\tilde e_{LI}},m_{\tilde
  e_{RJ}},m_{\tilde e_{RI}},m_{\tilde e_{RI}})) \nn
&+& c_W^2 (F_2(|M_2|,m_{\tilde e_{LJ}},m_{\tilde e_{LI}},m_{\tilde
  e_{RJ}},m_{\tilde e_{RJ}},m_{\tilde e_{RI}}) \nn
&+& F_2(|M_2|,m_{\tilde e_{LJ}},m_{\tilde e_{LI}},m_{\tilde
  e_{RJ}},m_{\tilde e_{RI}},m_{\tilde e_{RI}})))
\eea
\bea
X_{ZNL3}^{JI} &=& Y_l^{I*} (M_{LR}^2)_{JJ}^* \left( 2 v_1 (M_1^* s_W^2
E_0(|M_1|,|\mu|,m_{\tilde e_{LJ}},m_{\tilde e_{RJ}},m_{\tilde
  e_{RI}})\right. \nn
&-& c_W^2 M_2^* E_0(|M_2|,|\mu|,m_{\tilde e_{LJ}},m_{\tilde e_{RJ}},m_{\tilde
  e_{RI}})) \\
&-& s_W^2 (v_1 M_1^* + v_2 \mu) (F_2(|M_1|,|\mu|,m_{\tilde e_{LJ}},m_{\tilde
  e_{RJ}},m_{\tilde e_{RJ}},m_{\tilde e_{RI}})\nn
&+& F_2(|M_1|,|\mu|,m_{\tilde e_{LJ}},m_{\tilde e_{RJ}},m_{\tilde
  e_{RI}},m_{\tilde e_{RI}}) + F_2(|M_1|,|\mu|,|\mu|,m_{\tilde
  e_{LJ}},m_{\tilde e_{RJ}},m_{\tilde e_{RI}})) \nn
&+& c_W^2 (v_1 M_2^* + v_2 \mu) (F_2(|M_2|,|\mu|,m_{\tilde
  e_{LJ}},m_{\tilde e_{RJ}},m_{\tilde e_{RJ}},m_{\tilde e_{RI}}) \nn
&+& \left.F_2(|M_2|,|\mu|,m_{\tilde e_{LJ}},m_{\tilde
  e_{RJ}},m_{\tilde e_{RI}},m_{\tilde e_{RI}}) +
F_2(|M_2|,|\mu|,|\mu|,m_{\tilde e_{LJ}},m_{\tilde e_{RJ}},m_{\tilde
  e_{RI}})) \right) \nonumber
\eea
\bea
X_{ZNL4}^{JI} &=& \frac{e^2 (v_1^2 - v_2^2)}{\sqrt{2} s_W^2 c_W^2}
\left(s_W^4 D_0(|M_1|,|\mu|,m_{\tilde e_{LJ}},m_{\tilde e_{LI}}) +
c_W^4 D_0(|M_2|,|\mu|,m_{\tilde e_{LJ}},m_{\tilde e_{LI}}) \right.\nn
&+& 2 s_W^2 c_W^2 \re(M_1 M_2^*) E_0(|M_1|,|M_2|,|\mu|,m_{\tilde
  e_{LJ}},m_{\tilde e_{LI}}) \nn
&-& s_W^4 (E_2(|M_1|,|M_1|,|\mu|,m_{\tilde e_{LJ}},m_{\tilde e_{LI}})
+ E_2(|M_1|,|\mu|,|\mu|,m_{\tilde e_{LJ}},m_{\tilde e_{LI}})) \nn
&-& c_W^4 (E_2(|M_2|,|M_2|,|\mu|,m_{\tilde e_{LJ}},m_{\tilde e_{LI}})
+ E_2(|M_2|,|\mu|,|\mu|,m_{\tilde e_{LJ}},m_{\tilde e_{LI}}))\nn
&+& 2 s_W^2 c_W^2 E_2(|M_1|,|M_2|,|\mu|,m_{\tilde e_{LJ}},m_{\tilde
  e_{LI}}) \nn
&+& \frac{1}{2} s_W^4 (|\mu|^2 - |M_1|^2)
F_2(|M_1|,|M_1|,|\mu|,|\mu|,m_{\tilde e_{LJ}},m_{\tilde e_{LI}}) \nn
&+&\frac{1}{2} c_W^4 (|\mu|^2 - |M_2|^2)
F_2(|M_2|,|M_2|,|\mu|,|\mu|,m_{\tilde e_{LJ}},m_{\tilde e_{LI}}) \nn
&+& \left. s_W^2 c_W^2 (|\mu|^2 - \re(M_1 M_2^*))
F_2(|M_1|,|M_2|,|\mu|,|\mu|,m_{\tilde e_{LJ}},m_{\tilde e_{LI}})
\right)
\eea
\bea
X_{ZNL5}^{JI} &=& Y_l^{I*} (M_{LR}^2)_{II}^* \left(2 v_1 (s_W^2 M_1^*
E_0(|M_1|,|\mu|,m_{\tilde e_{LJ}},m_{\tilde e_{LI}},m_{\tilde e_{RI}})
\right. \nn
&-& c_W^2 M_2^* E_0(|M_2|,|\mu|,m_{\tilde e_{LJ}},m_{\tilde
  e_{LI}},m_{\tilde e_{RI}})) \nn
&-& s_W^2 (v_1 M_1^* + v_2 \mu) (F_2(|M_1|,|\mu|,m_{\tilde e_{LJ}},m_{\tilde
  e_{LI}},m_{\tilde e_{RI}},m_{\tilde e_{RI}}) \nn
&+& F_2(|M_1|,|\mu|,|\mu|,m_{\tilde e_{LJ}},m_{\tilde
  e_{LI}},m_{\tilde e_{RI}})) \nn
&+& c_W^2 (v_1 M_2^* + v_2 \mu) (F_2(|M_2|,|\mu|,m_{\tilde
  e_{LJ}},m_{\tilde e_{LI}},m_{\tilde e_{RI}},m_{\tilde e_{RI}})\nn
&+&\left. F_2(|M_2|,|\mu|,|\mu|,m_{\tilde e_{LJ}},m_{\tilde
  e_{LI}},m_{\tilde e_{RI}})) \right)\nn
&+& \sqrt{2}\,  (s_W^2 F_2(|M_1|,m_{\tilde e_{LJ}},m_{\tilde
  e_{LI}},m_{\tilde e_{LI}},m_{\tilde e_{RI}},m_{\tilde e_{RI}}) \nn
&+& c_W^2 F_2(|M_2|,m_{\tilde e_{LJ}},m_{\tilde e_{LI}},m_{\tilde
  e_{LI}},m_{\tilde e_{RI}},m_{\tilde e_{RI}}))
\left|(M_{LR}^2)_{II}\right|^2
\eea
\bea
%
X_{ZNR1}^{JI} & = & Y_l^I \left(2 v_1 M_1^* D_0(|M_1|,|\mu|,m_{\tilde
  e_{LI}},m_{\tilde e_{RJ}}) \right. \nn
&-& (v_1 M_1^* + v_2 \mu ) (E_2(|M_1|,|\mu|,m_{\tilde e_{LI}},m_{\tilde
  e_{LI}},m_{\tilde e_{RJ}}) \nn
&+& \left. E_2(|M_1|,|\mu|,|\mu|,m_{\tilde e_{LI}},m_{\tilde e_{RJ}}))
\right) \nn
&-& 2 \sqrt{2}\, E_2(|M_1|,m_{\tilde e_{LI}},m_{\tilde
  e_{LI}},m_{\tilde e_{RJ}},m_{\tilde e_{RI}}) (M_{LR}^2)_{II}^* 
\eea
%
%
\bea
X_{ZNR2}^{JI} & = & \frac{\sqrt{2} e^2(v_1^2 - v_2^2) }{c_W^2} \left(
|M_1|^2 E_0(|M_1|,|\mu|,|\mu|,m_{\tilde e_{RJ}},m_{\tilde e_{RI}})
\right. \nn
&+& E_2(|M_1|,|M_1|,|\mu|,m_{\tilde e_{RJ}},m_{\tilde e_{RI}})\nn
&-& \frac{1}{2}(|M_1|^2 - |\mu|^2) (F_2(|M_1|,|\mu|,|\mu|,m_{\tilde
  e_{RJ}},m_{\tilde e_{RJ}},m_{\tilde e_{RI}}) \nn
&+& F_2(|M_1|,|\mu|,|\mu|,m_{\tilde e_{RJ}},m_{\tilde
  e_{RI}},m_{\tilde e_{RI}}) \nn
&+& \left. 2 F_2(|M_1|,|\mu|,|\mu|,|\mu|,m_{\tilde e_{RJ}},m_{\tilde
  e_{RI}})) \right)
\eea
\bea
X_{ZNR3}^{JI} & = & Y_l^I (M_{LR}^2)_{II} \left( 2 v_1 M_1^*
E_0(|M_1|,|\mu|,m_{\tilde e_{LI}},m_{\tilde e_{RJ}},m_{\tilde e_{RI}})
\right. \nn
&-& (v_1 M_1^* + v_2 \mu ) (F_2(|M_1|,|\mu|,m_{\tilde
  e_{LI}},m_{\tilde e_{LI}},m_{\tilde e_{RJ}},m_{\tilde e_{RI}}) \nn
&+& \left. F_2(|M_1|,|\mu|,|\mu|,m_{\tilde e_{LI}},m_{\tilde e_{RJ}},m_{\tilde
  e_{RI}})) \right) \nn
&-& 2 \sqrt{2}\, F_2(|M_1|,m_{\tilde e_{LI}},m_{\tilde
  e_{LI}},m_{\tilde e_{RJ}},m_{\tilde e_{RI}},m_{\tilde e_{RI}})
\left|(M_{LR}^2)_{II}\right|^2
\eea
%
%
\bea
X_{ZNR4}^{JI} & = & -2 \sqrt{2}\, (F_2(|M_1|,m_{\tilde
  e_{LJ}},m_{\tilde e_{LJ}},m_{\tilde e_{LI}},m_{\tilde
  e_{RJ}},m_{\tilde e_{RI}}) \nn
&+& F_2(|M_1|,m_{\tilde e_{LJ}},m_{\tilde e_{LI}},m_{\tilde
  e_{LI}},m_{\tilde e_{RJ}} ,m_{\tilde e_{RI}})) (M_{LR}^2)_{JJ}
(M_{LR}^2)_{II}^*
\eea
\bea
X_{ZNR5}^{JI} & = & - Y_l^I (M_{LR}^2)_{JJ} \left(2 \mu v_2
E_0(|M_1|,|\mu|,m_{\tilde e_{LJ}},m_{\tilde e_{LI}},m_{\tilde e_{RJ}})
\right. \nn
&-& (v_1 M_1^* + v_2 \mu ) (F_2(|M_1|,|M_1|,|\mu|,m_{\tilde
  e_{LJ}},m_{\tilde e_{LI}},m_{\tilde e_{RJ}}) \nn
&+& \left. F_2(|M_1|,|\mu|,m_{\tilde e_{LJ}},m_{\tilde
  e_{LI}},m_{\tilde e_{RJ}},m_{\tilde e_{RJ}})) \right)
\eea

\subsection{CP-even Higgs-lepton vertex}
\label{app:lhmi}

The dominant MI terms in the effective CP-even Higgs - lepton
couplings (see~\eq{eq:lhdef}) can be split into four classes,
\bea
F_h^{IJK} = \frac{1}{(4\pi)^2} \left(F_{hnd}^{IJK} + F_{hY}^{IJK} +
F_{hdec}^{IJK}+ F_{hm}^{IJK}\right)\;,
\eea
defined as (below we give the sum of neutralino and chargino
contributions, the latter appearing only as single term depending on
sneutrino masses in eq.~(\ref{eq:ly}) and follow notation
of~\eq{eq:miform}):
\smallskip

\noindent
1. Contributions proportional to non-holomorphic $A_l'$ trilinear
terms\footnote{For comparison with commonly used notation of the Higgs
  mixing angles, note that\\ $(v_1 Z_R^{2K} - v_2 Z_R^{1K})/v_1 =
  \left\{\begin{array}{lcl}
\sin(\alpha-\beta)/\cos\beta & \mathrm{for} & K=1\\
\cos(\alpha - \beta)/\cos\beta & \mathrm{for} & K=2
\end{array}\right.\,. $}, non-decoupling for $M_{SUSY} \gg v$:
\bea
(F_{hnd~ALR}')^{IJK} &=& \frac{e^2(v_1 Z_R^{2K}- v_2 Z_R^{1K}
  )}{\sqrt{2}c_W^2 v_1} \sqrt{\bar M_{LR}^{IJ}}\; M_1^{\star}\;
C_0(\left|M_1\right|,m_{\tilde e_{LI}}, m_{\tilde e_{RJ}}) \\
(F_{hnd~LL})^{IJK} &=& \frac{e^2 (v_1 Z_R^{2K}- v_2
  Z_R^{1K})}{\sqrt{2}c_W^2 v_1} \bar M_{LL}^{IJ}\; M_1^{\star} D_0
(\left|M_1\right|,m_{\tilde e_{LI}}, m_{\tilde e_{LJ}}, m_{\tilde
  e_{RJ}}) A_L^{'JJ} \nn
(F_{hnd~RR})^{IJK} &=& \frac{e^2 (v_1 Z_R^{2K}- v_2 Z_R^{1K}
  )}{\sqrt{2}c_W^2 v_1} \bar M_{RR}^{IJ}\; M_1^{\star} \; D_0
(\left|M_1\right|,m_{\tilde e_{LI}}, m_{\tilde e_{RI}}, m_{\tilde
  e_{RJ}}) A_L^{'II}\nonumber
\label{eq:lnh}
\eea

\noindent
2. Contributions suppressed by the lepton Yukawa couplings, also
non-decoupling for $M_{SUSY} \gg v$:
\bea
(F_{hY~LL})^{IJK} &=& - \frac{e^2}{2\sqrt{2} v_1 c_W^2 s_W^2} \left(v_1
Z_R^{2K}- v_2 Z_R^{1K}\right) \left( s_W^2 M_1^{\star} \mu^{\star}
(D_0 (\left|M_1\right|,\left|\mu\right|, m_{\tilde e_{LI}}, m_{\tilde
  e_{LJ}}) \right. \nn
& +& 2 D_0(\left|M_1\right|,m_{\tilde e_{LI}}, m_{\tilde e_{LJ}},
m_{\tilde e_{RJ}}) ) - c_W^2 M_2^{\star} \mu^{\star} (
D_0(\left|M_2\right|,\left|\mu\right|, m_{\tilde e_{LI}}, m_{\tilde
  e_{LJ}}) \nn
& +& \left. 2 D_0(\left|M_2\right|, \left|\mu\right|, m_{\tilde
  \nu_I}, m_{\tilde\nu_J}) ) \right) \; \bar M_{LL}^{IJ}\; Y_L^{J} \nn
(F_{hY~RR})^{IJK} &=& - \frac{e^2}{\sqrt{2}v_1c_W^2} \left(v_1 Z_R^{2K}-
v_2 Z_R^{1K}\right) M_1^{\star}\mu^{\star} (D_0
(\left|M_1\right|,m_{\tilde e_{LI}},m_{\tilde e_{RI}}, m_{\tilde
  e_{RJ}}) \nn
&-& D_0 (\left|M_1\right|,\left|\mu\right|, m_{\tilde e_{RI}},
m_{\tilde e_{RJ}}) ) \; \bar M_{RR}^{IJ}\; Y_L^{I}
\label{eq:ly}
\eea

\noindent
3. Contributions decoupling as $v^2/M_{SUSY}^2$.  We neglect here
terms proportional to $\Delta_{LL}$, $\Delta_{RR}$, $\Delta_{LR}'$ as
they are dominated by non-decoupling contributions listed in points 1)
and 2). Only the terms proportional to $\Delta_{LR}^{IJ}$ and
$\Delta_{RL}^{JI*}$ are generated starting at order $v^2/M_{SUSY}^2$.
To simplify the expressions, below we also neglect terms additionally
suppressed by lepton Yukawa couplings (this approximation becomes
inaccurate for large $\mu$ and $\tan\beta\geq 30$, when the diagonal
LR elements of the slepton mass matrix proportional to $\mu Y_l$
become important).
\bea
(F_{hdec~ALR})^{IJK} &=& \left( \frac{e^4}{4 v_1 \sqrt{2} c_W^4 s_W^2}
\left((v_1 Z_R^{1K} - v_2 Z_R^{2K}) M_1^* (2 s_W^2 D_0(|M_1|,m_{\tilde
  e_{LI}},m_{\tilde e_{RJ}},m_{\tilde e_{RJ}}) \right.\right. \nn
&-& (2 s_W^2 - 1) D_0(|M_1|,m_{\tilde
  e_{LI}},m_{\tilde e_{LI}},m_{\tilde e_{RJ}}))\nn
&+& 2 (v_1 Z_R^{1K} + v_2 Z_R^{2K}) (c_W^2 (M_1^* + M_2^*)
E_2(|M_1|,|M_2|,|\mu|,m_{\tilde e_{LI}},m_{\tilde e_{RJ}}) \nn
&-& 2 s_W^2 M_1^* E_2(|M_1|,|M_1|,|\mu|,m_{\tilde e_{LI}},m_{\tilde
  e_{RJ}}))\nn
&+& 2 (v_2 Z_R^{1K} + v_1 Z_R^{2K}) \left(M_1^* \mu^* (c_W^2 M_2^*
E_0(|M_1|,|M_2|,|\mu|,m_{\tilde e_{LI}},m_{\tilde e_{RJ}})\right. \nn
&-& s_W^2 M_1^* E_0(|M_1|,|M_1|,|\mu|,m_{\tilde e_{LI}},m_{\tilde
  e_{RJ}}))\nn
&+& \mu (c_W^2 E_2(|M_1|,|M_2|,|\mu|,m_{\tilde e_{LI}},m_{\tilde
  e_{RJ}})\\
&-&\left. s_W^2 E_2(|M_1|,|M_1|,|\mu|,m_{\tilde e_{LI}},m_{\tilde
  e_{RJ}}))) \right) \nn
%
%
&-& \frac{e^2 v_1^2}{\sqrt{2}c_W^2} Z_R^{1K} M_1^*
\left(\left|A_l^{II}\right|^2 E_0(|M_1|,m_{\tilde e_{LI}},m_{\tilde
  e_{LI}},m_{\tilde e_{RJ}},m_{\tilde e_{RI}})\right.\nn
&+& \left. \left. \left|A_l^{JJ}\right|^2 E_0(|M_1|,m_{\tilde
  e_{LJ}},m_{\tilde e_{LI}},m_{\tilde e_{RJ}},m_{\tilde e_{RJ}})
\right) \right) \sqrt{\bar M_{LR}^{IJ}}\; \nn
(F_{hdec~BLR})^{IJK} &=& - \frac{e^2 v_1^2}{\sqrt{2}c_W^2} Z_R^{1K}
M_1^* E_0(|M_1|,m_{\tilde e_{LJ}},m_{\tilde e_{LI}},m_{\tilde
  e_{RJ}},m_{\tilde e_{RI}}) A_l^{II} A_l^{JJ} \sqrt{\bar
  M_{LR}^{JI}}\; \nonumber
\label{eq:ldec}
\eea

\noindent
4. Contributions decoupling as $M_{h(H)}^2/M_{SUSY}^2$. Here, we do
not show numerically small terms suppressed by lepton Yukawa couplings
or flavour-diagonal $A$ terms:
\bea
(F_{hm~ALR})^{IJK} = \frac{e^2 M_{H_0^K}^2}{\sqrt{2} c_W^2} Z_R^{1K}
M_1^* C_0'(|M_1|,m_{\tilde e_{RJ}},m_{\tilde e_{LI}}) \sqrt{\bar
  M_{LR}^{IJ}}\;
\label{eq:lmh}
\eea
where by $C_0'$ we denote the derivative of $C_0$ over the external
Higgs mass, $C_0' = \frac{\partial C_0}{\partial M_h^2}$
(see~\eq{eq:c0diff}).

\subsection{CP-odd Higgs-lepton vertex}
\label{app:lami}

For the processes considered in this article, the contribution from
the LFV CP-odd Higgs-lepton vertex can become important only in the
case of the three body charged lepton decays and only in the limit of
$M_{SUSY} \gg v$, when photon, $Z^0$ and box contributions
decouple. Thus, we give here only the dominant non-decoupling terms
for this vertex.
\bea
F_A^{IJ} = \frac{1}{(4\pi)^2} \left( F_{And}^{IJ} + F_{AY}^{IJ} +
F_{Am}^{IJ} \right)\;.
\eea
As for CP-odd Higgs vertices, we give the sum of the neutralino and
chargino contributions, the latter appearing only as single term
depending on sneutrino masses in~\eq{eq:lya}:
\smallskip

\noindent
1. Contributions proportional to non-holomorphic $A_l'$ terms:
\bea
(F_{And~ALR}')^{IJ} &=& -\frac{ie^2}{\sqrt{2}c_W^2\cos\beta}
\,\sqrt{\bar M_{LR}^{IJ}}\; M_1^{\star}\;
C_0(\left|M_1\right|,m_{\tilde e_{LI}}, m_{\tilde e_{RJ}}) \nn
(F_{And~LL})^{IJ} &=& -\frac{ie^2}{\sqrt{2}c_W^2 \cos\beta} \, \bar
M_{LL}^{IJ}\; M_1^{\star} D_0 (\left|M_1\right|,m_{\tilde e_{LI}},
m_{\tilde e_{LJ}}, m_{\tilde e_{RJ}}) A_L^{'JJ} \nn
(F_{And~RR})^{IJ} &=& -\frac{ie^2}{\sqrt{2}c_W^2 \cos\beta} \,\bar
M_{RR}^{IJ}\; M_1^{\star} \; D_0 (\left|M_1\right|,m_{\tilde e_{LI}},
m_{\tilde e_{RI}}, m_{\tilde e_{RJ}}) A_L^{'II}
\label{eq:lna}
\eea

\noindent
2. Contributions suppressed by lepton Yukawa couplings:
\bea
(F_{AY~LL})^{IJ} &=& \frac{i e^2}{2\sqrt{2} c_W^2
  s_W^2\cos\beta}\left( s_W^2 M_1^{\star} \mu^{\star} (D_0
(\left|M_1\right|,\left|\mu\right|, m_{\tilde e_{LI}}, m_{\tilde
  e_{LJ}}) \right.\nn
& +& 2 D_0(\left|M_1\right|,m_{\tilde e_{LI}}, m_{\tilde e_{LJ}},
  m_{\tilde e_{RJ}}) ) - c_W^2 M_2^{\star} \mu^{\star}
  (D_0(\left|M_2\right|,\left|\mu\right|, m_{\tilde e_{LI}}, m_{\tilde
    e_{LJ}}) \nn
&+&\left.  2 D_0(\left|M_2\right|, \left|\mu\right|, m_{\tilde \nu_I},
  m_{\tilde\nu_J}))\right) \; \bar M_{LL}^{IJ}\; Y_L^{J} \nn
(F_{AY~RR})^{IJ} &=& \frac{i e^2}{\sqrt{2}c_W^2\cos\beta}
  M_1^{\star}\mu^{\star} (D_0 (\left|M_1\right|,m_{\tilde
    e_{LI}},m_{\tilde e_{RI}}, m_{\tilde e_{RJ}}) \nn
&-& D_0 (\left|M_1\right|,\left|\mu\right|, m_{\tilde e_{RI}},
m_{\tilde e_{RJ}}) ) \; \bar M_{RR}^{IJ}\; Y_L^{I}
\label{eq:lya}
\eea

\noindent
3. Contributions proportional to $M_A^2/M_{SUSY}^2$
(see~\eq{eq:c0diff} for the definition of $C_0'$). As in~\eq{eq:lmh}
we do not show numerically small terms suppressed by lepton Yukawa
couplings or flavour-diagonal $A$ terms:
\bea
(F_{Am~ALR})^{IJK} = - \frac{i e^2 M_A^2\sin\beta}{\sqrt{2} c_W^2}
M_1^* C_0'(|M_1|,m_{\tilde e_{RJ}},m_{\tilde e_{LI}}) \sqrt{\bar
  M_{LR}^{IJ}}\;
\label{eq:lmA}
\eea

\subsection{4-lepton box diagrams}
\label{app:lboxmi}

All genuine box diagram contributions listed in
eqs.~(\ref{eq:lboxa}--\ref{eq:lboxd}) have negative mass dimension and
without any cancellations explicitly decouple like $v^2/M_{SUSY}^2$.
Thus, it is sufficient to expand them only in the lowest order in
chargino and neutralino mass insertions. Also the LR slepton mass
insertions are always associated with additional factors of
$v/M_{SUSY}$. Thus in the leading $v^2/M_{SUSY}^2$ order only LL and
RR slepton mass insertion can contribute to formulae for box diagrams.

Expressions listed below are valid only for $\Delta L=1$ processes,
i.e. excluding combinations of indices $I=J, K=L$ or $I=K, J=L$ - for
these one would also take into account flavour conserving diagrams. As
mentioned in Sec.~\ref{sec:llll}, we do not consider MI expanded
expressions for exotic $\Delta L=2$ processes.

The chargino diagrams contribute significantly only to the $B_{VLL}$,
all other contributions are at least double Yukawa suppressed and very
small. The $B_{VLL}$ term is:
\bea
(4\pi)^2 B_{VLLC}^{JIKL} & =& \frac{e^4}{4 s_W^4}
\left(E_2(|M_2|,|M_2|,m_{\tilde\nu_I},m_{\tilde\nu_J},m_{\tilde\nu_K})
\left(\delta^{KL} \Delta_{LL}^{JI} \bar M_{LL}^{IJ} + \delta^{JL}
\Delta_{LL}^{KI} \bar M_{LL}^{IK} \right) \right. \nn
&+& E_2(|M_2|,|M_2|,m_{\tilde\nu_J},m_{\tilde\nu_K},m_{\tilde\nu_L})
\left(\delta^{IK} \Delta_{LL}^{JL} \bar M_{LL}^{JL}
+\left. \delta^{IJ} \Delta_{LL}^{KL} \bar M_{LL}^{KL}
\right)\right)
\label{eq:cboxmi}
\eea
Contributions arising from neutralino box diagrams, both normal and
crossed added together, are listed below in
Eqs.~(\ref{eq:nboxmi1}-\ref{eq:nboxmi2}).  We do not give here
formulae for the neutralino contributions to $B_{SLL}, B_{SRR},
B_{TL}$ and $B_{TR}$, as they are also double Yukawa suppressed and
small.

\bea
(4\pi)^2 B_{VLLN}^{JIKL} &=& \frac{e^4}{16 s_W^4 c_W^4}
\left(\left(\delta^{KL} \Delta_{LL}^{JI} \bar M_{LL}^{IJ}\;
\right.
+ \delta^{IK} \Delta_{LL}^{JL} \bar M_{LL}^{JL}\; \right) (3
c_W^4 E_2(|M_2|,|M_2|,m_{\tilde e_{LI}},m_{\tilde e_{LJ}},m_{\tilde
  e_{LL}})\nn
&+& 3 s_W^4 E_2(|M_1|,|M_1|,m_{\tilde e_{LI}},m_{\tilde
  e_{LJ}},m_{\tilde e_{LL}}) \nn
&-& 2 c_W^4 D_0(|M_2|,m_{\tilde e_{LI}},m_{\tilde e_{LJ}},m_{\tilde
  e_{LL}}) - 2 s_W^4 D_0(|M_1|,m_{\tilde e_{LI}},m_{\tilde
  e_{LJ}},m_{\tilde e_{LL}}) \nn
&+& 4 s_W^2 c_W^2 \re(M_1 M_2^*) E_0(|M_1|,|M_2|,m_{\tilde
  e_{LI}},m_{\tilde e_{LJ}},m_{\tilde e_{LL}}) \nn
&+& 2 s_W^2 c_W^2 E_2(|M_1|,|M_2|,m_{\tilde e_{LI}},m_{\tilde
  e_{LJ}},m_{\tilde e_{LL}})) \nn
&+& \left(\delta^{JL} \Delta_{LL}^{KI} \bar M_{LL}^{IK} 
+ \delta^{IJ} \Delta_{LL}^{KL} \bar M_{LL}^{KL} \right) (3 c_W^4
E_2(|M_2|,|M_2|,m_{\tilde e_{LI}},m_{\tilde e_{LK}},m_{\tilde e_{LL}})
\nn
&+& 3 s_W^4 E_2(|M_1|,|M_1|,m_{\tilde e_{LI}},m_{\tilde
  e_{LK}},m_{\tilde e_{LL}})\nn
&-& 2 c_W^4 D_0(|M_2|,m_{\tilde e_{LI}},m_{\tilde e_{LK}},m_{\tilde
  e_{LL}}) - 2 s_W^4 D_0(|M_1|,m_{\tilde e_{LI}},m_{\tilde
  e_{LK}},m_{\tilde e_{LL}}) \nn
&+& 4 s_W^2 c_W^2 \re(M_1 M_2^*) E_0(|M_1|,|M_2|,m_{\tilde
  e_{LI}},m_{\tilde e_{LK}},m_{\tilde e_{LL}}) \nn
&+&\left. 2 s_W^2 c_W^2 E_2(|M_1|,|M_2|,m_{\tilde e_{LI}},m_{\tilde
  e_{LK}},m_{\tilde e_{LL}}))\right)
\label{eq:nboxmi1}
\eea
\bea
(4\pi)^2 B_{VRRN}^{JIKL} &=& - \frac{e^4}{c_W^4}
\left(\left(\delta^{KL} \Delta_{RR}^{IJ}
 \bar M_{RR}^{IJ} 
+ \delta^{IK} \Delta_{RR}^{LJ} \bar M_{RR}^{JL} \right) (2
D_0(|M_1|,m_{\tilde e_{RI}},m_{\tilde e_{RJ}},m_{\tilde e_{RL}})
\right. \nn
&-& 3 E_2(|M_1|,|M_1|,m_{\tilde e_{RI}},m_{\tilde e_{RJ}},m_{\tilde
  e_{RL}})) \nn
&+& \left(\delta^{JL} \Delta_{RR}^{IK} \bar M_{RR}^{IK} 
+. \delta^{IJ} \Delta_{RR}^{LK} \bar M_{RR}^{KL} \right) (2
D_0(|M_1|,m_{\tilde e_{RI}},m_{\tilde e_{RK}},m_{\tilde e_{RL}}) \nn
&-& \left.3 E_2(|M_1|,|M_1|,m_{\tilde e_{RI}},m_{\tilde
  e_{RK}},m_{\tilde e_{RL}}))\right)
\eea
\bea
(4\pi)^2 B_{VLRN}^{JIKL} &=& - \frac{e^4}{4c_W^4} ( \delta^{KL}
\Delta_{LL}^{JI} \bar M_{LL}^{IJ} (2 D_0(|M_1|,m_{\tilde
  e_{LI}},m_{\tilde e_{LJ}},m_{\tilde e_{RL}})\nn
&-& 3 E_2(|M_1|,|M_1|,m_{\tilde e_{LI}},m_{\tilde e_{LJ}},m_{\tilde
  e_{RL}})) \nn
&+& \delta^{IJ} \Delta_{RR}^{LK} \bar M_{RR}^{KL} (2 D_0(|M_1|,m_{\tilde
  e_{LI}},m_{\tilde e_{RK}},m_{\tilde e_{RL}})\nn
&-& 3 E_0(|M_1|,|M_1|,m_{\tilde e_{LI}},m_{\tilde e_{RK}},m_{\tilde
  e_{RL}}))
\eea
\bea
(4\pi)^2 B_{VRLN}^{JIKL} &=& - \frac{e^4}{4c_W^4} ( \delta^{KL}
\Delta_{RR}^{IJ} \bar M_{RR}^{IJ} (2 D_0(|M_1|,m_{\tilde
  e_{RI}},m_{\tilde e_{RJ}},m_{\tilde e_{LL}})\nn
&-& 3 E_2(|M_1|,|M_1|,m_{\tilde e_{RI}},m_{\tilde e_{RJ}},m_{\tilde
  e_{LL}})) \nn
&+& \delta^{IJ} \Delta_{LL}^{KL} \bar M_{LL}^{KL} (2 D_0(|M_1|,m_{\tilde
  e_{RI}},m_{\tilde e_{LK}},m_{\tilde e_{LL}})\nn
&-& 3 E_0(|M_1|,|M_1|,m_{\tilde e_{RI}},m_{\tilde e_{LK}},m_{\tilde
  e_{LL}}))
\eea
\bea
(4\pi)^2 B_{SLRN}^{JIKL} &=&  \frac{e^4}{2c_W^4} (\delta^{JL}
\Delta_{LL}^{KI} \bar M_{LL}^{IK} (2 D_0(|M_1|,m_{\tilde
  e_{LI}},m_{\tilde e_{LK}},m_{\tilde e_{RL}}) \nn
&-& 3 E_2(|M_1|,|M_1|,m_{\tilde e_{LI}},m_{\tilde e_{LK}},m_{\tilde
  e_{RL}})) \nn
&+& \delta^{IK} \Delta_{RR}^{LJ} \bar M_{RR}^{LJ} (2
D_0(|M_1|,m_{\tilde e_{LI}},m_{\tilde e_{RJ}},m_{\tilde e_{RL}}) \nn
&-& 3 E_2(|M_1|,|M_1|,m_{\tilde e_{LI}},m_{\tilde e_{RJ}},m_{\tilde
  e_{RL}})))
\eea
\bea
(4\pi)^2 B_{SRLN}^{JIKL} &=&  \frac{e^4}{2c_W^4} (\delta^{JL}
\Delta_{RR}^{IK} \bar M_{RR}^{IK} (2 D_0(|M_1|,m_{\tilde
  e_{RI}},m_{\tilde e_{RK}},m_{\tilde e_{LL}}) \nn
&-& 3 E_2(|M_1|,|M_1|,m_{\tilde e_{RI}},m_{\tilde e_{RK}},m_{\tilde
  e_{LL}})) \nn
&+& \delta^{IK} \Delta_{LL}^{JL} \bar M_{LL}^{LJ} (2
D_0(|M_1|,m_{\tilde e_{RI}},m_{\tilde e_{LJ}},m_{\tilde e_{LL}}) \nn
&-& 3 E_2(|M_1|,|M_1|,m_{\tilde e_{RI}},m_{\tilde e_{LJ}},m_{\tilde
  e_{LL}})))
\label{eq:nboxmi2}
\eea

\clearpage


\bibliographystyle{JHEP}

\bibliography{susylept}

\end{document}